\documentclass{article}

\usepackage{natbib}
\bibpunct{(}{)}{;}{a}{}{,}

\input epsf.tex

\usepackage[breaklinks,colorlinks,citecolor=black,linkcolor=black,urlcolor=black]{hyperref}

\newcommand*{\figref}[1]{Figure~\ref{#1}}
\newcommand*{\figrefs}[1]{Figures~\ref{#1}}
\newcommand*{\tabref}[1]{Table~\ref{#1}}

\newcommand*{\eqnref}[1]{Equation~\ref{#1}}

\newcommand*{\secref}[1]{Section~\ref{#1}}
\newcommand*{\secrefs}[1]{Sections~\ref{#1}}

\newcommand*{\unit}[1]{\ensuremath{\mathrm{\, #1}}}
\newcommand*{\mysub}[2]{\ensuremath{#1_{\mathrm{#2}}}}

\newcommand*{\Msun}{\ensuremath{\, M_{\odot}}}
\newcommand*{\erg}{\unit{erg}}
\newcommand*{\keV}{\unit{keV}}
\newcommand*{\eV}{\unit{eV}}
\newcommand*{\second}{\unit{s}}
\newcommand*{\cm}{\unit{cm}}
\newcommand*{\km}{\unit{km}}
\newcommand*{\Mpc}{\unit{Mpc}}
\newcommand*{\Kelvin}{\unit{K}}

\newcommand*{\kmps}{\ensuremath{\km\second^{-1}}}
\newcommand*{\ergpcmsqps}{\ensuremath{\erg\cm^{-2}\second^{-1}}}

\newcommand*{\E}[1]{\ensuremath{\times 10^{#1}}}

\newcommand*{\ltsim}{\ {\raise-.75ex\hbox{$\buildrel<\over\sim$}}\ }
\newcommand*{\gtsim}{\ {\raise-.75ex\hbox{$\buildrel>\over\sim$}}\ }
\newcommand*{\proptosim}{\ {\raise-.75ex\hbox{$\buildrel\propto\over\sim$}}\ }

\newcommand*{\Omegam}{\mysub{\Omega}{m}}
\newcommand*{\Omegab}{\mysub{\Omega}{b}}
\newcommand*{\Omegac}{\mysub{\Omega}{c}}
\newcommand*{\Omegal}{\ensuremath{\Omega_{\Lambda}}}

\newcommand*{\LCDM}{\ensuremath{\Lambda}CDM}
\newcommand*{\fgas}{\mysub{f}{gas}}
\newcommand*{\Mgas}{\mysub{M}{gas}}
\newcommand*{\Mtot}{\mysub{M}{tot}}
\newcommand*{\rhoc}{\mysub{\rho}{cr}}
\newcommand*{\rhom}{\mysub{\bar{\rho}}{m}}
\newcommand*{\wet}{\mysub{w}{et}}
\newcommand*{\zt}{\mysub{z}{t}}
\newcommand*{\Mnu}{\ensuremath{\sum m_\nu}}

\newcommand*{\dA}{\mysub{d}{A}}
\newcommand*{\ns}{\mysub{n}{s}}
\newcommand*{\Pm}{\mysub{P}{m}}
\newcommand*{\deltac}{\mysub{\delta}{c}}

\newcommand*{\Lx}{\mysub{L}{X}}
\newcommand*{\Tx}{\mysub{T}{X}}
\newcommand*{\Yx}{\mysub{Y}{X}}

\newcommand*{\mproton}{\mysub{m}{p}}

\newcommand*{\Chandra}{{\it{Chandra}}}
\newcommand*{\XMM}{{\it{XMM-Newton}}}
\newcommand*{\Suzaku}{{\it{Suzaku}}}
\newcommand*{\ROSAT}{{\it{ROSAT}}}
\newcommand*{\WMAP}{{\it{WMAP}}}
\newcommand*{\Planck}{{\it{Planck}}}

\usepackage{bm}

\newcommand*{\thetaCosmB}{\ensuremath{\bm{ \theta_\Omega}}}
\newcommand*{\thetaCosm}{\ensuremath{\bm{ \theta}}}
\newcommand*{\alphaCosm}{\ensuremath{\bm{ \alpha}}}

\newcommand*{\kmsmpc}{\ensuremath{\km\second^{-1}\Mpc^{-1}}}

\newcommand*{\msol}{\Msun}
\newcommand*{\msun}{\Msun}

\newcommand*{\Mobs}{\mysub{M}{obs}}

\newcommand*{\zest}{\mysub{z}{est}}

\newcommand*{\dln}{d \, {\rm ln}\,}
\newcommand*{\dlnM}{\dln M}
\newcommand*{\dndlnm}{\frac{dn}{\dln M}} 
\newcommand*{\dndlnM}{\frac{dn}{\dln M}} 
\newcommand*{\Pprim}{P_{\rm prim}}
\newcommand*{\rhocrit}{\rhoc}
\newcommand*{\omegaDE}{\mysub{\Omega}{DE}}
\newcommand*{\OmegaDE}{\omegaDE}
\newcommand*{\DeltaRsq}{\ensuremath{\Delta_{\cal R}^2}}

\newcommand*{\wmap}{\WMAP}
\newcommand*{\gpc}{\unit{Gpc}}
\newcommand*{\mpc }{\unit{Mpc}}
\newcommand*{\kpc }{\unit{kpc}}
\newcommand*{\hinv}{\hbox{$\, h^{-1}$} }
\newcommand*{\kms}{\ensuremath{\km\second^{-1}}}

\newcommand \Svec {\mathbf{S}}
\newcommand \svec {\mathbf{s}}
\newcommand \sbar {\mathbf{\bar{s}}}

\newcommand \mubar {\bar{\mu}}
\newcommand \sigmuone {\sigma_{\mu 1}}
\newcommand \sigmutwo {\sigma_{\mu 2}}
\newcommand \mvec {\mathbf{m}}

\newcommand*{\Ngal}{\hbox{$N_{\rm gal}$}}

\newcommand*{\sigmagal}{\hbox{$\sigma_{\rm gal}$}}

\newcommand*{\siggal}{\mysub{\sigma}{gal}}

\newcommand*{\fICM}{\mysub{f}{ICM}}
\newcommand*{\eg}{e.g.}

\begin{document}

\title{Cosmological Parameters from Observations of Galaxy Clusters}

\markboth{Allen, Evrard \& Mantz}{Cosmological Constraints from Galaxy Clusters}

\author{
  Steven W. Allen\vspace{-0.5ex}\\
  {\small Kavli Institute for Particle Astrophysics and Cosmology,}\vspace{-0.75ex}\\ {\small Department of Physics, Stanford University, 452 Lomita Mall, Stanford, CA 94305-4085, USA}\vspace{-0.75ex}\\ {\small and SLAC National Accelerator Laboratory, 2575 Sand Hill Road, Menlo Park, CA 94025, USA}\vspace{-0.75ex}\\ {\small email: \href{mailto: swa@stanford.edu}{swa@stanford.edu}}\vspace{1ex}\\
  August E. Evrard\vspace{-0.5ex}\\
  {\small Departments of Physics and Astronomy and Michigan Center for Theoretical Physics,}\vspace{-0.75ex}\\{\small  University of Michigan, Ann Arbor, MI 48109, USA}\vspace{-0.75ex}\\ {\small email: \href{mailto: evrard@umich.edu}{evrard@umich.edu}}\vspace{1ex}\\
  Adam B. Mantz\vspace{-0.5ex}\\
  {\small NASA Goddard Space Flight Center, Code 662, Greenbelt, MD 20771, USA}\vspace{-0.75ex}\\ {\small email: \href{mailto: adam.b.mantz@nasa.gov}{adam.b.mantz@nasa.gov}}
}
\date{June 9, 2011}

\maketitle


\begin{abstract}
\noindent

Studies of galaxy clusters have proved crucial in helping to establish
the standard model of cosmology, with a universe dominated by dark
matter and dark energy. A theoretical basis that describes clusters as
massive, multi-component, quasi-equilibrium systems is growing in its
capability to interpret multi-wavelength observations of expanding
scope and sensitivity.  We review current cosmological results,
including contributions to fundamental physics, obtained from
observations of galaxy clusters.  These results are consistent with
and complementary to those from other methods.  We highlight several
areas of opportunity for the next few years, and emphasize the need
for accurate modeling of survey selection and sources of systematic
error. Capitalizing on these opportunities will require a
multi-wavelength approach and the application of rigorous statistical
frameworks, utilizing the combined strengths of observers, simulators
and theorists.
\end{abstract}

\setcounter{tocdepth}{2}
\tableofcontents

\section{INTRODUCTION}\label{sec:intro}

The statistical character of our sky's population of clusters of
galaxies, viewed from radio to gamma-ray wavelengths, is sensitive to
models of cosmology, astrophysics, and large-scale gravity.  Galaxy
clusters are cosmographic buoys that signal locations of peaks in the
large-scale matter density.  The population is shallow and finite.
Surveys in the coming decades will definitively map our universe's
terrain as defined by the highest $\sim \! 10^5$ peaks.  Current maps
have advanced to the stage where Abell 2163, a cluster at redshift $z
\sim 0.2$ with a plasma virial temperature $kT= 12.27 \pm 0.90 \keV$
\citep{Mantz0909.3099} and galaxy velocity dispersion $\siggal = 1434
\pm 60 \kms$ \citep{Maurogordato08}, has been nominated a candidate
for \emph{the} most massive cluster in the universe
\citep{Holz1004.5349}, the cosmic equivalent of Mount Everest.

Physically, galaxy clusters are manifested in the most massive of the
bound structures -- termed {\sl halos} (or {\sl haloes}) -- that
emerge in the cosmic web of large-scale structure (LSS).  
The LSS web is a gravitationally amplified descendant of a weak noise
field seeded by quantum fluctuations during an early, inflationary
epoch \citep{Bond96}.  Its evolutionary dynamics have been well
studied into the non-linear regime by N-body simulations
\citep{Bertschinger98}.  Locally bound regions (the halos) emerge,
initially via coherent infall within a narrow mass range and,
subsequently, via a combination of infall and hierarchical merging
that widen the dynamic range, pushing to increasingly
larger halo masses.  The merging process is of considerable
interest for cluster studies, driving astrophysical signatures that
can test physical models from the nature of dark matter
\citep{Clowe0608407} to the magnetohydrodynamics of hot, dilute
plasmas \cite[\eg][]{Kunz1003.2719}.  But merging also potentially confuses
cosmological studies, by creating close halo pairs that may appear as one 
cluster in projection and by introducing variance into observable signals.

Halos are multi-component systems consisting of dark matter and
baryons in several phases: black holes; stars; cold, molecular gas;
warm/hot gas; and non-thermal plasma.  After decades of study via
N-body and hydrodynamic simulation and related methods \cite[see recent
review by][]{Borgani0906.4370}, models for the detailed evolution of
the baryons in clusters are growing in capability to describe an
increasingly large and rich volume of observations.  What is clear
empirically is that the galaxy formation process is globally
inefficient: a recent study by \citet{Giodini09} finds that stellar
mass accounts for only $12 \pm 2$ percent of the total baryon budget
in the most massive halos.  Radiative cooling of gas is overcome by
feedback from various sources, including mechanical and radiative
input from supernova winds and black hole jets, thermal conduction and
other plasma processes, and ablation and harassment during
gravitational encounters.

While the hierarchical nature of structure formation implies that
galaxy and cluster formation are deeply intertwined and, therefore,
that detailed understanding of cluster structure and evolution {\sl
requires} that we understand galaxy formation, the scales separating
the most massive clusters from the largest galaxies -- roughly a
factor of 100 in length and 1000 in mass -- allow progress to be made
by approximate physical treatments.  The dark matter kinematic
structure, including remnant, fine-scale \emph{sub-halos}
\citep{Moore98, Springel0012055}, as well as the morphology and
scaling behaviors of the hot, intracluster medium (ICM) that dominates
the baryonic component \citep{Evrard90, Navarro95, Bryan9710107}, are examples of
areas where direct simulations made good, early progress.

A key aspect of their multi-component nature is the fact that clusters
offer multiple, observable signals across the electromagnetic spectrum
\citep[e.g.][]{Sarazin88}.  At X-ray wavelengths, the hot ICM emits thermal
bremsstrahlung and line emission from ionized metals injected into the
plasma by stripping and feedback processes.  Stellar emission from
galaxies and intracluster light dominates the optical and
near-infrared.  At millimeter wavelengths, inverse Compton scattering 
within clusters distorts the spectrum of the cosmic microwave background
(CMB). Gravitational lensing offers a unique probe into the total matter
distributions in clusters.  Synchrotron emission from relativistic
electrons is visible at radio frequencies.  These and other signatures
discussed below provide physically coupled, and often observationally
independent, lines of evidence with which to test astrophysical models
of cluster evolution.  A challenge to cluster cosmology  is the construction of accurate
statistical models that address survey observables explicitly 
while incorporating intrinsic property covariance.  

\subsection{Clusters as Cosmological Probes} \label{sec:intro_cosmo}

The use of clusters to study cosmology has a history dating to Zwicky's discovery of dark matter in the Coma Cluster \citep{Zwicky33}.   Brightest cluster galaxies were later employed as standard candles to study the local expansion history of the universe; \citet{Hoessel80} actually derived (with low significance) a negative deceleration parameter using this approach, implying accelerated expansion consistent with present findings.  In the 1980's, measurement of the enhanced spatial clustering of clusters relative to galaxies supported the model of Gaussian random initial conditions expected from inflation \citep{Bahcall83}.   In the early 1990's, an apparent discrepancy between local baryon fraction measurements of clusters \citep{Fabian91, Briel92} with primordial nucleosynthesis expectations helped rule out a model with critical matter density \citep{White93}.  The revelation of hot clusters at high redshift later that decade  \citep{Donahue9707010, Bahcall9803277} presaged the ultimate discovery  of dark energy from Type Ia supernova (SNIa) surveys.  The turn of the millennium witnessed a flurry of activity aimed at 
measuring the amplitude of the matter power spectrum from cluster counts. 
X-ray studies in particular showed that the amplitude was lower than had 
been accepted previously 
\cite[\eg][]{Borgani2001ApJ...561...13B,Reiprich0111285,Seljak0111362,Pierpaoli0210567,Allen0208394,Schuecker0208251}, a result 
later confirmed by CMB and cosmic shear measurements.  These studies also 
exposed the importance of understanding systematic effects associated with 
the use of directly observable quantities as proxies for mass 
\citep{Henry0809.3832}.

Recent studies have used cluster counts or the ICM mass fraction in very massive systems (both methods described in more detail below) to constrain cosmological parameters.  These studies are consistent with other observations that find a universe dominated by dark energy ($73\%$), with sub-dominant  dark matter ($23\%$), and a small minority of baryonic material  ($4.6\%$)  \citep{Komatsu1001.4538}.   A detailed pedagogical treatment of how cluster studies helped establish this reference cosmology is given in the review of \citet{Voit0410173}.  \citet{Rosati0209035} review X-ray studies of clusters from the ROSAT satellite era.

Explaining the nature of the dark energy and dark matter are core problems of physics.   The consensus `concordance' cosmological model, $\Lambda$CDM, postulates that dark energy (DE) is associated with a small, non-zero vacuum energy, equivalent to a cosmological constant term in Einstein's equations.  Another possibility is that DE arises from a light scalar field (or fields) that evolves over cosmic time.  A third option is that DE is essentially an apparition, not a source term of Einstein's general relativistic equations but a reflection of their breakdown at length and time scales of cosmic dimensions  \citep[\eg][]{Copeland0603057}.
Sky surveys of cosmic systems, from supernovae to galaxies to clusters of galaxies, provide the means to discriminate among these alternatives.   

\begin{figure}[t]
  \vspace{-0.5truecm}
  \centerline{
   \epsfxsize=10cm
   \epsfbox{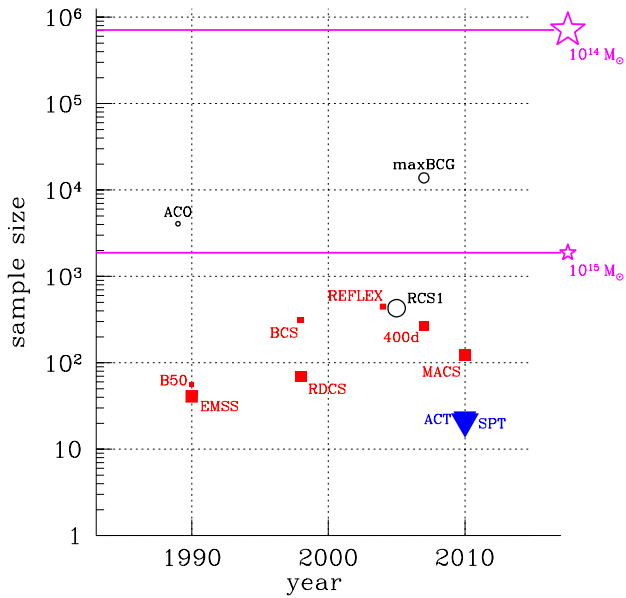}
  }
  \vspace{-1.0truecm}
  \caption{Yields from modern surveys of clusters used for
cosmological studies are shown, with symbol size proportional to
median redshift.  Samples selected at optical (circles), X-ray (red
squares), and mm (blue triangles) wavelengths are discussed in
\secref{sec:obs_surveys} .  Stars and horizontal lines show full sky counts of halos expected in the reference \LCDM\ cosmology (see \secref{sec:theory}) with masses above $10^{15}$ and $10^{14} \msol$.  Such halo samples have median redshifts of $0.4$ and $0.8$, respectively.}
  \label{fig:surveyHistory}
\end{figure}

Forthcoming cluster surveys at mm, optical/near-infrared, and X-ray wavelengths, discussed in \secref{sec:new_surveys}, have the 
potential to find hundreds of thousands of groups and clusters.  \figref{fig:surveyHistory} puts these efforts into historical perspective, by plotting size against year of publication for cluster samples that generated cosmological constraints discussed in this review.   Symbol size is proportional to median sample redshift, and symbol types encode the selection method.   The stars at far right show theoretical estimates of the all-sky number and median redshift of halos with masses above $10^{14} \msol$ and $10^{15} \msol$.  The former mass limit roughly marks the transition from galaxy groups to galaxy clusters, while the latter marks the deepest potential wells with ICM temperatures $kT \gtsim 5 \keV$.  Current surveys have made good progress, but the full population of clusters remains largely undiscovered.

Optical and X-ray surveys have the longest histories, but these traditional methods are being complemented by new approaches.  Space-based surveys in the near-infrared extend optical methods to $z >1$ \citep{Eisenhardt0804.4798, Demarco1002.0160}, and the first few clusters identified by their gravitational lensing signature have been published \citep{Wittman0507606}.   Ongoing mm surveys have released the first sets of clusters discovered through the Sunyaev-Zel'dovich (SZ) effect \citep{Marriage10, Vanderlinde1003.0003, Plancksurvey11}, with the promise of much more to come.

Panoramic, multi-wavelength surveys of common sky areas offer profound improvements to our understanding of clusters as astrophysical systems, which in turn further empowers their  use for cosmological studies.  And while considerable challenges to interpretation and modeling of survey data certainly exist, a halo model framework, discussed in \secref{sec:theory}, is rising to meet this task.  

\subsection{Cosmic Calibration via Simulations} \label{sec:intro_calibration}

A feature common to many techniques that study DE is the nature of the
input data, which consists of catalogs of properties, $\mathbf{x}$, of
discrete objects that lie along our past light-cone.  Upcoming
wide-field  surveys will generate
$\mathbf{x}$-catalogs of large dimension
that will be distilled to constrain perhaps tens of cosmological and
astrophysical parameters.  Such catalogs may contain internal support
through the use of complementary methods: besides galaxy clusters (CL),
the same data set can be analyzed for baryon acoustic
oscillations (BAO) and, for optical surveys, weak 
lensing (WL). (In the case of repeat
observations, optical surveys can also be analyzed for 
Type Ia supernovae and gravitational time delay
signatures.)  Science processing leads to a compressed set of
statistical signals, $\mathbf{y}_i$, where $i$ indicates an 
aforementioned method.  For large cluster surveys, $\mathbf{y}$ might
consist of counts of clusters binned by sky area, redshift and
detected signal.

Extracting accurate constraints on a set of cosmological model parameters, $\thetaCosm$, from these surveys requires sophisticated likelihood analyses.  The critical ingredient is $p(\mathbf{y}_i | \thetaCosm)$, the underlying likelihood that the CL (and other) statistics of the observed sky would be realized within a particular universe.   Key capabilities that enable such likelihood analysis are:
\vspace{-2.0truept}
\begin{enumerate}
\item to predict statistical expectations, $p(\mathbf{y}_i | \thetaCosm)$, for {\sl many} universes, $\thetaCosm$; \\
\vspace{-19.0truept}
\item to extract {\sl unbiased} statistical signals from the raw catalog, $\mathbf{y}_i(\mathbf{x})$; \\
\vspace{-19.0truept}
\item to understand the expected signal covariance, COV($\mathbf{y}_i, \mathbf{y}_j).$
\vspace{-2.0truept}
\end{enumerate}

Genuine understanding of cosmological models from observed cluster
data is dependent on the degree to which theory and simulation can
provide robust predictions for the observed signals.  While numerical
simulations of LSS can predict catalog-level yields for a given cosmology
\citep[\eg][]{Springel0604561},
such predictions necessarily entail additional astrophysical
assumptions, meaning $p(\mathbf{y}_i | \thetaCosm)$ is actually
$p(\mathbf{y}_i | \thetaCosm, \alphaCosm)$, where $\alphaCosm$
represents degrees of freedom introduced by an assumed astrophysical
model.  Recovering cosmological information from survey data therefore
necessitates marginalization over a reasonable range of astrophysical
assumptions.  On the other hand, as cosmological constraints from all
methods improve, the cluster community can potentially invert the
problem, recovering constraints on astrophysical models after
marginalizing over cosmology.

We begin this review by describing the theoretical basis for cluster
cosmology (\secref{sec:theory}), and include there an opening
discussion of important sources of systematic error.  Key
observational windows are described in \secref{sec:observations}, and
recent cosmological constraints are reviewed in
\secref{sec:cosmo_constraints}.  In \secref{sec:physics}, cluster
contributions to particle physics and gravity are examined. In
\secref{sec:opportunities}, we highlight opportunities for important,
near-term progress.  In closing, we emphasize some essential
considerations in survey modeling and analysis (\secref{sec:modeling})
before presenting our conclusions (\secref{sec:conclusions}).

\section{THEORETICAL BASIS} \label{sec:theory}

This section sketches a theoretical description of the halo framework that supports cluster cosmology.  There is considerable richness to the galaxy formation problem that we omit here; the recent review of \citet{Benson1006.5394} provides substantial detail.  Simulation studies of halo evolution into the strongly non-linear regime are becoming increasingly powerful, but finite resolution and uncertainty in astrophysical treatments limit predictive power.  While not yet robust enough to offer sharp prior characterization of the astrophysics required for cosmological studies, simulations offer key insights into the structure and physics sensitivity of the functions that relate  observable signals to halo mass and epoch.  

\subsection{LSS and Halo Formation from Inflation} 
Ample evidence now supports the picture that LSS formed via gravitational amplification of initially small density fluctuations, $\delta \equiv (\rho-\bar{\rho})/\bar{\rho}$.  Cosmic microwave background  anisotropy measurements are consistent with expectations from a large class of basic inflationary models \citep[\eg][]{Baumann0810.3022}.  Such models are characterized by an instantaneous primordial power spectrum, $\Pprim(k) \sim | \delta_k(a)^2| \sim  k^{\ns}$, with spectral index, $\ns$, expected to be close to unity.   Here, $\delta_k$ is the Fourier transform of the density fluctuations, $\delta(\mathbf{x})$.

After inflation ceases, fluctuations in the coupled photon-baryon-dark matter fluid evolve in ways that are now well understood from linearized Boltzmann treatments  \citep{Seljak0306052}.  For the standard case of adiabatic fluctuations, and on scales above the baryon Jeans mass, the post-recombination matter (dark matter and baryons) power spectrum exhibits a growing mode that scales with the cosmic expansion parameter, $a$, as 
\begin{equation}
\Pm(k,a,\thetaCosmB) = G^2(a, \thetaCosmB) \, T^2(k,\thetaCosmB) \, P_{\rm prim}(k) .
\label{eq:powspect} 
\end{equation} 
Here, $T(k,\thetaCosmB)$ is a transfer function that encapsulates evolution before recombination at $z \sim 1100$, $G(a, \thetaCosmB)$ is the density perturbation growth factor from linear theory, and $\thetaCosmB$ is the controlling parameter set of the background cosmological model.   Dark energy models that involve modifications to general relativity may introduce $k$-dependence into the growth function above.  

The power spectrum in the reference \LCDM\ model is set by present-epoch energy densities, $\thetaCosmB = \{ \Omegab h^2, \Omegac h^2, \Omegal \}$, where $h = H_0/100 \kmsmpc$ is the dimensionless Hubble constant and $\Omega_X \equiv \rho_X/\rhocrit$ is the density of component $X$ relative to the critical density, $\rhocrit = 3H_0^2/8 \pi G$.    The curvature density, $1- \sum_X \Omega_X$, is zero to within $\pm 0.007$ \citep{Komatsu1001.4538}, consistent with a flat spatial metric on cosmic scales.   Our notation uses `b' for baryons, `c' for cold, dark matter (CDM), and `m' for all matter: $\Omegam = \Omegab + \Omegac$.   In the minimal model, the dark energy is a vacuum energy with equation of state, $p = - \rho c^2$.  We use $\Omegal$ for this case and employ $\OmegaDE$ when referring to models wherein the dark energy equation of state, $w = p / (\rho c^2)$, differs from $-1$.   

\begin{table}
  \center
  \caption{Flat \LCDM\ Parameters$^a$ from \wmap+BAO+$H_0$} \vspace{1.0ex}
  \label{tab:WMAPcosm}
  \begin{tabular}{cc}
    \hline\hline \vspace{-2.5ex}\\
    Parameter & Value  \\
    \hline \vspace{-2.25ex}\\
    $\Omegal$  & $0.725 \pm 0.016$   \vspace{0.5ex} \\
    $\Omegac$    & $0.229 \pm 0.015$   \vspace{0.5ex} \\
    $\Omegab$    & $0.0458 \pm 0.0016$  \vspace{0.5ex} \\
    $h$  & $0.702 \pm 0.014$ \vspace{0.5ex} \\
    $\ns$  & $0.968 \pm 0.012$ \vspace{0.5ex} \\
    $10^{10} \DeltaRsq(k_0)$ & $0.2430 \pm 0.0091$ \vspace{0.5ex} \\
     $\sigma_8$ & $0.816 \pm 0.024$ \vspace{0.5ex} \\
    \hline 
  $^a$ From \cite{Komatsu1001.4538}.
  \end{tabular}
\end{table}

Current constraints for a flat \LCDM\ model from CMB measurements, combined with angular clustering of red galaxies and local measurements of $H_0$, are shown in Table~\ref{tab:WMAPcosm} \citep{Komatsu1001.4538}.  The parameter $\DeltaRsq(k)$ is the  variance in density fluctuations evaluated at horizon crossing, which is independent of $k$ for $\ns = 1$, and the wavenumber $k_0 = 0.002 \mpc^{-1}$ corresponds to a large comoving length scale, $\sim \pi/k_0 = 1.6 \gpc$.  

In the minimal model, the matter density fluctuations filtered within a sphere of comoving radius $R$ are Gaussian distributed with zero mean.  The comoving radius defines a mass, $M = (4\pi/3) \rhocrit R^3$,  
of matter within that radius in the young universe, when $\Omegam(a) = 1$.  Early observations that the variance in galaxy counts is near unity on a scale of $R= 8\hinv\mpc$ led to this as a  conventional choice of scale at which to quote the fluctuation amplitude (see Table~\ref{tab:WMAPcosm} ).  The corresponding mass, $M = 0.59 \times 10^{15} \hinv \Msun$, is characteristic of rich clusters of galaxies.  

The variance of linearly-evolved, CDM fluctuations, filtered on mass scale $M$, has the form 
\begin{equation}
\sigma^2(M,a)  =  \int \, \frac{d^3k}{(2\pi)^3} \ W^2(kR) \ \Pm(k,a) .
\label{eq:sigmasq} 
\end{equation} 
where the filter function is $W(y) = 3[\sin(y)/y^3 - \cos(y)/y^2]$ for the typical case of sharp (or top-hat) spatial filtering within radius $R$.  
Evaluating \eqnref{eq:sigmasq} at $8 \hinv \mpc$ and $a=1$ produces the oft-quoted matter power spectrum normalization parameter, $\sigma_8$.   We will see below that $\sigma(M,a)$ serves as a similarity variable for expressing model-independent forms of the halo space density and clustering.  

The evolution of the fluctuation spectrum, \eqnref{eq:powspect}, is valid at early times or at scales sufficiently large so that $\sigma(M,a) \ll 1$ at all times.  On small scales, where CDM power spectra are generically maximum, fluctuation growth produces $\delta \ge 1$, and linear theory breaks down.  Mode-mode coupling terms become important to the dynamics, and solutions in Fourier space become difficult.   
While higher-order perturbation theory solutions can extend analytic evolution to later times than linear theory \citep[\eg][]{Bernardeau0112551, Crocce0509418}, the full problem is typically treated using N-body simulations, discussed below.  

A recent analytical advance considers LSS as an effective fluid.  \citet{Baumann10042488} show that integrating out small-scale, non-linear structures renormalizes the cosmological background and introduces dissipative terms, of order $v^2/c^2$, into the dynamics of large-scale modes, with $v$ the typical velocity dispersion of collapsed halos.  Since even the most massive halos have $v < 0.01c$, the magnitude of these effects is very small.  Furthermore, \citet{Baumann10042488} show that virialized halos decouple completely from large-scale dynamics, at all orders in the post-Newtonian expansion.

\subsubsection{HALO MODEL DESCRIPTION OF LSS} \label{sec:theory:halomodel}

Astrophysical structures, from the first stars at high redshift to galaxy clusters at low redshift, tend to emerge from local maxima of the filtered density field.  While density peaks are generally non-spherical \citep{BBKS86}, a first-order description considers them spherical and isolated from their surroundings.  Birkhoff's theorem then implies that the expansion histories of radial mass shells within a peak follow trajectories perturbed from the overall background, with sufficiently dense shells expanding to a maximum size and then contracting.  The traditional \emph{ansatz} assumes collapse by a radial factor of two \citep{GunnGott72}, after which a quasi-virialized and quasi-hydrostatic structure -- a perfectly spherical \emph{halo} -- is born.   

The collapse criterion is that the \emph{linearly-evolved} perturbation amplitude reach a critical value, $\delta(a) = \deltac$, with $\deltac = 1.686$ the conventional choice.  Applying this idea to the CDM spectrum, \eqnref{eq:sigmasq}, leads to a characteristic mass scale, $M_\ast(a)$, defined by $\sigma[M_\ast(a), a] = \deltac$.  At a given epoch, a spectrum of halo masses exist, with masses above (below) $M_\ast(a)$ forming from perturbations with amplitudes above (below) the \emph{rms} level of the filtered Gaussian spectrum.  
Considerable literature \citep[\eg][and many others]{PressSchechter74, Bond91, BondMyers96, ShethTormen9901122} has established this picture as the \emph{halo model} of large-scale structure.  We review here only aspects relevant for cluster cosmology; a more thorough review can be found in \citet{Cooray0206508}.  

The basic element of the halo model is the population mean space density, $n(M,z)$, in units of number per unit comoving volume, commonly referred to as the \emph{mass function}.   Expressed as a differential function of mass, it takes the form
\begin{equation}
\dndlnm =  \frac{\rhom}{M} \, \left| \frac{\dln \sigma}{\dln M} \right| \, f(\sigma) , 
\label{eq:massFtnBasic} 
\end{equation} 
where $\rhom = \Omegam \rhocrit$ is the comoving mean matter density and $f(\sigma)$ is a model-dependent function of the filtered perturbation spectrum, \eqnref{eq:sigmasq}.  Analytic forms for $f(\sigma)$ capture much, but not all, of the behavior seen in N-body simulations, as discussed below.  

The spatial clustering of halos is described by a modified version of the matter power spectrum.  On large spatial scales, or low wavenumbers, the halo autocorrelation power spectrum is modified, 
\begin{equation}
P_{\rm hh}(k,a) =  b^2(M,a) \, \Pm(k,a) , 
\label{eq:corrFtnBasic} 
\end{equation} 
where $b(M,a)$, the halo \emph{bias function}, is independent of $k$, for the case of Gaussian fluctuations, but dependent on mass and epoch.  While this expression applies to the spatial autocorrelation of systems with fixed mass $M$, it generalizes to the cross-correlation between sets of halos at different masses, $P_{\rm h1h2}(k,a) =  b(M_1,a) b(M_2,a) \, \Pm(k,a)$.   The theory of peaks in Gaussian random fields expresses the bias as a function of the normalized peak height, $\nu = \deltac / \sigma(M,a)$ \citep{Kaiser84, BBKS86}.   

Below, we show that N-body simulations support the forms of equations~(\ref{eq:massFtnBasic}) and (\ref{eq:corrFtnBasic}),  but a precise fit to the mass function requires that $f(\sigma)$ be adjusted to include explicit redshift dependence, $f(\sigma,z)$.   There are subtleties to the definition of mass in simulations that must also be taken into account.

\subsubsection{ASTROPHYSICAL PROCESSES} \label{sec:theory_astrophysics}

Various astrophysical processes play out within the photon-baryon components of the evolving cosmic web, including hydrodynamic, magnetohydrodynamic and radiative transfer effects; star and black hole formation with associated feedback of momentum, energy and entropy; and so on.  Except for the immediate vicinity of black holes, these processes involve classical physics that is largely known.  But the fully three-dimensional and non-linear nature of the problem, the wide dynamic range in length and time scales, and the non-trivial couplings among the constituent physical processes introduce tremendous complexity into baryon evolution.  Galaxy formation is truly a Grand Challenge computational problem.  We touch on select issues relevant to the observable features of galaxy clusters.

{\sl Shocks and turbulent MHD heating.}   During halo formation, gravitational potential energy in the baryonic component is thermalized via shocks.  The highest Mach numbers, of tens or more, should occur in the accretion shocks at the edges of clusters \citep[\eg][]{Pfrommer06}.  While these strong shocks are expected to be efficient particle accelerators, recent observations place tight limits on the volume-averaged pressure contributions from relativistic particles \citep{Ackermann10}.     Shocks with Mach numbers of a few are also associated with major mergers: a spectacular example is  the narrow radio relic in the cluster CIZA J2242.8+5301, for which \citet{vanWeeren1010.4306} use multi-frequency radio and polarization observations to infer a Mach number $4.6 ^{+1.3}_{-0.9}$ in a shock located $1.5 \mpc$ from the cluster center.  Most of the energy thermalized during cluster formation, however, is dissipated in weak shocks that are persistently driven by dissipating sub-structures and ongoing minor mergers.  Shocks are also driven by jets from AGNs and, at earlier times, by winds from star forming galaxies.  

Details of the nano-parsec scale physics that drive thermalization remain under active study, especially the roles of magnetic fields, turbulence and plasma instabilities \citep[\eg][]{Kunz1003.2719}.   Observations and simulations discussed below indicate that thermalization is efficient; thermal pressure supplies the bulk of support against gravity within the halo potential except during brief periods near periapsis of major mergers.  

{\sl Radiative cooling.} Since the intracluster plasma
(and, to a lesser extent, its interstellar counterpart) is optically thin at most wavelengths, radiation loss is the primary cooling mechanism for the baryonic component of halos.  Indeed, the classic criterion for setting an upper bound on galaxy size comes from balancing the gas cooling time against the halo dynamical time \citep{WhiteRees78}.  
The first generation of stars form at $z \sim 30$, aided by molecular hydrogen line emission, within halos of mass $\sim \! 10^6 \msun$ \citep{Abel0112088, Bromm0905.0929}.  By $z \sim 10$, atomic line cooling in halos with virial temperatures above $10^4$~K produces the first generation of galaxies, which grow hierarchically for a time determined by the large-scale environment.  Proto-cluster regions have more efficient cooling at high redshift than do proto-voids, but the feedback from vigorous, early production of compact sources helps to quench star formation before a large fraction of baryons are converted to stars.  

The cooling timescale of the gas in massive halos is typically longer than a Hubble time, except for a subset of systems that exhibit cool cores.  The central $\sim 100 \kpc$ region of such systems tends to be X-ray bright and typically contains a dominant elliptical galaxy.  We discuss aspects of cool core phenomenology in \secref{sec:future:coreevol}.   

{\sl Star and black hole formation.}  Cold, molecular gas fuels star formation.  The star formation rate can roughly be considered as proportional to the local rate of gas cooling below $10^4$~K, but there are other considerations.  Different venues for star formation exist, ranging from quiescent disks to the bulges of tidally-triggered starburst galaxies, and it is not yet clear whether a single model based on local gas conditions captures the full range of observed behavior.    
Supermassive black hole (SMBH) growth occurs through mergers and accretion in galactic cores, and these central engines drive quasar and radio jet activity  \citep[\eg][]{diMatteo0502199}.  Sloan Digital Sky Survey 
(SDSS) quasar studies indicate that SMBHs of mass $\sim 10^{9} \msun$ exist at $z =7$ \citep{Fan06}, and processes for forming such large black holes in the first few hundred million years of the universe have been proposed \citep{Volonteri1003.4404}.

In Gaussian random fields, small-scale peaks are more abundant when embedded within large-scale peaks, so the largest galaxies and quasars at high redshift represent the progenitors of massive galaxies observed in low redshift clusters.  

{\sl Feedback from compact sources.}  Feedback of mass, momentum, and entropy from stellar/SMBH sources is important at all stages of the LSS hierarchy.  Photoionization and supernova-driven winds serve to limit cooling and star formation in low-mass halos \citep{DekelSilk86}.   Jets driven by accretion onto the central SMBH appear to be required to limit the maximum size of galaxies \citep[\eg][]{Croton0508046, Cattaneo0907.1608}.   Formulations for this feedback typically tie the energy input to the mass accretion rate which, in turn, is governed by the local rate of cooling and/or cold accretion.  

\begin{figure}
  \centerline{
    \epsfxsize=9cm
    \epsfbox{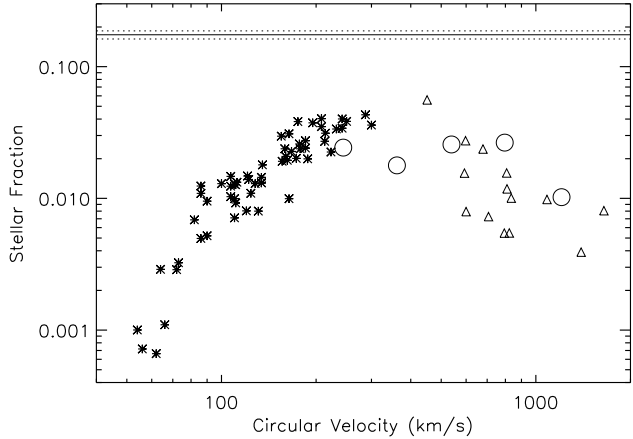}
  }
  \caption{The observed stellar mass fraction as a function of halo circular velocity for systems ranging from galaxies to rich clusters indicates that star formation efficiency peaks in halos of mass $\sim \! 10^{13} \hinv \msol$.  From \citet{Dai0911.2230}.
  }
  \label{fig:dai2010fstarvcirc}
\end{figure}

The end result of this competition between cooling and heating is that heating largely wins.  The overall efficiency of star formation is small, and peaks in halos of roughly galactic scale \citep[e.g.][]{Moster0903.4682}.  \figref{fig:dai2010fstarvcirc} shows a recent compilation of stellar mass fraction ($M_{\rm star}/M$) measurements as a function of halo circular velocity, $v_{\rm circ} = \sqrt{GM/r}$, with $M$ the total halo mass and $r$ its radius \citep{Dai0911.2230}.  The horizontal lines show the cosmic baryon fraction, $\Omegab/\Omegam =  0.171 \pm 0.009$, derived from Wilkinson Microwave Anisotropy Probe (\WMAP{}) data analysis \citep{Dunkley0803.0586}.  

The stellar mass fraction is maximized at a few tens of percent of the cosmic mean in halos with $v_{\rm circ} \sim 300 \kms$, equivalent to a mass of $10^{13} \hinv\msol$ at $z=0$.  In cluster-sized halos, the stellar fraction declines with mass, taking on values $\sim 10\%$ of the global baryon fraction at the highest masses.  Yet, the largest galaxies are found in the cores of massive clusters, and their very old stellar populations produce a characteristically narrow `red sequence' in a color-magnitude diagram of cluster members.  

{\sl Dynamical and thermodynamical equilibrium.}   In the context of the evolving cosmic web, the processes above must  contend with conditions imposed by halo merging.  At any given time, major mergers, such as those involving progenitor pair mass ratios larger than $0.3$, occur in $\sim 10\%$ of the population, concentrated toward the highest masses.  These rare events can drive the mass contents of a halo considerably out of equilibrium.  

Minor mergers, while much more frequent, are also less damaging.  Current simulations and observations indicate that the dynamical and thermodynamical response of halos is quite fast.  Hydrostatic and virial equilibrium assumptions are typically valid to within roughly ten percent for the majority of the cluster population \citep[\eg][]{Rasia06, Nagai0609247}.  

All this astrophysical evolution offers a treasure trove of observational possibilities.  
Uniquely in massive clusters, all of the matter is readily observed, allowing a
complete census to be taken.  Stars make up 1--3\% \citep{Lin04, Gonzalez07, Giodini09}; $\sim \! 15\%$
resides in the hot, diffuse, intergalactic gas \citep{Allen0706.0033,
Simionescu10}; and the rest is in the form of non-baryonic CDM
(Section~\ref{sec:physics_dm}).

\subsection{Cosmological Tests with Massive Halos}

As tracers of massive halos, galaxy clusters provide a number of signatures that are sensitive to the underlying cosmology.  We review here the principles underlying key methods.   A typical set of cosmological parameters for such studies might consist of the primordial spectrum amplitude and slope, the present-epoch densities of the three energy components dominant at late times, the dimensionless Hubble constant, and the DE equation of state parameters, 
\begin{equation}
 \thetaCosm \ = \ \left\{ \ns, \DeltaRsq, \Omegab h^2, \Omegac h^2, \OmegaDE, h, w_0, w_a \right\} , 
\label{eq:thetaCosm} 
\end{equation} 
where the last two parameters define a linearly-evolving DE equation of state,
\begin{equation}
  w(a) = w_0 + w_a (1-a).
  \label{eq:wwa}
\end{equation}
This particular set is meant to be illustrative.  There is considerable variation in the literature, and many works restrict analysis to a flat cosmology, which removes one degree of freedom from the above through the condition $\Omegab + \Omegac + \OmegaDE = 1$.

\subsubsection{HALO COUNTS AND CLUSTERING} \label{sec:theory:countsclustering}

The yield of upcoming cluster surveys will be sufficiently large to enable disaggregation by angular position, redshift, and the observed signal, $S$.  (Note the latter is also referred to in the literature as the \emph{mass proxy}, or sometimes the \emph{observable mass}, $\Mobs$.) 
Complications associated with the signal--mass likelihood and with redshift estimation are discussed below.  As a starting point, consider a perfect tracer of mass, $S=M$, with error-free redshifts, $\zest = z$.   Within a given survey, the expected number of halos, $\bar{N}_{ai}$, in a cell 
described by mass bin $a$ and redshift bin $i$ with solid angle $\Delta\Omega$ is 
\begin{equation}
\bar{N}(M_a, z_i)  \equiv  \bar{N}_{ai} =  \frac{\Delta\Omega}{4 \pi}  \int_{z_i}^{z_{i+1}} dz \, \frac{dV}{dz}  \ \int_{\ln M_a}^{\ln M_{a+1}} \, \dlnM \ \dndlnM .
\label{eq:mbar} 
\end{equation} 
Cosmology enters this expression through the mass function and the volume element, $dV/dz$.  

The counts in each large spatial bin will deviate from the mean by an excess number, $b(M_a,z_i) \delta(\mathbf{x})$,  determined by the local large-scale density field, $\delta(\mathbf{x})$.  Following  \citet{Cunha1003.2416}, the spatial covariance of the counts is 
\begin{equation}
C^a_{ij}  = \left\langle \left(N_{ai} - \bar{N}_{ai}\right)\left(N_{aj} - \bar{N}_{aj}\right) \right\rangle =  \bar{N}_{ai} \bar{N}_{aj} \xi^a_{ij} ,
\label{eq:Sij} 
\end{equation} 
where $ \xi^a_{ij}$ describes the spatial correlation between 
  mass-redshift bins,
\begin{equation}
\xi^a_{ij}  = \int \, \frac{d^3k}{(2\pi)^3} \left|W_i(\mathbf{k}) W_j(\mathbf{k})\right| f\left({ \mathbf{k} \cdot \bm{\Delta} \mathbf{x} }\right) \, b_{ai} b_{aj} \, \Pm(k,z) .
\label{eq:xipixel} 
\end{equation} 
Here, $W_i$ is the window function for cell $i$ (that, when present, can include the effects of redshift estimate uncertainties) and $f$ is a geometric term that depends on the comoving separation, $\bm{\Delta} \mathbf{x}$, between cells $i$ and $j$.   When cells $i$ and $j$ sample different redshifts, an accurate approximation uses their geometric mean to evaluate $\Pm(k,z)$ \citep{Cunha1003.2416}.

Combining the spatial clustering with a diagonal shot noise term forms the full covariance for a survey sample.  Derivatives of the mean counts and covariance with respect to model parameters form the Fisher information matrix used in survey forecasts.   Expressions for the Fisher matrix can be found in \citet{Hu0602147}.   

Equations~(\ref{eq:mbar}) through (\ref{eq:xipixel}) serve as the foundation of likelihood analysis of large cluster surveys.  To be useful in practice, these expressions must undergo a number of modifications, including: transformation from mass to the signal used for cluster detection, $p(S|M,z)$; inclusion of counting errors arising from incompleteness (missed sources) and impurities (false sources); and inclusion of photometric uncertainties, $p(\zest | z)$.   We discuss these issues in \secref{sec:theorytopractice} and summarize current results in \secref{sec:constraints:growth}.

\subsubsection{BARYON FRACTION AS A STANDARD QUANTITY} \label{sec:theory:fgas}

The mass fraction of hot gas, $\fgas$, measured within a characteristic radius of a halo at redshift $z$ can be written as 
\begin{equation} \label{eq:constraints:fgas_model}
  \fgas(z) = \Upsilon(z) \left( \frac{\Omegab}{\Omegam} \right),
\end{equation}
where $\Upsilon(z)$ accounts for star formation and other baryon
effects within that radius.  At large radii in the most massive halos,
where the hot ICM dominates the baryon budget and the impacts of
feedback processes are modest, baryon losses are small and
$|1-\Upsilon| \ltsim 0.1$ is a reasonable expectation.

Motivated by the growing body of measurements of $\fgas$ from the \ROSAT{} X-ray satellite, \citet{Sasaki9611033} and \citet{Pen9610090} recognized that a mismatch in the dependence on metric distance, $d$, between gas mass ($\propto d^{5/2}$) and total mass ($\propto d$) measured from X-ray observations implied that gas fraction measurements in massive clusters could be exploited as a distance estimator, with $\fgas(z) \propto d(z)^{3/2}$.   Like Type Ia supernovae, massive clusters serve as standard calibration sources that test the expansion history of the universe.   Key benefits, relative to survey counts, are the ability to perform this test with a relatively small number of clusters and the relative insensitivity to cluster selection.   We summarize results from this exercise in \secref{sec:constraints:fgas}.

\subsubsection{DISTANCES FROM JOINT X-RAY AND SZ OBSERVATIONS} \label{sec:theory:xsz}

In a similar vein, \citet{WhiteSilk78} noted that X-ray and SZ
measurements could be combined to determine distances to clusters.  
The CMB spectral shift is governed by the Compton y-parameter, a measure of the electron pressure along the line of sight, $y \propto \int dx \, n_{\rm e}(x) T(x)$.  
Given an observed SZ signal, $y_{\rm obs}$, and a predicted signal based 
on X-ray measurements of the ICM density and temperature,
$y_{\rm pred}$, the angular diameter distance scales as
\begin{equation}
  \dA \propto \left( \frac {y_{\rm obs}}{y_{\rm pred}} \right)^2.
  \label{eq:dAXraySZ}
\end{equation}
The cosmological constraint originates from the distance
dependence of the X-ray measurements, $y_{\rm pred}(z) \propto
d(z)^{1/2}$, and the requirement that $y_{\rm pred}=y_{\rm
obs}$. Accurate SZ and X-ray flux and temperature calibration are
particularly important to this method, referred to below as XSZ.

\subsubsection{ANGULAR THERMAL SZ POWER SPECTRUM } \label{sec:theory:tsz}

The thermal and kinetic SZ signals from clusters (\secref{sec:obs_sz}) cause distortions in the CMB at small
angular scales ($\ell \! \sim \!  1000$).  If the distortion pattern from a single halo of mass $M$ at redshift $z$ is described by an angular Fourier transform, $\tilde{y}(M,z,\ell)$, then the full halo population will generate a fluctuation spectrum \citep{Shaw1006.1945}
\begin{equation}
C_\ell  \propto   \int \, dz \, \frac{dV}{dz}  \ \int \, \dlnM \ \dndlnM \ \tilde{y}^2(M,z,\ell) .
\label{eq:Cell} 
\end{equation} 
Adding halo spatial correlations gives a small correction to this estimate \citep{Komatsu0205468}.   This approach to testing cosmology is limited by degeneracy with astrophysical assumptions, as the interplay between $\tilde{y}(M,z,\ell)$ and $dn/\dlnM$ makes clear. 

\subsubsection{BULK FLOWS} \label{sec:bulk_flow}

Measurements of the cosmic peculiar velocity field contain additional
cosmological information (e.g. \citealt{Strauss9502079} and references
therein). The kinetic SZ effect (Section~\ref{sec:obs_sz}) in
principle offers a way to measure the peculiar velocities of galaxy
clusters. Although some initial results based on such measurements
have been reported \citep[e.g.][]{Kashlinsky0809.3734,
Kashlinsky0910.4958, Keisler0910.4233, Osborne1011.2781}, the
technique has not yet reached the maturity of those discussed
above and is not discussed further in this review.

\subsection{Halo Model Calibration via Simulations}

N-body simulations of a single, collisionless dark matter fluid offer the means to investigate non-linear evolution of LSS under an implicit `light-traces-mass' assumption.   The technology supporting such simulations has advanced to the state where $N=10^{12}$ is available (but not yet realized) on peta-scale computational platforms \citep{Pope2010}.  Employing larger-$N$ simply to model bigger volumes is a natural mode of growth, since parallelization is relatively simple (large-volume domain decomposition minimizes the particle transfer among computational nodes), the number of timesteps is independent of $N$, and the light-traces-mass assumption is easier to justify under modest mass and force resolution.  Large-volume simulations produce generous halo population realizations with which to calibrate the mass function and clustering of halos, and current state-of-the-art studies employ ensembles of $10^{9-10}$-particle simulations.  

Coupled N-body and gas dynamic simulation methods enable multi-fluid studies that break free of the light-traces-mass assumption.  Indeed, the first  application of this class of codes tested the possible separation of baryons and neutrinos within clusters formed in a universe dominated by massive neutrinos  \citep{EvrardDavis88}.   The field has advanced considerably since then, and we refer the reader to \citet{Borgani0906.4370} for a  recent review.  We discuss primarily dark matter simulations here, with some relevant multi-fluid simulations results presented in the next section.

\subsubsection{MEASURES OF HALO MASS} \label{sec:theory:masses}

Through the mass function, halo mass provides the critical measure that connects observables to  the underlying cosmology.   But halos are complex, dynamic structures that confound attempts at a  unique definition of mass.  

In the model of spherical collapse applied to initial density peaks, the halo edge and interior mass  are readily defined by the outermost caustic in dark matter or by the location of the shock in cold baryonic accretion  \citep{Bertschinger85}.
In both cases, this radius marks an abrupt transition in the mean radial velocity, separating a nearly hydrostatic interior from an infall-dominated exterior.  Halos forming in 3-D simulations deviate from this ideal case in important ways, some of which can be described by higher-order analytic approaches to peak evolution \citep{BondMyers96}.  The collapse process is more ellipsoidal than spherical, and merging competes with smooth accretion as the dominant mode of halo growth \citep[e.g.,][]{Fakhouri0906.1196}.   Defining centers and boundaries in this complex environment has become a matter of convention.

Two common algorithmic conventions have emerged: (i) percolation, also known as friends-of-friends (FOF), and  (ii) spherical overdensity (SO).  FOF first links all pairs of particles within a given distance, $b$, then merges them into groups based on a shared link condition (`a friend of a friend is a friend').   The SO approach first filters the particle field to identify peaks, then grows spheres around peaks with sizes determined by an interior density threshold, $3M(<r_\Delta ) /(4 \pi r_\Delta^3) = \Delta \rho_{\rm t}$.  The threshold density $\rho_{\rm t}$ is typically chosen to be either the background matter density, $\rho_{\rm t} = \rhom(z)$, or the critical density, $\rho_{\rm t} = \rhocrit(z)$. Unless otherwise specified, we adopt the latter convention in this article.

\begin{figure}
  \centerline{
    \epsfxsize=4.75cm
    \epsfbox{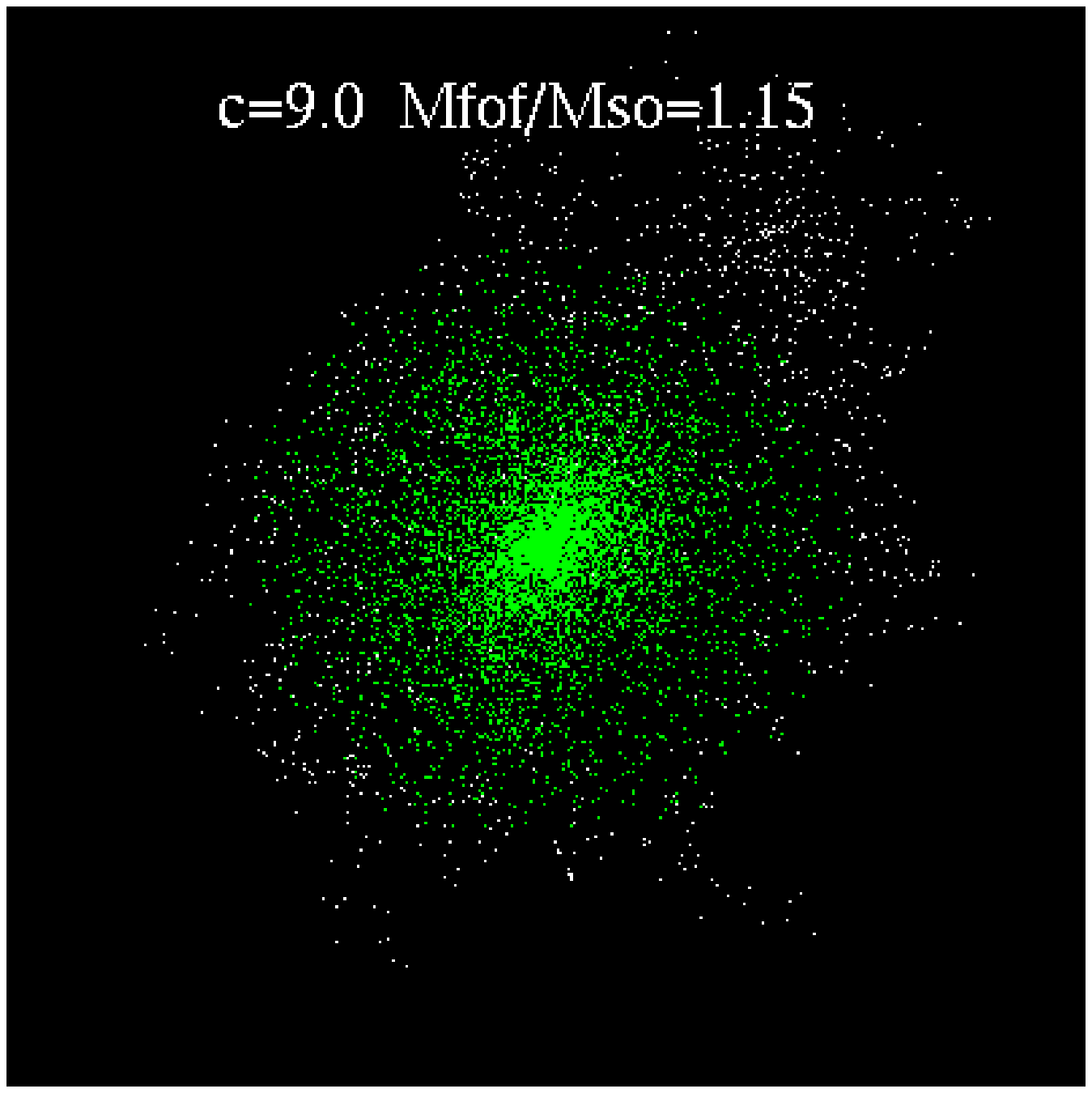}
    \epsfxsize=4.75cm
    \epsfbox{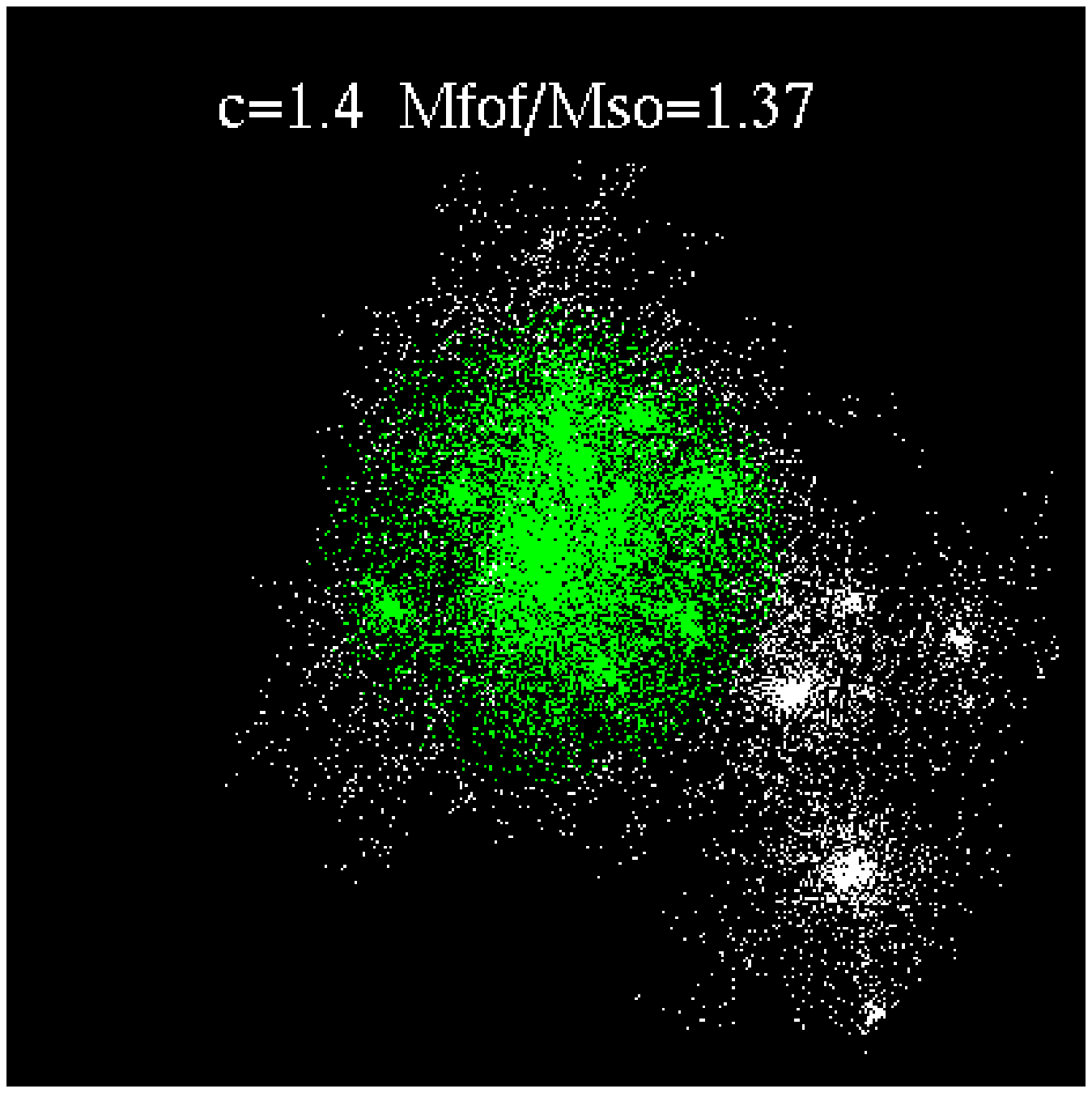}
    \epsfxsize=4.75cm
    \epsfbox{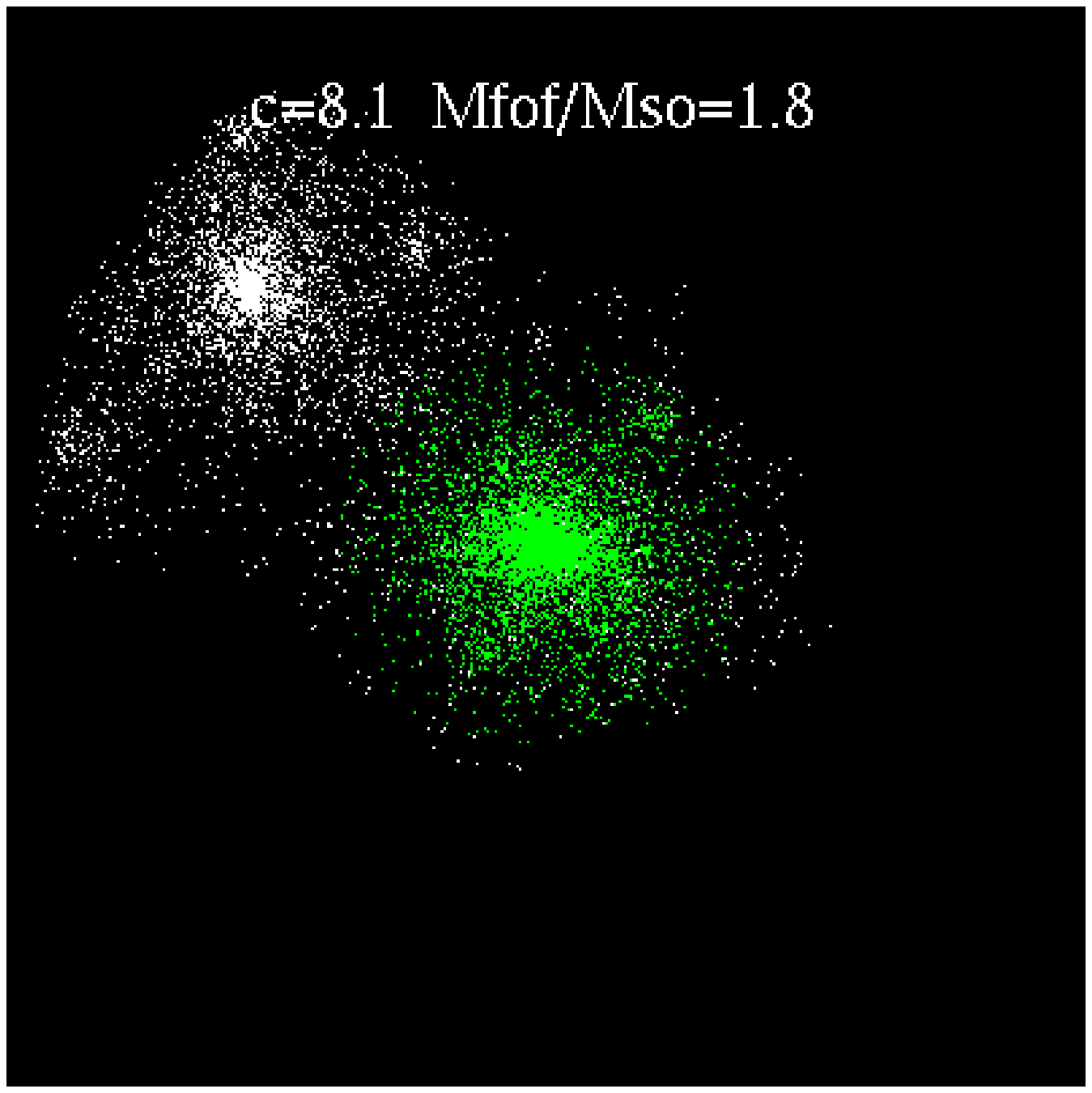}
  }
  \caption{Three examples of halos identified under both FOF (white and green colored particles) and SO (green only) algorithms.   The mass ratio for each case is given, as is the concentration parameter, $c$, derived from a radial density fit.  See text for details.  Adapted from \citet{Lukic0803.3624}.}
  \label{fig:lukic09}
\end{figure}

Several studies discuss the relative merits of these approaches and argue values for the parameters $b$ and $\Delta$ \citep[\eg][]{ColeLacey96, White0011495, Lukic0803.3624}.  \figref{fig:lukic09} provides a visualization of three simulated halos spanning a range of dynamical and morphological behaviors.  In each panel, white particles are members of the FOF halo with $b = 0.2 (V/N)^{1/3}$ while green are SO members using $\Delta = 200$ against $\rhocrit(z)$.  These are typical of parameter values used in the literature.  The left panel shows a relatively isolated system where the two methods give fairly consistent results.  The other panels show two discrepant cases; in the middle is a highly-structured, active merger while, at right, percolation across a filamentary bridge links two similarly sized systems that are just beginning to merge.  

The discrepant cases do not dominate in number, but neither are they uncommon.  For cosmological studies, what is important is to establish an accurate accounting process to enumerate observable halo features.  Roughly speaking, observers viewing the systems in \figref{fig:lukic09} would be likely to identify one dominant cluster in the left and middle panels, and two in the right.  An FOF accounting system would need to admit a non-unitary condition (one halo maps to two clusters) when converting mass to observable signals.  In contrast, SO masses map to integrated aperture observations more directly.  For this reason, SO masses see more frequent use for survey data analysis.

\subsubsection{HALO MASS FUNCTION AND CLUSTERING} 

The original multiplicity function paper of \cite{PressSchechter74} used the clustering of particles in N-body experiments with $N=1000$ to support their analytic form for $f(\sigma)$ in \eqnref{eq:massFtnBasic}.   Later, \citet{ShethTormen9901122} used $N=10^7$ simulations to set free parameters of their $f(\sigma)$ model derived using an ellipsoidal, rather than spherical, collapse approximation.  Using a suite of simulations of open and flat cosmologies with $\Omegam$ ranging from  $0.3$ to 1, \cite{Jenkins0005260} found a unique, three-parameter form for $f(\sigma)$ that produced a mass function accurate to $\sim \! 30\%$ across the suite of models.  

A recent study by \citet{Tinker0803.2706} employs 22 large ($N \sim 10^9$) simulations produced with three independent N-body codes to calibrate a functional form motivated by \citet{ShethTormen9901122}, 
\begin{equation}
f(\sigma) = A \left[  \left( \frac{\sigma}{b} \right)^{-a} + 1 \right] \, e^{-c/\sigma^2}  .
\label{eq:massFtnTinker} 
\end{equation} 
This study was the first to open the density threshold degree of freedom; their fitting parameters are published as functions of $\Delta$ (against $\rhom(z)$) for $ \Delta \in [200, 3200]$.   With the high statistical power of their simulation ensemble, \citet{Tinker0803.2706} achieve a fit with $5\%$ statistical precision in halo number at $z = 0$ for a \LCDM\ cosmology.  Maintaining this precision for redshifts $z \le 2.5$ requires the introduction of mild redshift dependence into the fit parameters, $A(z)$, $a(z)$ and $b(z)$.  The theoretically expected halo counts above masses $M_{200} = 10^{14}$ and $10^{15} \msol$ in the reference \LCDM\ cosmology, shown in Figure~\ref{fig:surveyHistory}, are based on the Tinker form for threshold $\Delta = 200$ against the mean mass density \citep[see fitting formulae in][]{Mortonson1011.0004}. 

On the other hand, the bias function measured in the same simulation ensemble shows no need for such redshift-dependent corrections.  Framed in terms of the normalized linear perturbation amplitude, $\nu \! \propto \! \sigma(M)^{-1}$, \citet{Tinker1001.3162} find a robust fit of the form
\begin{equation}
b(\nu) =  1 - D \frac{\nu^d}{\nu^d + \deltac^d} + E  \nu^e  + F \nu^f , 
\label{eq:biasTinker} 
\end{equation} 
with a single set of parameters $\{d, e, f, D, E, F\}$ that are written only as functions of $\Delta$.   For the case $\nu = 3$ (i.e., $3 \sigma$ peaks), the value of the bias is large, $b \sim 6$, for $\Delta = 200$.  The cluster power spectrum, \eqnref{eq:corrFtnBasic}, can be enhanced by factors of several tens over the mass power spectrum.   

The very massive end of the FOF mass function was recently revised by \citet{Crocce0907.0019} using $2048^3$-particle simulations in \LCDM\ cubic volumes up to $7680 \hinv \mpc$ in scale.  Above $10^{15} \hinv \msol$, their fit lies up to $30\%$ above prior calibrations \citep{Jenkins0005260, Warren0506395}.

\subsubsection{INTERNAL HALO STRUCTURE} 

Gravitational relaxation drives the phase-space structure of halos to a common structure that applies from small galactic satellites to the most massive galaxy clusters.  The form of the radial density,
\begin{equation} 
\rho(r) = {{ \rhocrit(z) \, A_c} \over {
({r/r_{\rm s}}) \, \left(1+{r/r_{\rm s}} \right)^2}}, 
\label{eq:NFW}
\end{equation}
is known as the Navarro-Frenk-White (NFW) profile \citep{Navarro95}.  
Here, $r_{\rm s}$ is the scale radius, $c$ is the concentration parameter
(with $c=r_{200}/r_{\rm s}$) and $A_c = {200 c^3 / 3 \left[
{{\rm ln}(1+c)-{c/(1+c)}}\right]}$.

Simulations show that concentration and mass are weakly correlated.  In the mass range of galaxies to clusters, $c \propto M^{-\zeta}$, with $\zeta \sim 0.14$ at $z=0$ and $\zeta \rightarrow 0$ at $z \gtsim 3$   \citep[e.g.,][]{Gao08}.  That study finds that a fixed concentration, $c \sim 4 \pm 1$, applies in the mean to high mass halos, independent of redshift.   Tracking the mass accretion histories of halos in simulations, \citet{Wechsler0108151} find a common functional form, and show that the formation epoch correlates strongly with concentration.   The concentration--mass relation can be understood as a result of adiabatic contraction of differently-shaped peaks in the linear density field \citep{Dalal1010.2539}.   

\subsection{From Halos to Clusters: Mass Proxies, Scaling Relations and Projection Effects}

Cluster cosmology originates from phenomena observed on the sky, in the 2+1 space of angular coordinates and redshift.  The observables employed for a likelihood analysis must be predicted under a set of combined cosmological and astrophysical parameters, $\{ \thetaCosm, \alphaCosm \}$.  For constraints based on cluster counts, the mass function, $n(M,z)$, written in terms of spherical overdensity or percolation measures from simulations needs to be translated into a signal function, $n(\Svec,z)$, for one or more signals, $S_i$.    We use the terms \emph{signal} and \emph{observable} interchangeably, and generically they refer to bulk measures at mm (SZ decrement $Y$), optical (richness, $\Ngal$, or velocity dispersion, $\sigmagal$), or X-ray (luminosity, $\Lx$; temperature, $\Tx$; gas mass, $\Mgas$; and/or gas thermal energy, $\Yx=k\Tx \Mgas$) wavelengths (see \secref{sec:observations}).  An ideal experiment would measure all of these observables within apertures optimally matched to the underlying halo sizes, $r_\Delta(M,z)$.  This ideal is often frustrated by signal-to-noise constraints and confused by projection effects and foreground/background contamination.  

\subsubsection{OBSERVABLE SIGNAL LIKELIHOOD FROM MULTIVARIATE SCALING RELATIONS  } \label{sec:theory:mvscaling}

Scaling relations for cluster signals, based on assumptions of virial equilibrium and self-similar internal structure, were first published by \citet{Kaiser1986MNRAS.222..323K}.  In this model, halos at fixed mass and redshift are identical, and scalings with mass and redshift follow calculated power-law behaviors. Observations generally support power-law behavior, but not always with the self-similar slope  (\secref{sec:constraints:scaling}).  We describe here a non-self-similar model that incorporates arbitrary mass scaling and allows for variations at fixed mass and redshift.  

For compactness of notation, let $s_i=\ln(S_i)$, for each of the $N$ observables, $S_i$, and let $\mu = \ln M$.  The power-law assumption transforms to log-linear scaling  
\begin{equation}
\sbar(\mu,z)  \ = \ \mvec \mu \ + \  \mathbf{b}(z), 
\label{eq:sbarmu}
\end{equation}
where the average is over a very large cosmic volume. The elements of $\mvec$ are the slopes of the individual mass-observable relations, and the intercepts $\mathbf{b}(z)$ reflect the evolution at fixed mass.  At a fixed epoch, we can always choose units such that $b_i(z) = 0$ (as we do below).   For cosmological studies, a measure of merit is the equivalent mass scatter in each signal, $\sigma_{\mu i} \equiv \sigma_i/m_i$.  

Various processes, including different formation histories and the stochastic nature of mergers, generate deviations from the mean.  Taking these as Gaussian in the log leads to a form for the \emph{conditional signal likelihood}, 
\begin{equation}
p\left(\svec | \mu, z\right) \ = \ \frac{1}{(2\pi)^{N/2} | \Psi |^{1/2}} \exp \left\{ -\frac{1}{2} \left[\svec-\sbar(\mu,z)\right]^\dagger \mathbf{\Psi}^{-1} \left[ \svec-\sbar(\mu,z)\right] \right\} .
\label{eq:pofs}
\end{equation}
The elements of the covariance matrix, $\Psi_{ij} \equiv \left\langle  (s_i-\bar{s}_i) (s_j-\bar{s}_j) \right\rangle$, could have mass or redshift dependence, but a first-order approach considers them as constants.    

When the mass variance of signals is small, $\sigma_{\mu i}^2 \ll 1$, then the above expressions can be convolved with a locally power-law approximation to the mass function, $n(\mu,z) = A e^{-a \mu}$, to obtain the local  \emph{signal space density} function,
\begin{equation}
n(\svec, z) \  = \  \frac{A \Sigma} { (2\pi)^{(N-1)/2} |\Psi|^{1/2}} \exp \left[ -\frac{1}{2} \left( \svec^\dagger \Psi^{-1} \svec - \frac{\mubar^2}{\Sigma^2}  \right)\right] ,
\label{eq:nofs}
\end{equation}
where $\Sigma^2  = \  ( \mvec^\dagger \Psi^{-1} \mvec )^{-1}$ is the variance about the mean log-mass selected by the set of signals $\svec$, 
\begin{eqnarray}
\mubar(\svec, z)   \ =  \   \frac{\mvec^\dagger \Psi^{-1} \svec } {\mvec^\dagger \Psi^{-1} \mvec } - a \Sigma^2  \  \equiv  \  \mubar_0(\svec, z) -  a \Sigma^2  .
\label{eq:mubar}
\end{eqnarray}
The first term above is the mean mass for the case of a flat mass function, $a = 0$.  The second term, represents the (Eddington) mass bias induced by asymmetry in the mass function convolution.  Upscattering of low-mass systems dominates when $a > 0$, and the high-mass end of the  \LCDM\  mass function is steep, $a \gtsim 3$   \citep{Mortonson1011.0004}.    These equations make explicit the degeneracy between cosmology (\eg\ $A$ and $a$) and astrophysics (\eg\ $\mvec$ and $\Psi$) inherent in cluster counts.  They provide the means to compute biases, relative to a mass complete sample, associated with signal-limited cluster samples (discussed further in \secref{sec:theory:selection}).

\begin{figure*}
  \centerline{
    \epsfxsize=6cm
    \epsfbox{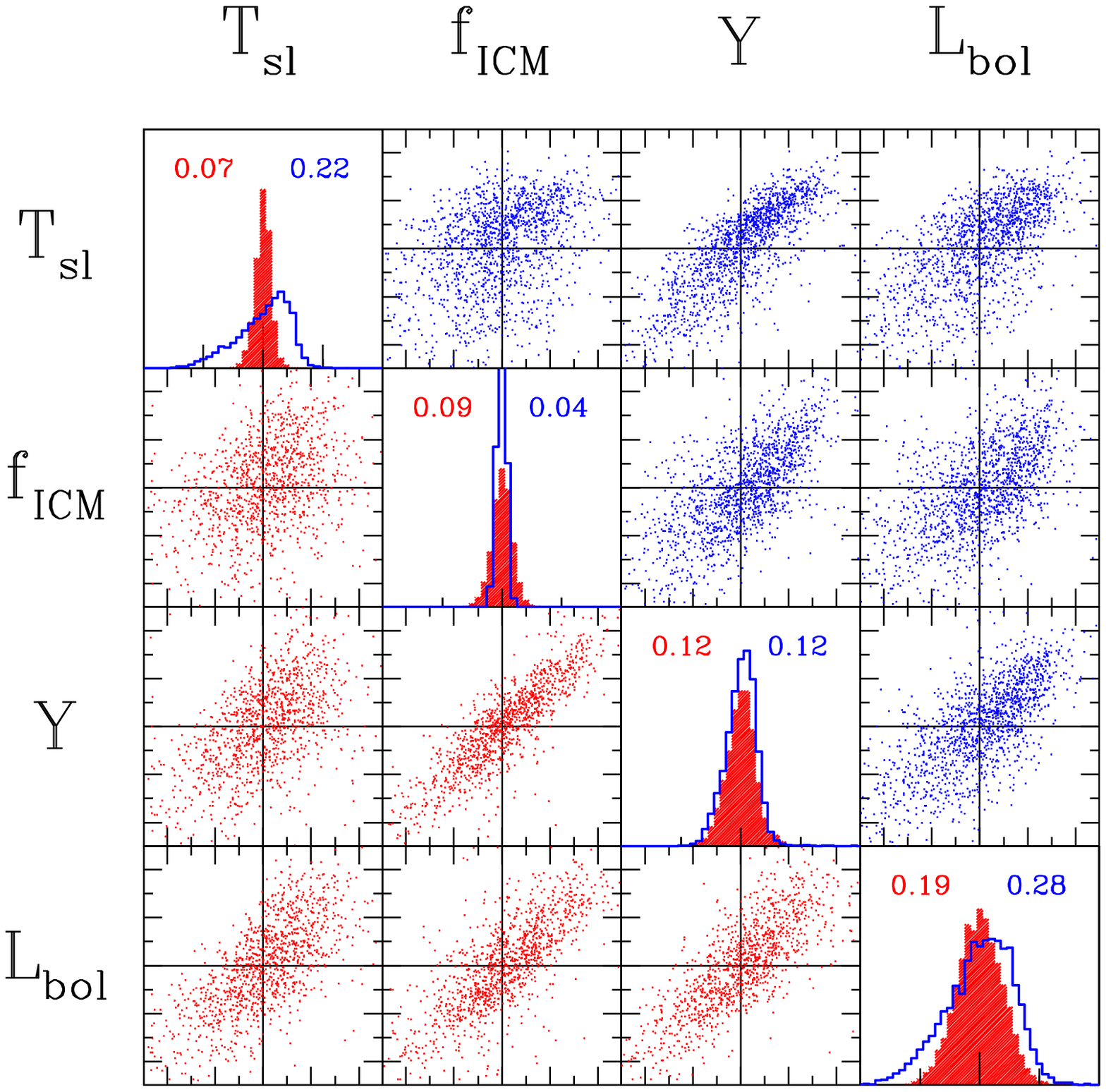}
    \epsfxsize=6cm
    \epsfbox{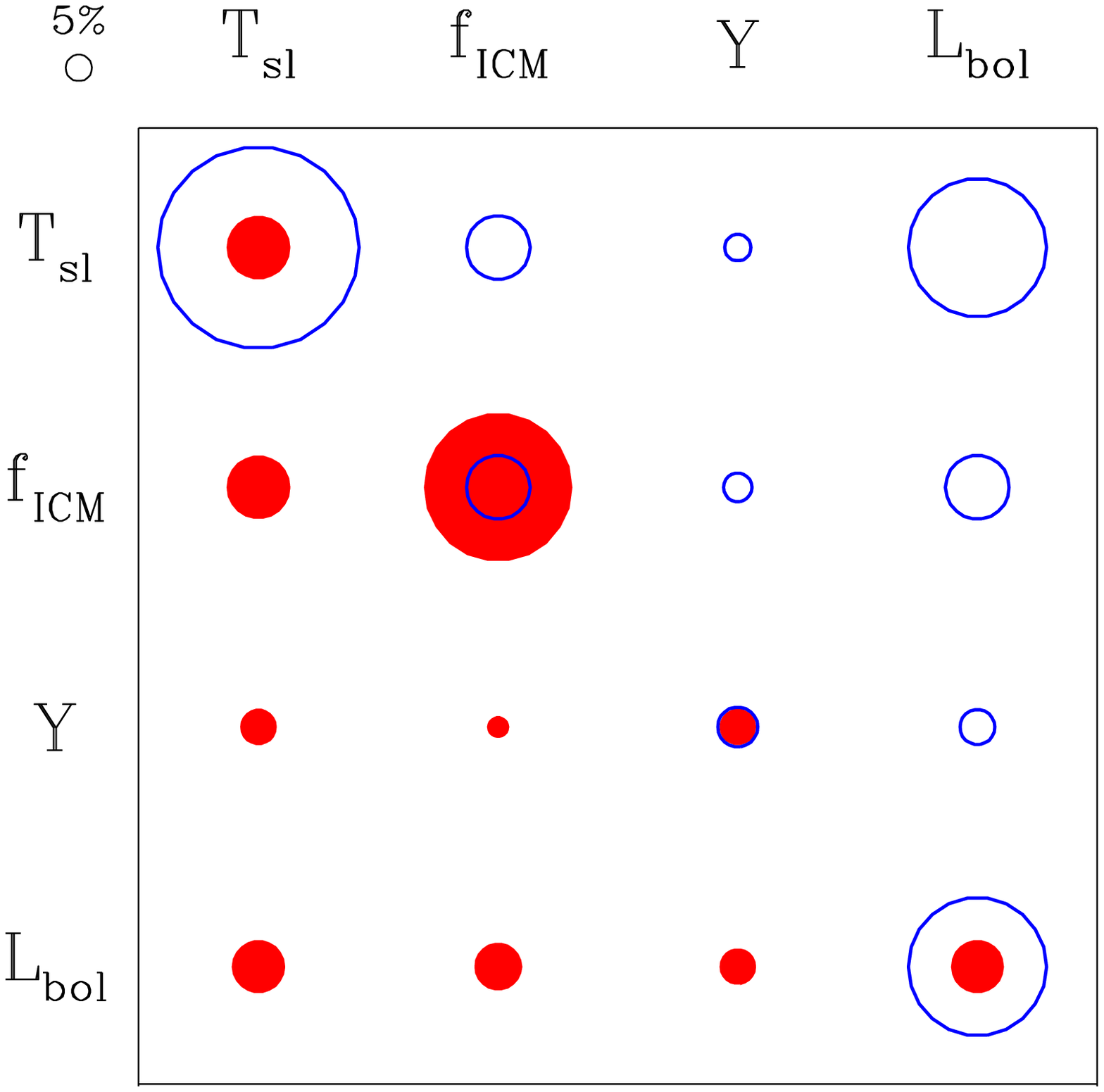}
    }
    \caption{{\bf Left:} Covariance of internal properties of $> \!4500$ halos with $M_{200} > 5 \times 10^{13} \hinv \msun$ extracted from Millennium Gas Simulations produced under two different physical treatments.  Off-diagonal panels show normalized ($(s_i-\bar{s}_i)/\sigma_i$) pairwise deviations under preheating (PH, lower) and gravity-only (GO, upper) treatments;  large tickmarks are separated by unity.   Diagonal panels show the distribution of ln(property) deviations for PH (red) and GO (blue) models, with dispersions given in the legend.  {\bf Right:} Visual representation of the mass variance, \eqnref{eq:Sigma2_2D}, obtained using the property pairs at left.  The radii scale with $\Sigma$, and a $5\%$ reference is shown in the upper left.  Adapted from \citet{Stanek0910.1599}.
 }
 \label{fig:stanek10}
\end{figure*}

Figure~\ref{fig:stanek10} provides support for this model from Millennium Gas Simulation analysis  \citep{Stanek0910.1599}.  The left panel shows deviations about the mean behavior of four intrinsic (3-dimensional) properties measured within $r_{200}$ for $> \! 4500$ halos with mass $M_{200} > 5 \times 10^{13} \hinv \msol$ at $z=0$.   The lower diagonal and red histograms show results from a cooling and preheating (PH) treatment of the baryons, where the entropy is instantaneously raised to $200 \keV \, {\rm cm}^2$ at $z = 4$.  Only a small fraction of baryons cool into stars in this model \citep{Young1007.0887}.  The upper diagonal and blue histograms are from a gravity-only (GO) treatment, where the gas is heated only by shocks and does not cool.    

The internal properties generally have modest variance, and pairs tend to be positively correlated with typical correlation coefficient $r \sim 0.4-0.8$.   Halos identified by a pair of properties will  have mass variance 
\begin{equation}
\Sigma^{2} \ = \ (1-r^2) \left( \sigmuone^{-2} + \sigmutwo^{-2} - 2r \sigmuone^{-1}\sigmutwo^{-1} \right)^{-1},
\label{eq:Sigma2_2D}
\end{equation} 
shown by the areas of the off-diagonal circles in the right panel of Figure~\ref{fig:stanek10}.  Individual properties lie along the diagonal.  The intrinsic gas thermal energy, $Y$, selects mass with $7\%$ dispersion, the best individual measure for both physics cases.  This level is also seen in the  simulations of \citet{Nagai0512208} which include cooling, star formation and feedback.   
Pairs of intrinsic measurements always improve mass selection, and the strong correlation between $\fICM$ and $Y$ combines with the large mass variance of $\fICM$ to achieve mass selection with $4\%$ scatter in the PH model.  

Applying Bayes' theorem to this model allows one to write the likelihood of mass and an observable, $s_2$, for a sample selected on observable $s_1$.  When the two signals are correlated, one can show that the scaling with mass of the non-selection signal will be 
\begin{equation}
\bar{s}_2(s_1) \ = \ m_2 \left[ \mubar(s_1)  + \alpha r \sigmuone \sigmutwo \right]  ,
\label{eq:s2barGs1}
\end{equation}
which is biased relative to the naive expectation of $m_2 s_1 /m_1$. The intrinsic correlation between signals at fixed mass is relatively challenging to constrain from current data, but first measurements have been made for samples selected using optical \citep{Rozo0809.2794} and X-ray  \citep{Mantz0909.3099} observations.

\subsection{From Theory to Practice: Sources of Systematic Error} \label{sec:theorytopractice}

Clusters on the sky relate to halos through selection on one or more
observables.  Matching cluster detections (which originally reside in
a 2+1 space of angular position and signal-to-noise) to halos
can sometimes be complex; two halos along nearly the same line of sight may be blended into a single cluster, or a single halo may be fragmented into more than one cluster. The frequency of these occurrences is typically not large, $\ltsim \! 10\%$, but the exact values are sensitive to a number of factors, particularly detection method and mass, and so are best modeled via direct sky realizations \citep[\eg][]{Sehgal1010.1025}.  

The selection observable can be distorted from its intrinsic value (\eqnref{eq:pofs}) by triaxiality, by additional sources along the line-of-sight, by mis-centering and/or mis-estimation of the radial scale, and by other effects.  Telescope/instrument calibration and data processing methods also contribute to the error budget.  For upcoming studies using cluster counts, photometric redshift errors have an important, but not dominant, effect (Section~\ref{sec:new_surveys}).   

\subsubsection{SAMPLE SELECTION} \label{sec:theory:selection}

Testing cosmology with halo counts and clustering requires that the theoretical mass function be transformed, via the scaling relations and a model of the selection process, into a prediction for the distribution of clusters in the space of survey observables (\eg{} redshift and X-ray flux).  The scaling relation parameters set the space density portion of the survey yield (\eqnref{eq:nofs}) in terms of the (cosmologically dependent) local amplitude, $A(\mu, z)$, and logarithmic slope, $a(\mu, z)$, of the mass function.  Sample selection must be well understood to avoid perturbing $A(\mu, z)$ and $a(\mu, z)$ from their true values, biasing cosmological results.  Fortunately, such effects can be mitigated by survey self-calibration \citep{Majumdar0305341} or by calibration using follow-up observations, as discussed below. 

The task of empirically constraining the scaling relations is complicated by the fact that the clusters targeted for follow-up observations are themselves subject to selection effects related to their original discovery.  In an X-ray flux-limited sample, for example, higher X-ray luminosity at a given mass leads to  a larger probability of detection (commonly known as Malmquist bias). The effects of selection bias must therefore be accounted for in the calibration of scaling relations, much as in the cosmological analysis (e.g. \citealt{Stanek0602324,Sahlen09}).

\begin{figure}[t]
  \centerline{
    \epsfxsize=6cm
    \epsfbox{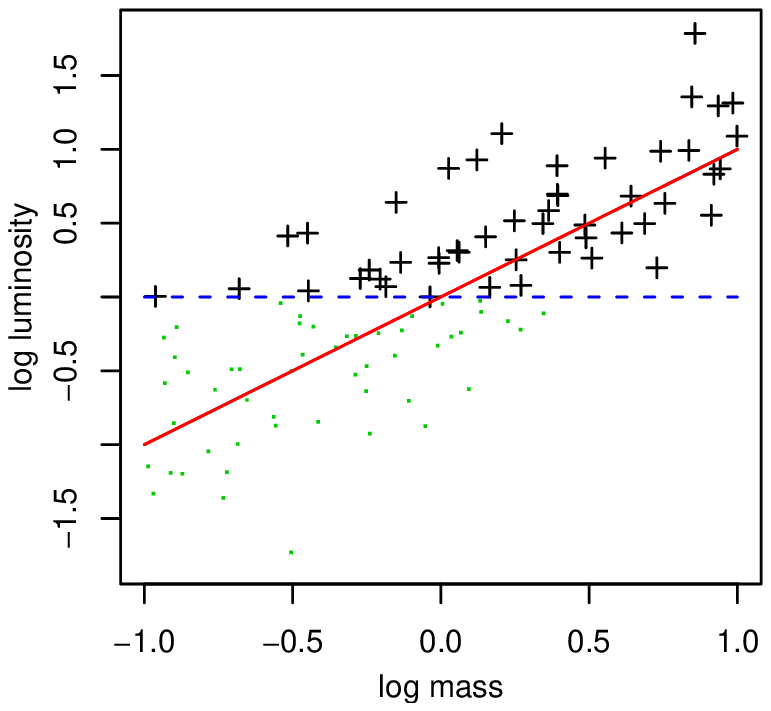}
    \hspace{7mm}
    \epsfxsize=6cm
    \epsfbox{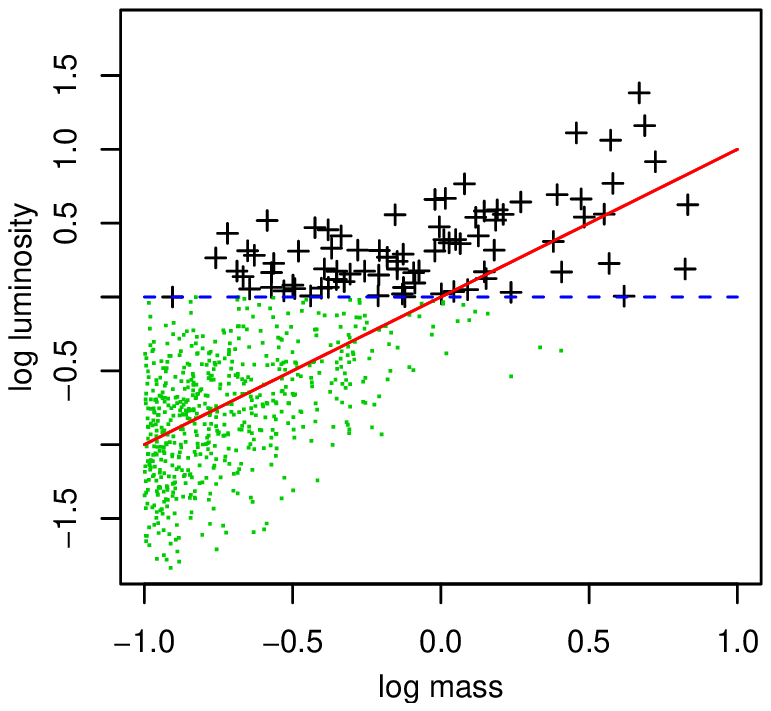}
  }
  \caption{Cartoon illustrating generically how the distribution of observed scaling relation data (black crosses) do not reflect the underlying scaling law (red line) due to selection effects (e.g. a luminosity threshold; blue, dashed line). Green dots indicate undetected sources. The left panel shows an unphysical case in which cluster log-masses are uniformly distributed, while the mass function in the right panel is a more realistic, steeper power-law (normalized to produce roughly the same number at high masses). The steepness of the mass function has a clear effect on the degree of bias in the detected sample. To recover the correct scaling relation, an analysis must account for both the selection function of the data and the underlying mass function of the cluster population. Adapted from \citealt{Mantz0909.3099}.}
  \label{fig:constraints:selection}
\end{figure}

\figref{fig:constraints:selection} illustrates the influence of selection on observed scaling relation data in a cartoon case. The full population (black crosses and green points) obeys a scaling law (red line) with non-trivial intrinsic scatter. In the simple case where detection requires a particular threshold luminosity (the dashed, blue line), it can be seen that, even if every detected cluster is followed up to obtain precise measurements of the mass and luminosity, the resulting data set will be a biased representation of the full population.  While complete at the highest masses, the sample is increasingly incomplete at low masses, with the low-luminosity systems absent.

A closely related consideration is the effect of the underlying mass function on the observed scaling relation data. The distribution of the relation's independent variable(s) (in this case cluster masses) within the full population generically influences constraints on scaling laws (e.g. \citealt{Gelman2004BayesianDataAnalysis,Kelly0705.2774}). Neglecting to account for this influence corresponds to the assumption of uniformly distributed independent variables; often this approximation is sufficient, but the exponentially steep slope of the cluster mass function suggests that we should take the issue seriously in the context of cluster cosmology \citep{Mantz0909.3098}. \figref{fig:constraints:selection} illustrates how the steepness of the mass function influences the fraction of the observed data which are strongly biased relative to the underlying scaling relation. Given the need to solve for both the slope and scatter of the scaling relation, accounting for the disparity in the number of high-mass and low-mass systems is critical.  We note that simply conditioning the sampling distribution on cluster detection, as some authors have done, is not sufficient to rigorously recover all the scaling information.

Note that this effect has a floor set by non-zero intrinsic scatter in the scaling relations, but the effect can in principle be enhanced by measurement error.  However, measurement errors in current X-ray and optical cluster surveys are typically smaller than the intrinsic dispersion, even at the survey limit.  Thus, re-measurement of the survey observables through deeper, follow-up observations (e.g. to improve the signal-to-noise of X-ray or SZ flux) does not circumvent the issue of selection bias in the scaling relation analysis. 
 
\begin{figure}[t]
  \centerline{
    \epsfxsize=6cm
    \epsfbox{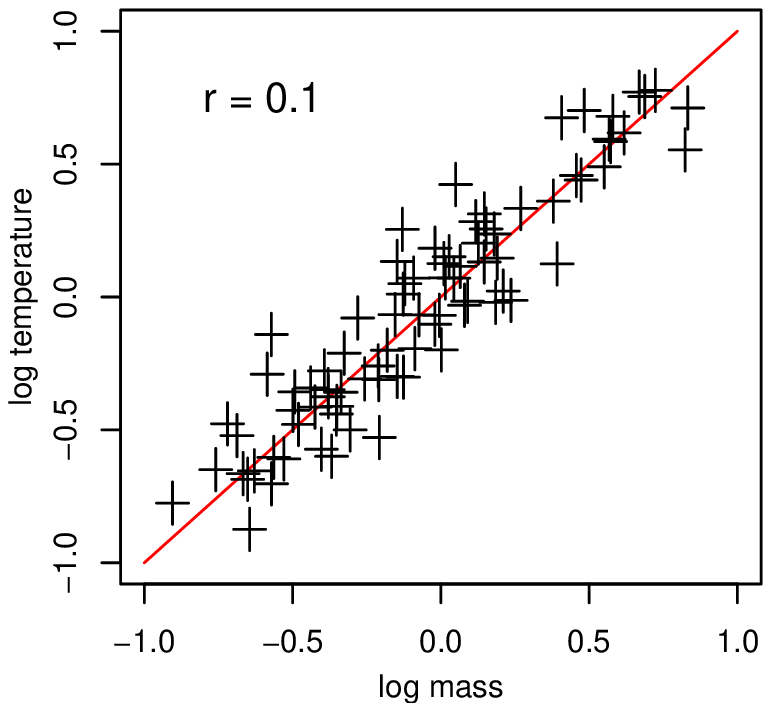}
    \hspace{7mm}
    \epsfxsize=6cm
    \epsfbox{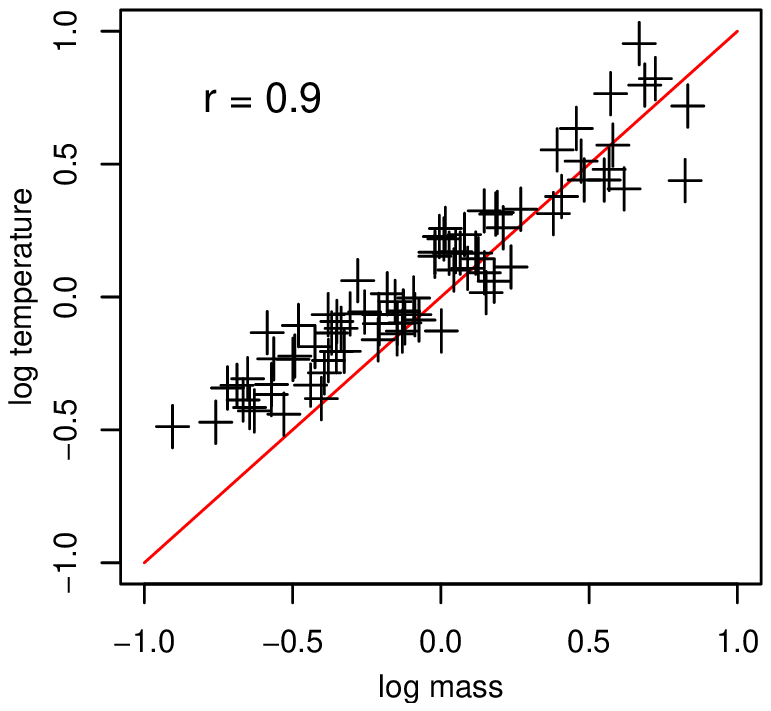}
  }
  \caption{Cartoon scaling relations where the observable of interest is not the basis of cluster selection. In both panels, the red line indicates the true scaling relation, and the black crosses correspond to the detected clusters in the right panel of \figref{fig:constraints:selection}. The marginal scatter in this relation is chosen to be smaller than that in \figref{fig:constraints:selection}, consistent with measured values of the luminosity--mass and temperature--mass intrinsic scatters (\secref{sec:obs_masses_proxies}). The intrinsic temperature--luminosity correlation at fixed mass is relatively small in the left panel ($r=0.1$) and large in the right panel ($r=0.9$); in the latter case, the observed data are significantly influenced by selection bias despite the fact that the selection was made using a different observable.}
  \label{fig:constraints:selection2}
\end{figure}

While selection bias clearly influences scaling relations involving the selection observable, it also influences relations of other signals with which the selection observable has non-zero intrinsic correlation (\eqnref{eq:s2barGs1}). This is illustrated in \figref{fig:constraints:selection2}, for a signal which is correlated with the selection observable with coefficient 0.1 (left panel) and 0.9 (right panel). The red line shows the true scaling law and the points shown correspond to the detected clusters from \figref{fig:constraints:selection} (right panel). With relatively mild intrinsic correlation, as has been found for temperatures and soft X-ray flux detection \citep{Mantz0909.3099}, the distribution of data points closely follows the underlying relation; for more extreme values of the correlation coefficient, as might be expected, e.g., between temperature and SZ signal, deviations due to selection bias become evident. Note that the severity of the effect also depends on the covariance of the signals rather than only on the correlation coefficient (i.e. the size of the marginal scatter in each signal is also important).

Cluster samples are often characterized in terms of \emph{completeness} and \emph{purity} \citep{WhiteKochanek02}.  Completeness is used in many ways, but its simplest form for cluster cosmology refers to the fraction of halos above mass $M$ at redshift $z$ that are identified in a survey with some observable limit, $S_{\rm lim}(z)$.  Completeness of unity is achievable at high masses when the survey limit, $S_{\rm lim}(z)$, lies sufficiently far in the signal likelihood's negative tail.   Impurity is a measure of false positive sources in the sample.  Fewer conventions for its definition exist in the literature.  Generically, one can write the observed counts above some signal limit $S$ as a sum, $N_{\rm obs}(>S) = N_{\rm true}(>S) + N_{\rm false}(>S)$, where the first term represents genuine cluster systems -- manifestations of a single massive halo along the line of sight -- and the second expresses detections of other origin.  Zero impurity, $N_{\rm false}(>S) = 0$, is a desired goal.

\subsubsection{PROJECTION EFFECTS} 

Telescopes aimed at a distant halo necessarily collect photons that originate elsewhere along the multi-gigaparsec sightline than within the target system.  Due to their softer angular profiles, SZ, lensing and optical cluster signals can be blended more readily than X-ray.  Chance orientations of two or more halos within local supercluster regions create an asymmetric tail to high signal values.  Considered in terms of mass selection, the effect produces a tail to low masses in the distribution of halo mass selected at a given signal \citep[\eg][]{Cohn0706.0211}.  

Since the matter components of halos are generally ellipsoidal rather than spherical, orientation variations also produce scatter in signals observed in halos of fixed mass.  Signals are generally maximized when viewed along the long axis and minimized along the short axis.  Orientation can affect cluster selection, with prolate systems oriented along the line-of-sight being preferentially included.  Since its collisional nature drives the X-ray emitting gas toward equipotential surfaces, it tends to be rounder than the dark  matter and so less susceptible to orientation bias.  

As discussed in \secref{sec:observations}, the density squared dependence of the X-ray emissivity means that X-ray selection is less prone to projected confusion.   Optical richness measurements roughly trace mass density and are therefore more easily confused by projection and orientation effects.  SZ measurements are intermediate, since the SZ effect depends on electron pressure, the product of density and temperature.

\subsection{Non-Standard Scenarios} 

It is important to keep in mind that theory offers many potential deviations from the reference \LCDM\ cosmology sketched above.   Key model assumptions -- that the dark matter is a weakly interacting massive particle, that inflation produced a Gaussian spectrum of initial density fluctuations with a power-law initial spectrum, that small-amplitude metric perturbations are well described by Newtonian, weak field expansions in general relativity, and so on -- need to be rigorously tested.  In \secref{sec:physics}, we discuss ways in which clusters can be used to test a number of proposed modifications to the reference model.

\section{OBSERVATIONAL TECHNIQUES} \label{sec:observations}

In this section we review briefly the physics underlying
multiwavelength observations of galaxy clusters.  We summarize efforts
to construct cluster catalogs, with an emphasis on surveys that have
led to cosmological constraints. We discuss techniques used to measure
the masses of clusters, and observable proxies that correlate tightly
with mass.

\subsection{Multiwavelength Measurements of Galaxy Clusters} \label{sec:obs_observations}

\subsubsection{X-RAY OBSERVATIONS}
\label{sec:obs_xray}

\begin{figure}
  \centerline{
    \epsfxsize=4.5cm
    \epsfbox{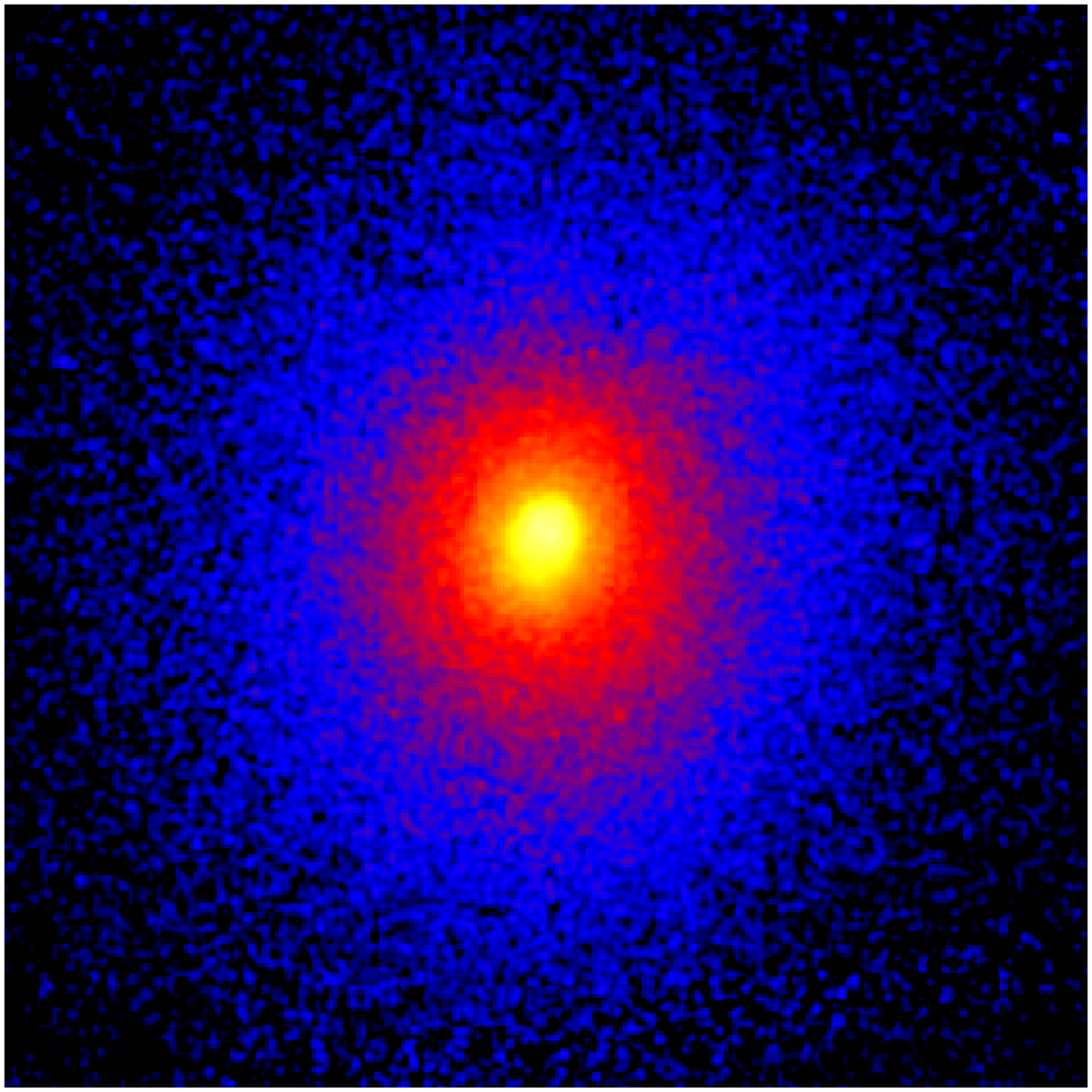}
    \epsfxsize=4.5cm
    \epsfbox{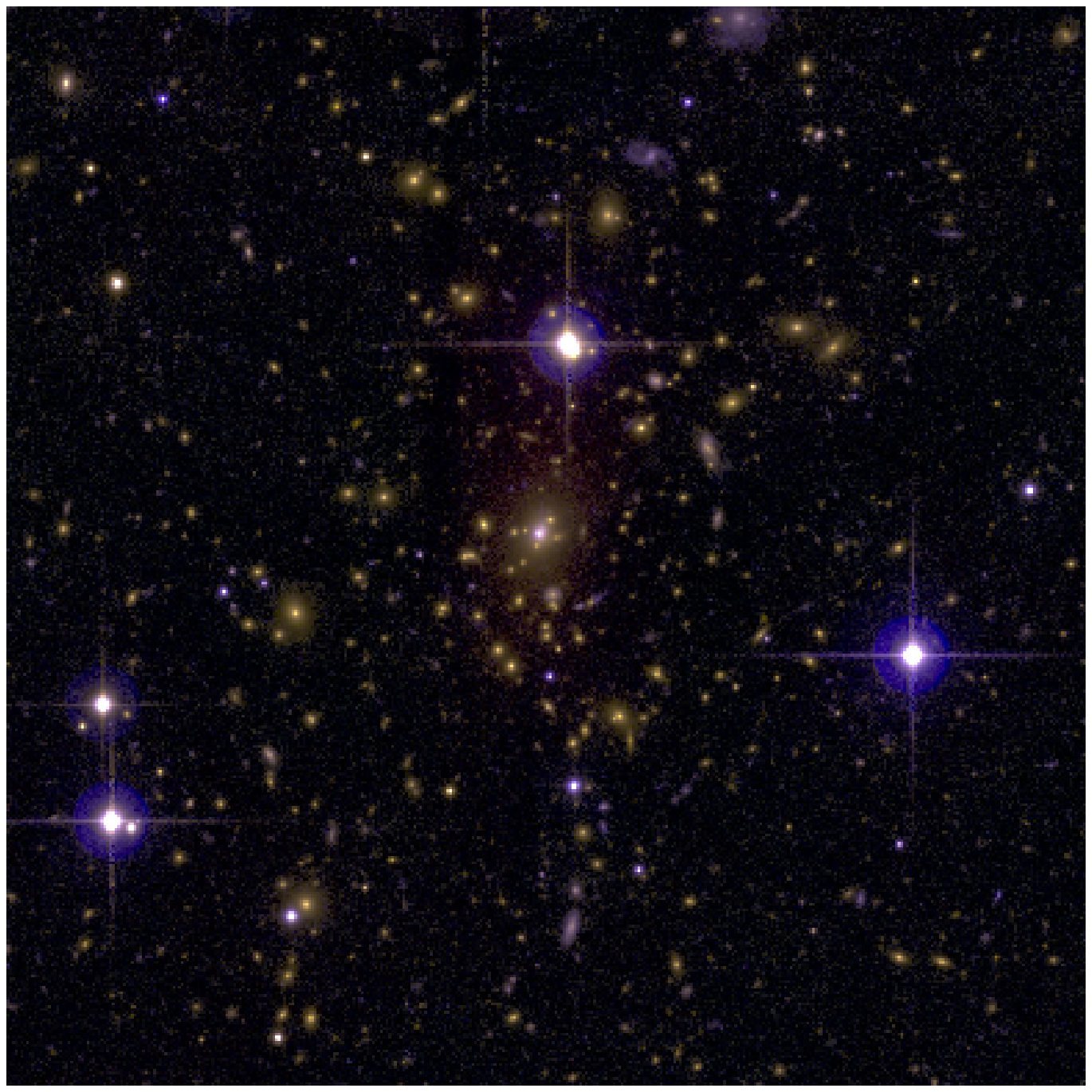}
    \epsfxsize=4.5cm
    \epsfbox{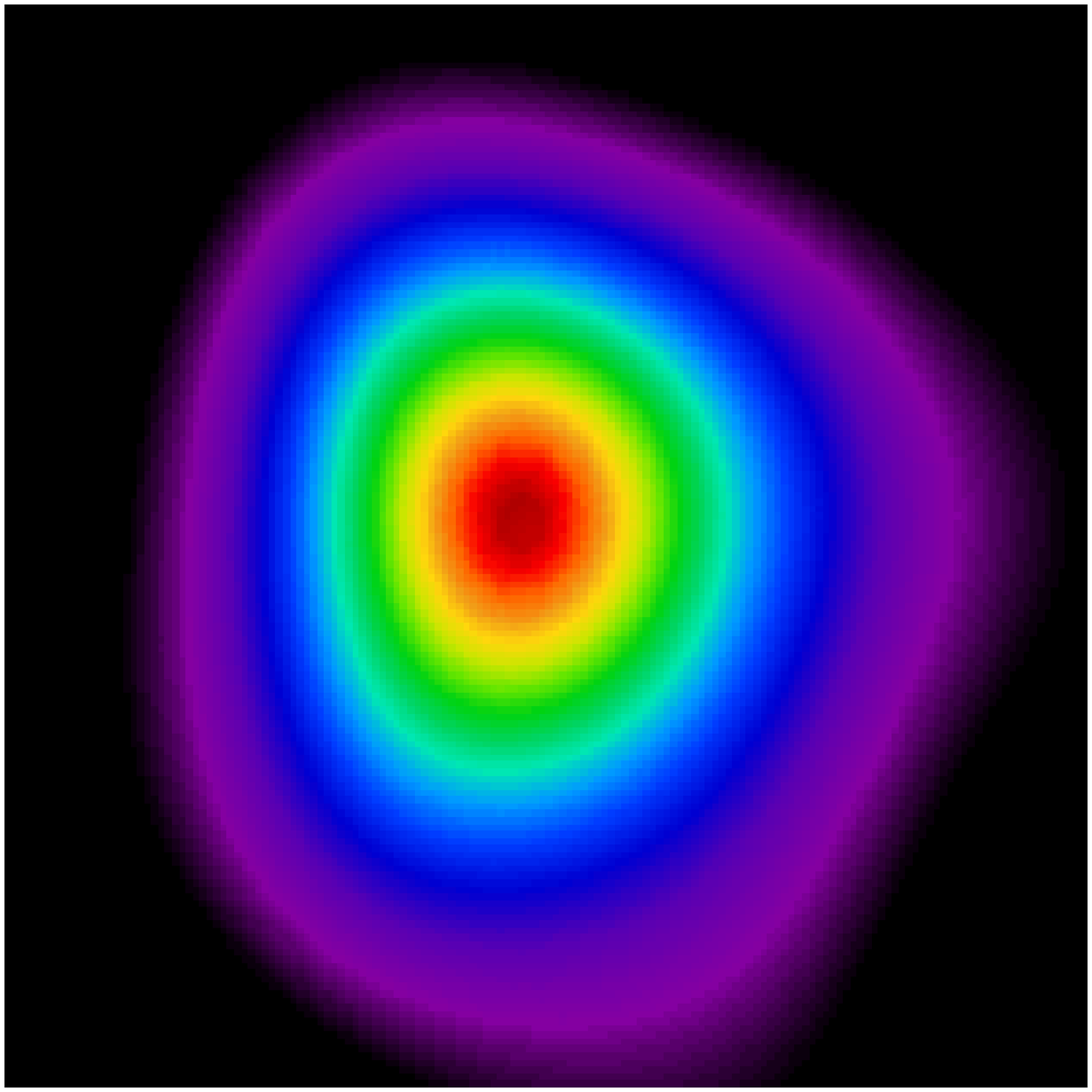}
   }
  \caption{Images of Abell~1835 ($z=0.25$) at X-ray, optical and mm wavelengths, exemplifying the regular multi-wavelength morphology of a massive, dynamically relaxed cluster. All three images are centered on the X-ray peak position and have the same spatial scale, 5.2~arcmin or $\sim 1.2\Mpc$ on a side (extending out to $\sim r_{2500}$; \citealt{Mantz0909.3099}). Figure credits: {\it Left}: X-ray: Chandra X-ray Observatory/A. Mantz; {\it Center}, Optical: Canada France Hawaii Telescope/A. von der Linden et al.; {\it Right}, SZ: Sunyaev Zel'dovich Array/D. Marrone.}
  \label{fig:multiwavelength}
\end{figure}

Most of the baryons in the Universe are in diffuse gas.  Typically,
this gas is very difficult to observe. Within galaxy clusters,
however, gravity squeezes the gas, heating it to virial temperatures
of $10^7$--$10^8\Kelvin$, which causes it to shine brightly in
X-rays. Galaxy clusters therefore `light up' at X-ray wavelengths as
luminous, continuous, spatially-extended sources
(\figref{fig:multiwavelength}).

The primary X-ray emission mechanisms from the diffuse ICM are
collisional: free-free emission (bremsstrahlung); free-bound
(recombination) emission; and bound-bound emission (mostly line
radiation). The emissivities of these processes are proportional to
the square of the electron density, which ranges from $\sim 10^{-1}
\cm^{-3}$ in the centers of bright `cool core' clusters to $\sim
10^{-5}\cm^{-3}$ in cluster outskirts. At these low densities, the
X-ray emitting plasma is optically thin and in the coronal limit,
which makes modeling straightforward.

For survey observations, the primary X-ray observables are flux,
spectral hardness and spatial extent. Using deeper, follow-up
observations of individual clusters, modern X-ray satellites allow the
spatially-resolved spectra of clusters to be determined precisely,
permitting measurements of the density, temperature and metallicity
profiles of the ICM, and a host of derived thermodynamic quantities.
For reviews of the principles underlying X-ray observations of
clusters see, e.g., \citet{Sarazin88} and \citet{Boehringer09}.

\subsubsection{OPTICAL AND NEAR INFRARED OBSERVATIONS}
\label{sec:obs_optical}

The optical and near-IR emission from galaxy clusters is predominantly
starlight. The galaxy populations of clusters are dominated by
ellipticals and lenticulars (i.e. early-type galaxies). This is
particularly true in the central regions, where the largest and most
luminous galaxies are found (\figref{fig:multiwavelength}).

The old and relatively homogeneous nature of their stellar populations
leads to the majority of the galaxies in clusters occupying relatively
tight loci in color-magnitude diagrams \citep[e.g.][]{Bower92}.  This
characteristic has proved important to modern cluster finding
algorithms.

For optical surveys of clusters, the main observables are the richness
(i.e. the number of galaxies within the detection aperture),
luminosity and color. For follow-up observations of individual
clusters, aimed in particular at measuring their masses, the primary
observables are the galaxy number density, luminosity, and velocity
dispersion profiles. Typical velocity dispersions for large clusters
are of order 1000\kmps.

For reviews of optical studies of galaxy clusters including
discussions of the development of the field, see \citet{Bahcall77} and
\citet{Biviano00}.
      
\subsubsection{SZ OBSERVATIONS}
\label{sec:obs_sz}

As CMB photons pass through a galaxy cluster they have a
non-negligible chance to inverse Compton scatter off the hot ICM
electrons.  This scattering boosts the photon energy and gives rise to
a small but significant frequency-dependent shift in the CMB spectrum
observed through the cluster known as the thermal Sunyaev-Zel'dovich
(hereafter SZ or tSZ) effect \citep{Sunyaev72}. The magnitude of the
effect is proportional to the line of sight integral of the product of
the gas density and temperature.  The kinetic SZ (kSZ) effect is an
additional, smaller distortion of the CMB spectrum due to the peculiar
motion of a cluster with respect to the Hubble Flow (i.e. the CMB rest
frame). The magnitude of the kSZ effect is proportional to the
peculiar velocity. For a review see \citet{Carlstrom02}.

\subsubsection{GRAVITATIONAL LENSING}
\label{sec:obs_lensing}

According to general relativity, the gravity associated with a mass
concentration will bend light rays passing near to it in a phenomenon
known as gravitational lensing. This can both magnify and distort the
images of background galaxies.
With modern data, gravitational lensing can be detected clearly in the
statistical appearance of background galaxies observed through
clusters (weak lensing), and in the field (often termed cosmic
shear). Occasionally, lensing can also lead to strong distortions and
multiple images of individual sources (strong lensing). For a galaxy
cluster and background galaxies of known redshifts, the measured
gravitational shear can be used to infer the cluster mass. For a
recent review of gravitational lensing, see \citet{Bartelmann10}.

\subsection{Constructing Cluster Catalogs}
\label{sec:obs_surveys}

A well-designed cluster survey should meet requirements in terms of
angular scale, flux sensitivity and redshift coverage. The survey
should be as complete (i.e. not have missed clusters that it should
have detected) and pure (i.e. not have detected spurious clusters) as
possible and the selection function describing the completeness and
purity as a function of signal, position and redshift should be known
precisely. The survey observables should correlate as tightly as
possible with mass.  In tension with these requirements, surveys must
also be constructed within the context of limited
resources.

\subsubsection{X-RAY SURVEYS}
\label{sec:obs_surveys_xray}

\begin{figure}
 \centerline{
   \epsfxsize=6cm
   \epsfbox{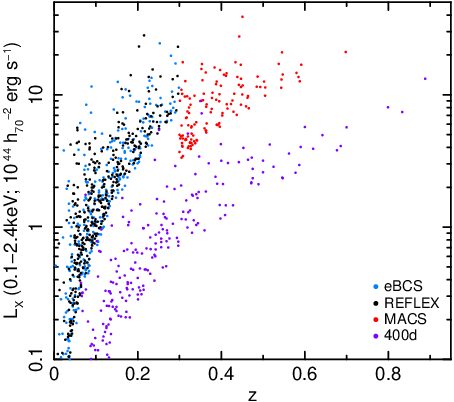}
  }
  \caption{X-ray luminosities and redshifts for four \ROSAT{} cluster
catalogs -- extended BCS (blue; \citealt{Ebeling1998MNRAS.301..881E,Ebeling0003191}),
REFLEX (black; \citealt{Bohringer0405546}), MACS (red;
\citealt{Ebeling0009101,Ebeling0703394,Ebeling1004.4683}), and 400d
(purple; \citealt{Burenin0610739}) -- sub-samples of which have been
used in cosmological studies.}
  \label{fig:Xray_surveys}
\end{figure}

X-ray observations currently offer the most mature and powerful
technique for constructing cluster catalogs. The primary advantages of
X-ray surveys are their exquisite purity and completeness, and the
tight correlations between X-ray observables and mass.

Galaxy clusters are simple to identify at X-ray wavelengths, being the
only X-ray luminous, continuous, spatially extended, extragalactic
X-ray sources. Clusters have typical soft X-ray band luminosities of
$10^{44}$ erg/s or more, and spatial extents of several arcmin or
larger, even at high redshifts. Given modest angular resolution,
e.g. $\Delta \theta \sim 1$~arcmin (full width half maximum, FWHM;
easily achievable) and tens of detected counts, the X-ray emission
from galaxy clusters can be detected against a background populated
otherwise only sparsely with point-like active galactic nuclei.

The first X-ray cluster catalogs constructed for cosmological work
(\citealt[][often called the `Brightest 50' or B50 catalog]{Edge90}; 
\citealt{Gioia90}) were based on the Ariel V and HEAO-1 all-sky surveys, and
pointed observations made with the {\it Einstein Observatory} and {\it
EXOSAT} (see also \citealt{Lahav89}). These catalogs provided early evidence for evolution in the
X-ray luminosity function of clusters \citep{Edge90,Gioia90, Henry92}
and were used subsequently in a series of pioneering cosmological
works \citep[e.g.][]{Henry91, Viana96, Kitayama9604141, Henry97,
Eke9802350}.  

These catalogs were eventually superseded by surveys carried out with
the \ROSAT{} satellite.  This mission, launched in June 1990, had two
main parts: the \ROSAT{} All-Sky Survey \citep[RASS;][]{Voges99},
spanning the first 6 months; and pointed observations, which took
place over the next 8 years.  The main instrument aboard \ROSAT{}, the
Position Sensitive Proportional Counter, had a modest point spread
function (PSF; $\sim 1$~arcmin FWHM in survey mode), but low
background and a wide field of view ($\sim 2$ degree diameter).

The main cluster catalogs constructed from the RASS and used in
cosmological studies include the ROSAT Brightest Cluster Sample
\citep[BCS;][]{Ebeling1998MNRAS.301..881E}, which covered the northern
hemisphere at high Galactic latitudes and low redshifts ($z<0.3$) to a
flux limit of $4.4 \times 10^{-12}$ \ergpcmsqps (0.1--2.4 keV); the
ROSAT-ESO Flux-Limited X-ray Galaxy Cluster Survey
\citep[REFLEX;][]{Bohringer0405546}, which covered the southern sky at
low redshifts to a flux limit of $3.0 \times 10^{-12}$ \ergpcmsqps~ in
the same band; the HIFLUGCS sample \citep{Reiprich0111285} of the
X-ray brightest clusters at high Galactic latitudes, with $F_{\rm X}>
2.0 \times 10^{-11}$ \ergpcmsqps (0.1--2.4keV); and the Massive Cluster
Survey \citep[MACS;][]{Ebeling1004.4683}, which extended this work to
higher redshifts $0.3<z<0.5$ and slightly fainter fluxes $F_{\rm
X}>2.0 \times 10^{-12}$ \ergpcmsqps. Other cluster surveys have been
constructed from the RASS, or are in the process of being constructed,
but have not yet been used to derive rigorous cosmological
constraints.

A number of X-ray cluster catalogs have also been constructed based on
serendipitous discoveries in the pointed phase of the ROSAT mission.
Notable among these are the ROSAT Deep Cluster Survey \citep[RDCS;][]
{Rosati98} and the 400 Square Degree ROSAT PSPC Galaxy Cluster
Survey \citep[400d;][]{Burenin0610739}, which have been used to derive
cosmological constraints. These catalogs cover much smaller areas than
the RASS, but reach an order of magnitude or more fainter in flux
(\figref{fig:Xray_surveys}).

A second major advantage of X-ray surveys is the observed strong
correlation between X-ray luminosity and mass across the entire flux
and redshift range of interest. These quantities follow a simple power
law relation (Section~\ref{sec:constraints:scaling}), with a
dispersion in luminosity at a given mass of $\sim 40\%$ and no
significant outliers \citep{Mantz0909.3099}. The density-squared
dependence also makes the X-ray survey signal from clusters relatively
insensitive to projection effects. Thus, an X-ray survey of sufficient
depth can be translated straightforwardly into statistical knowledge
of the distribution of massive halos.

In principle, X-ray surveys could be constructed using even
lower-scatter mass proxies (Section~\ref{sec:obs_masses_proxies}) such
as temperature or center-excised luminosity as the survey
observable. However, given the ease and depth to which total X-ray
luminosity can be measured, these lower-scatter mass proxies are
typically used as auxiliary data \citep[e.g.][]{Mantz0909.3098,Wu10}.

The primary disadvantage of X-ray cluster surveys is that they can
only be carried out from space, which makes their construction
relatively expensive.

\subsubsection{OPTICAL SURVEYS}
\label{sec:obs_surveys_optical}

The first extensive cluster catalog was constructed at optical
wavelengths by George Abell \citep{Abell58} based on visual inspection
of photographic plates from the Palomar Observatory Sky Survey.  Abell
identified clusters as concentrations of 50 or more galaxies in a
magnitude range m3 to m3+2 (where m3 is the magnitude of the third
brightest cluster member) and radius $R_\mathrm{A} = 1.5 h^{-1}$ Mpc (with
distance estimated based on the magnitude of the tenth brightest
galaxy). Clusters were further characterized into richness and
distance classes. Abell's catalog was updated and extended to the
southern sky by \citet{Abell89} (hereafter ACO). The final ACO sample
has more than 4000 clusters. An additional, early optical
cluster catalog extending to poorer systems was compiled by Zwicky and
collaborators \citep[see e.g.][]{Zwicky61}, although the search
criteria were less strict than Abell's.

\citet{HuchraGeller82} applied a percolation algorithm to an early CfA
redshift catalog to identify a set of 92 nearby groups and clusters.
Using 4-m class telescopes and a mix of photographic plate and CCD
observations, \citep{GunnHoesselOke86} opened high redshift cluster
studies by identifying 418 systems over $\sim \!  150$ deg$^2$
extending out to $z=0.92$.  Spatial and photometric matched filter
methods \citep[\eg][]{Postman96} as well as the introduction of N-body
simulations to calibrate projection effects
\citep[\eg][]{vanHaarlem97} marked the beginning of the modern era of
optical cluster cosmology.

Because the cores of galaxy clusters are dominated by red, early-type
galaxies, an effective way to reduce the impact of projection effects
is to use color information to select for overdensities of red
galaxies \citep[\eg][and references therein]{Gladders05}.   The Red-Sequence Cluster Survey (RCS), a sample of 956 clusters identified with a single ($R_c - z^\prime$) color, provided the first modern cosmological constraints using optical selection \citep{Gladders0603588}.  

To cover a broad range of redshifts, multi-color photometry is needed to track the
intrinsic 4000 angstrom break feature of old stellar populations as it
reddens.  The five-band photometry of the Sloan Digital Sky Survey
(SDSS) enabled such selection.  The maxBCG catalog
\citep{Koester0701265} of 13,823 clusters with optical richness $\Ngal
\ge 10$ was produced using $g-r$ colors and spans the redshift range
$0.1<z<0.3$.  Cosmological constraints from this sample
\citep{Rozo0902.3702} are discussed below.  Recently, larger SDSS
clusters samples have become available, identified using photo-z
clustering \citep{Wen0906.0803}, a Gaussian mixture modeling extension
of the maxBCG method \citep{Hao1010.5503}, and an adaptive matched
filtering approach \citep{Szabo1011.0249}.  These catalogs contain
between 40000 and 69000 clusters spanning $z \ltsim 0.6$, and cover
roughly 8000 deg$^2$ of sky.

A primary challenge to cosmological analysis using such catalogs is
the definition of robust mass proxies that possess minimal and
well-understood scatter across the full mass and redshift ranges of
interest.  Projection of filamentary structures and small groups along
the line of sight has a greater impact on optical cluster catalogs
than X-ray, and these effects introduce a degree of skewness into the
mass-observable relations \citep{Cohn0706.0211}.  Uncertainty in
modeling this and other selection effects currently limits the
constraining power offered by the large sample sizes of optical
cluster catalogs.

\subsubsection{SZ SURVEYS}
\label{sec:obs_surveys_sz}

The first large catalogs of galaxy clusters selected from observations
of the SZ effect are currently under construction, using measurements
made with the South Pole Telescope
\citep[SPT;][]{Vanderlinde1003.0003,Carlstrom09}, the Atacama
Cosmology Telescope \citep[ACT;][]{Kosowsky06,Marriage10} and the
\Planck{} satellite \citep{Bartlett08,Plancksurvey11}.  The primary
advantage of SZ surveys is that, in contrast to X-ray and optical
measurements, the SZ signal of a cluster does not undergo surface
brightness dimming. SZ surveys are therefore well-suited, in
principle, to searches for massive clusters at high redshifts. The
surveys mentioned above are each expected to produce catalogs of
hundreds of massive systems at intermediate-to-high redshifts.
Challenges for these projects include determining the optimal
observables (i.e. the best mass proxies) to measure from the
survey data in the current, low signal-to-noise ratio regime;
calibrating the mass scaling for these observables; and understanding
in detail the impact of contamination by radio and infrared sources
\citep{Sehgal10}.  Projection effects are also expected to be
more significant for SZ surveys than for X-rays \citep{Shaw08}.

\subsection{Mass Measurements and Mass Proxies} 
\label{sec:obs_masses}

\subsubsection{X-RAY MASSES} \label{sec:obs_masses_xray}

Accurate measurements of cluster masses provide a cornerstone of
cosmological work. X-ray mass measurements are based on the assumption
of hydrostatic equilibrium (HSE) in the ICM. For a spherically
symmetric system in HSE, the measured gas density and temperature
profiles can be related to the total mass \citep[e.g.][]{Sarazin88},
\begin{equation}
M(r) = -\frac{r\,kT(r)}{G \mu \mproton}\left[ \frac 
{\dln n}{\dln r} + \frac 
{\dln T}{\dln r} \right],
\end{equation}
where $M(r)$ is the mass within radius $r$, $T(r)$ is the ICM
temperature, $n(r)$ is the gas particle density, $G$ is Newton's
constant, $k$ is the Boltzmann constant, and $\mu \mproton$ is the
mean molecular weight. Note that the mass within radius $r$ depends
more strongly on the temperature than the density at that radius.

Hydrostatic equilibrium requires that the gravitational potential
remain stationary on a sound crossing time; that all motions in the
gas be subsonic; and that forces other than gas pressure and gravity
are unimportant. The hydrostatic method can therefore not be applied
robustly to systems undergoing major merger events, nor to regions of
otherwise relaxed clusters where these assumptions break down, e.g. in
their central regions where strong AGN feedback effects are commonly
observed \citep{Fabian03,Forman05,McNamara0709.2152}.

Out to intermediate radii, measurements of the gas temperature and
density profiles with \Chandra{} or \XMM{} are straightforward. At
large radii ($r\gtsim r_{500}$), however, where the X-ray emission is
faint, such measurements become challenging. Recent advances in this
regard have been made with the \Suzaku{} satellite, and opportunities
for additional progress remain
(\secref{sec:future_outskirts}). Potentially increased levels of
non-thermal pressure support
\citep[e.g.][]{Nagai0609247,Pfrommer07,Mahdavi0710.4132} and gas
clumping \citep{Simionescu10} can also complicate measurements at
large radii.

A number of approaches have been used in implementing the hydrostatic
method. The most common, which employs relatively strong priors, uses
parameterized fits to the observed, projected surface brightness and
temperature profiles; these are then used to calculate the appropriate
partial derivatives at each radius to determine the mass profile
\citep[e.g.][]{Cavaliere76,Jones84,Pratt02,Vikhlinin06}. A second, arguably
preferable, approach
employs a non-parametric deprojection of the brightness and
temperature data, but assumes that the mass distribution follows a
well-motivated parameterized form (e.g. Equation~\ref{eq:NFW}; \citealt{Allen01}, \citealt{Schmidt07}); this approach
simultaneously provides a framework for testing the validity of
various mass models. In the case of very high quality X-ray data, a fully
non-parametric deprojection of the surface brightness and temperature
data can be employed, without additional, regularizing assumptions
\citep[e.g.][]{Nulsen10}.

X-ray mass measurements are relatively insensitive to triaxiality
\citep{Gavazzi05}. For dynamically relaxed clusters, and for
measurements out to intermediate radii, simulations indicate that
hydrostatic X-ray masses should exhibit modest scatter ($\ltsim 10$\%)
and be biased low by $\sim 10-15$\%
\citep[e.g.][]{Evrard90,Nagai0609247,Meneghetti10}, due primarily to kinetic
pressure arising from residual gas motions.

\subsubsection{OPTICAL MASSES} \label{sec:obs_masses_optical}

Like the X-ray method, optical-dynamical mass measurements are based
on the assumption of dynamical equilibrium, with the galaxies used as
test particles in the cluster.  The mass enclosed within radius $r$ is
given by the Jeans equation \citep[e.g.][]{Binney87,Carlberg97}

\begin{equation}
M(r) = -{r\, \sigma_{\rm r}^2(r)\over G}
        \left[{{d \ln{\sigma_{\rm r}^2}}\over{d\ln{r}}} +
        {{d\ln{\nu}}\over{{d\ln{r}}}} +2\beta\right],
\label{eq:jeans}
\end{equation}

\noindent Where $\nu(r)$ is the galaxy number density, $\sigma_{\rm
r}(r)$ the 3-dimensional velocity dispersion, and $\beta$ the velocity
anisotropy parameter. These quantities can be determined under model
assumptions from the projected galaxy number density and velocity
dispersion profiles.

An advantage of the optical dynamical method over the X-ray method is
that it is insensitive to several forms of non-thermal pressure
support that affect X-ray mass measurements (e.g. magnetic fields,
turbulence, and cosmic ray pressure). The galaxy population can also
be observed at high contrast out to large radii. However, where the
X-ray gas is a collisional fluid that returns rapidly to equilibrium
following a disruption, the galaxies are collisionless and relax on a 
longer timescale \citep{White1005.3022}. Whereas
X-ray mass measurements are relatively insensitive to triaxiality, the
galaxy velocity anisotropy must be accounted for. The precision of
optical dynamical measurements is also limited by the finite number of
galaxies. While identifying the center of a cluster is straightforward
at X-ray wavelengths, at optical wavelengths this can be a source of
significant uncertainty.

The infall regions of clusters form a characteristic trumpet-shaped
pattern in radius-redshift phase-space diagrams, the edges of which
are termed caustics. The identification of these caustics enables
mass measurements out to larger radii (up to $10 h^{-1}$ Mpc), to 
an accuracy of $\sim 50$\% \citep[e.g.][]{Rines06}.

\subsubsection{LENSING MASSES} \label{sec:obs_masses_lensing}

In contrast to X-ray and optical dynamical methods, gravitational
lensing offers a way to measure the masses of clusters that is free of
assumptions regarding the dynamical state of the gravitating matter
\citep{Bartelmann10}. Weak lensing methods have an important role in
cosmological work: while triaxiality is expected to introduce scatter
in individual (deprojected) mass measurements at the level of tens of
per cent \citep{Corless07,Meneghetti10},
for statistical samples of clusters and using suitable mass
estimators, working over optimized radial ranges and with good
knowledge of the redshift distribution of the background population,
weak lensing measurements are expected to provide almost unbiased
results on the mean mass \citep{Clowe04,Corless09,Becker10}.

The most common technique employed in weak lensing mass measurements
is fitting the observed, azimuthally-averaged gravitational shear
profile with a simple parameterized mass model
\citep[e.g.][]{Hoekstra07}.  Stacking analyses of clusters detected in
survey fields have also proved successful in calibrating the mean
mass--observable scaling relations down to relatively low masses
\citep[e.g.][]{Johnston07,Rykoff0802.1069,Leauthaud10}.

Strong lensing enables precise measurements of the projected masses
through regions enclosed by gravitational arcs. In combination with
weak lensing, strong lensing constraints can improve significantly the
absolute calibration of projected mass maps
\citep[e.g.][]{Bradac05,Bradac0608408,Meneghetti10}. Deprojected
strong lensing measurements are particularly sensitive to triaxiality
\citep[e.g.][]{Gavazzi05,Oguri05}. Nonetheless, detailed
strong lensing studies of small samples of highly relaxed clusters
have reported mass measurements in good agreement with X-ray results
\citep[e.g.][]{Bradac0806.2320,Newman09,Newman11}.

Some recent works have used clusters with multiple systems of
strongly-lensed arcs to constrain the geometry of the Universe
\citep[][and references therein]{Jullo10}. This approach is related to
the shear ratio test also discussed for weak lensing measurements with
clusters \citep{Taylor07}.

\subsubsection{MASS PROXIES}
\label{sec:obs_masses_proxies}

A good mass proxy should be straightforward to measure and correlate
tightly with mass, exhibiting minimal dispersion across the mass and
redshift range of interest. Robust, low-scatter mass proxy information
for just a small fraction (appropriately selected) of the clusters in
a survey can boost significantly its constraining power with respect
to self-calibration alone \citep[e.g.][]{Mantz0909.3098,Wu10}.

The total X-ray luminosity has an intrinsic dispersion at fixed mass
of $\sim 40\%$ \citep{Mantz0909.3099, Vikhlinin0805.2207}. The scatter
in optical richness at fixed mass for the MaxBCG catalog is also $\sim
40\%$, with a modest inferred non-Gaussian tail toward low masses
\citep{Rozo0902.3702}. The scatter in the projected, integrated SZ
flux at fixed mass is expected to be somewhat smaller ($20-30$\%;
\citealt{Hallman07,Shaw08}), although this is yet to be measured
robustly from data. (Note that the scatter in the observed, projected
SZ flux is significantly larger than for the predicted, intrinsic
signal; \figref{fig:stanek10}.) Since total X-ray luminosity, optical
richness and integrated SZ flux are all survey observables, their
scatter versus mass tends to play a significant role in tests based on
cluster counts even when more precise mass measurements for a fraction of
the clusters are available from follow-up data.

Although not the basis for cluster surveys, other observables that
correlate tightly with mass can provide powerful mass proxies. For the
most massive clusters, the X-ray emitting gas mass is strongly
correlated with total mass, with an observed scatter $<10\%$ at fixed
mass \citep{Allen0706.0033}.  The X-ray temperature, and the product
of gas mass and temperature ($\Yx=kT\Mgas$), have observed scatters
$\ltsim 15\%$ (e.g. \citealt{Arnaud0709.1561,Vikhlinin0805.2207,Mantz0909.3099}); it is thought that the
tightness of the $\Yx$--$M$ relation may extend to lower masses than
for either the $\Tx$--$M$ or $\Mgas$--$M$ relations
\citep{Kravtsov0603205}. Center-excised X-ray luminosity also traces
mass extremely well, with an observed scatter $<10\%$
(\citealt{Mantz0909.3099}; see \secref{sec:future:coreevol}). Weak
lensing mass measurements exhibit larger scatter (tens
of per cent). However, the minimal predicted bias in the mean mass for
statistical samples of lensing measurements
(Section~\ref{sec:obs_masses_lensing}) makes the combination
with X-ray measurements particularly promising.

While the relation between the three-dimensional dark matter velocity
dispersion and mass is predicted to be tight \citep{Evrard08},
velocity anisotropy and projection effects cause the one-dimensional
velocity dispersion to exhibit larger scatter ($10-15$\%, implying a
scatter in mass at fixed velocity dispersion of $\sim 40\%$;
\citealt{White1005.3022}).  The relationship between galaxy and dark
matter velocity dispersion, the velocity bias, also remains uncertain.

\section{CURRENT COSMOLOGICAL CONSTRAINTS} \label{sec:cosmo_constraints}

The past decade has seen marked improvements in the data and analysis techniques employed in cluster cosmology. \Chandra{} observations of relaxed clusters provided important measurements of the distance scale, confirming the recent acceleration of the cosmic expansion. Improvements in mass measurements led to a convergence in estimates of $\sigma_8$ from cluster counts, and X-ray surveys extending to $z\gtsim0.5$ provided the first constraints on dark energy from the growth of structure. We begin this section with a review of the latest results from measurements of the number and growth of clusters, as well as the closely related problem of constraining scaling relations. We then review the state of  other cluster-based probes of cosmology. We conclude by summarizing the latest constraints on dark energy from the combination of galaxy cluster data with other, independent probes.  The application of cluster measurements to other areas of fundamental physics are discussed in \secref{sec:physics}.

\subsection{Counts and Clustering} \label{sec:constraints:growth}

\subsubsection{STATISTICAL FRAMEWORK} 
\label{sec:constraints:procedure}

 Because the cosmological model influences predictions for scaling relation observations, and vice versa (\secref{sec:theory:selection}), the two must be constrained simultaneously. The statistical framework needed for this approach, in which a subset of detected clusters are targeted for more detailed observations including mass estimation, is described by \citet{Mantz0909.3098}. Without reproducing the details here, we enumerate the components of the statistical model:
\begin{enumerate}
\item The mass function and expansion history, which together predict the number of clusters as a function of mass and redshift, $d^2\bar{N}/dz\,d\ln M=(dn/d\ln M)(dV/dz)$.
\item Stochastic scaling relations which describe the (multivariate) distribution of observables as a function of mass and redshift.
\item Measurements for each cluster and associated sampling distributions, describing the probability of obtaining particular measured values given true cluster properties (i.e. a model for the measurement errors), accounting for any covariances in measured quantities. We note that all quantities entering the scaling relations need not be measured for every cluster, apart from (necessarily) those determining cluster selection (e.g., mass follow-up need not be complete).
\item The selection function, quantifying the probability of a cluster being detected and included in the sample.
\end{enumerate}
Using the likelihood function corresponding to this formalism (see \citealt{Mantz0909.3098}), constraints on cosmological and scaling relation parameters are obtained by marginalizing over the true, unknowable properties of each cluster, constrained by the measured values. Straightforward adaptations of this approach, for example incorporating a cluster bias model (\secref{sec:theory:countsclustering}) or stacked gravitational lensing data for poor clusters and groups, are possible.

\subsubsection{LOCAL ABUNDANCE AND EVOLUTION} \label{sec:constraints:growth_results}

We concentrate our review on three recent, independent results based on measurements of the local abundance and evolution of clusters from X-ray and optical surveys, as well as preliminary results from new SZ surveys.

\citet{Mantz0709.4294,Mantz0909.3098} studied the BCS \citep{Ebeling1998MNRAS.301..881E}, REFLEX \citep{Bohringer0405546} and Bright MACS \citep{Ebeling1004.4683} cluster samples compiled from the RASS. This survey strategy, covering a large fraction of the sky to relatively shallow depth, is optimized to the task of finding the largest clusters at the expense of depth in redshift. The Mantz et~al. data set consists of 238 clusters with masses $M_{500}>2.7\E{14}\Msun$ distributed over the redshift range $0<z<0.5$. Pointed \Chandra{} or \ROSAT{} follow-up observations were used to measure $\Mgas$ for 94 clusters, while the remaining 144 had measurements only of the cluster redshift and survey flux. The \Mgas--\Mtot{} relation was marginalized over using hydrostatic mass estimates for relaxed clusters from \citet{Allen0706.0033}. The data set was used to simultaneously constrain cosmological parameters and the $\Lx$--$M$ and $\Tx$--$M$ relations using the method described above and detailed in \citet{Mantz0909.3098}.

\citet{Vikhlinin0812.2720} also used X-ray selected clusters, but pursued a different survey strategy. Their data set consists of disjoint low-redshift and high-redshift samples, with 49 clusters (originally culled from several RASS samples) at redshifts $0.025<z<0.22$ (nearly all $<0.1$)  and an additional 36 clusters serendipitously detected in the 400 Square Degree Survey \citep{Burenin0610739}  at $0.35<z<0.9$. This survey strategy, covering a smaller area of the sky to greater depth, naturally finds fewer of the most massive clusters, but extends to lower masses ($M_{500}>1.3\E{14}\Msun$) and higher redshifts. All 85 clusters in the full data set were followed up with \Chandra{}, and their masses were estimated using either $\Mgas$ or $\Yx$ as a proxy; the proxy--mass relations were calibrated using hydrostatic mass estimates for a sample of well observed, low-redshift clusters. The analysis includes empirical constraints on the scaling relations and corrections for selection effects, though not using exactly the approach described in \secref{sec:constraints:procedure}.

\begin{figure}
  \centerline{
    \hspace{1mm}
    \epsfxsize=5.63cm
    \epsfbox{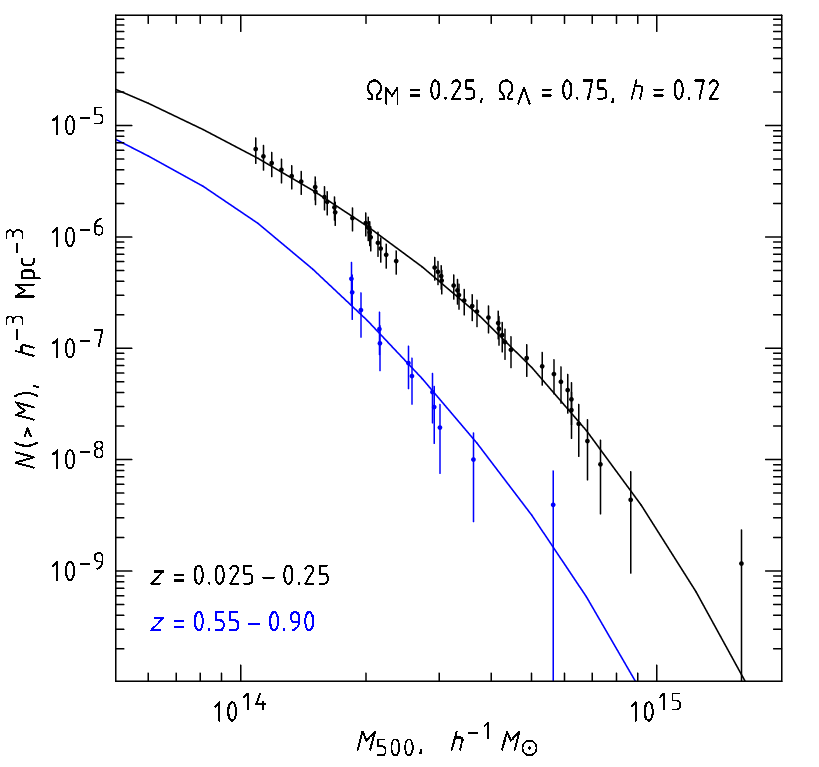}
    \hspace{10mm}
    \epsfxsize=5.6cm
    \epsfbox{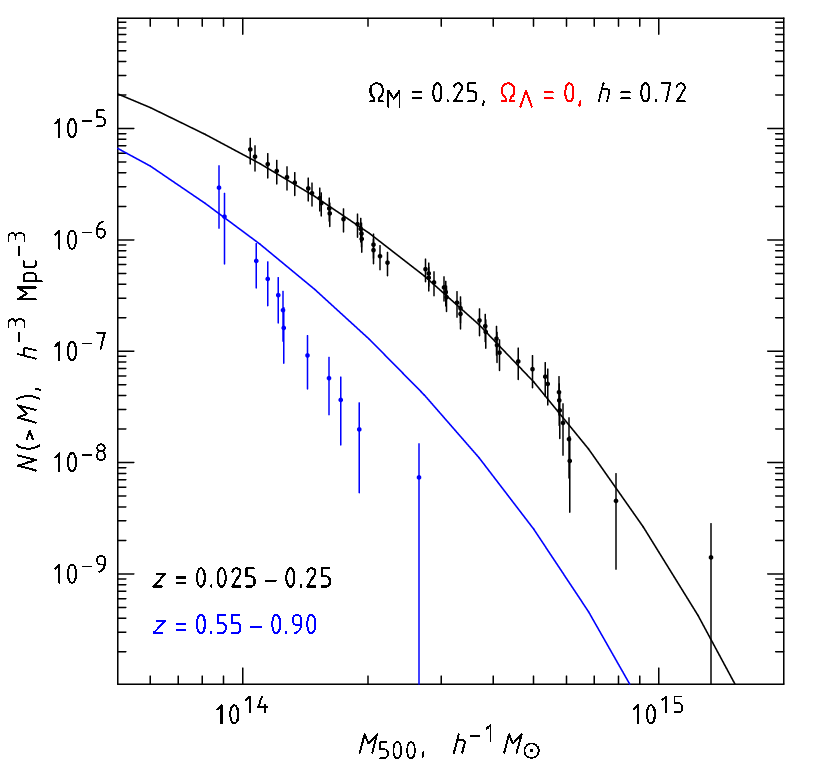}
  }
  \vspace{5mm}
  \centerline{
    \epsfxsize=6cm
    \epsfbox{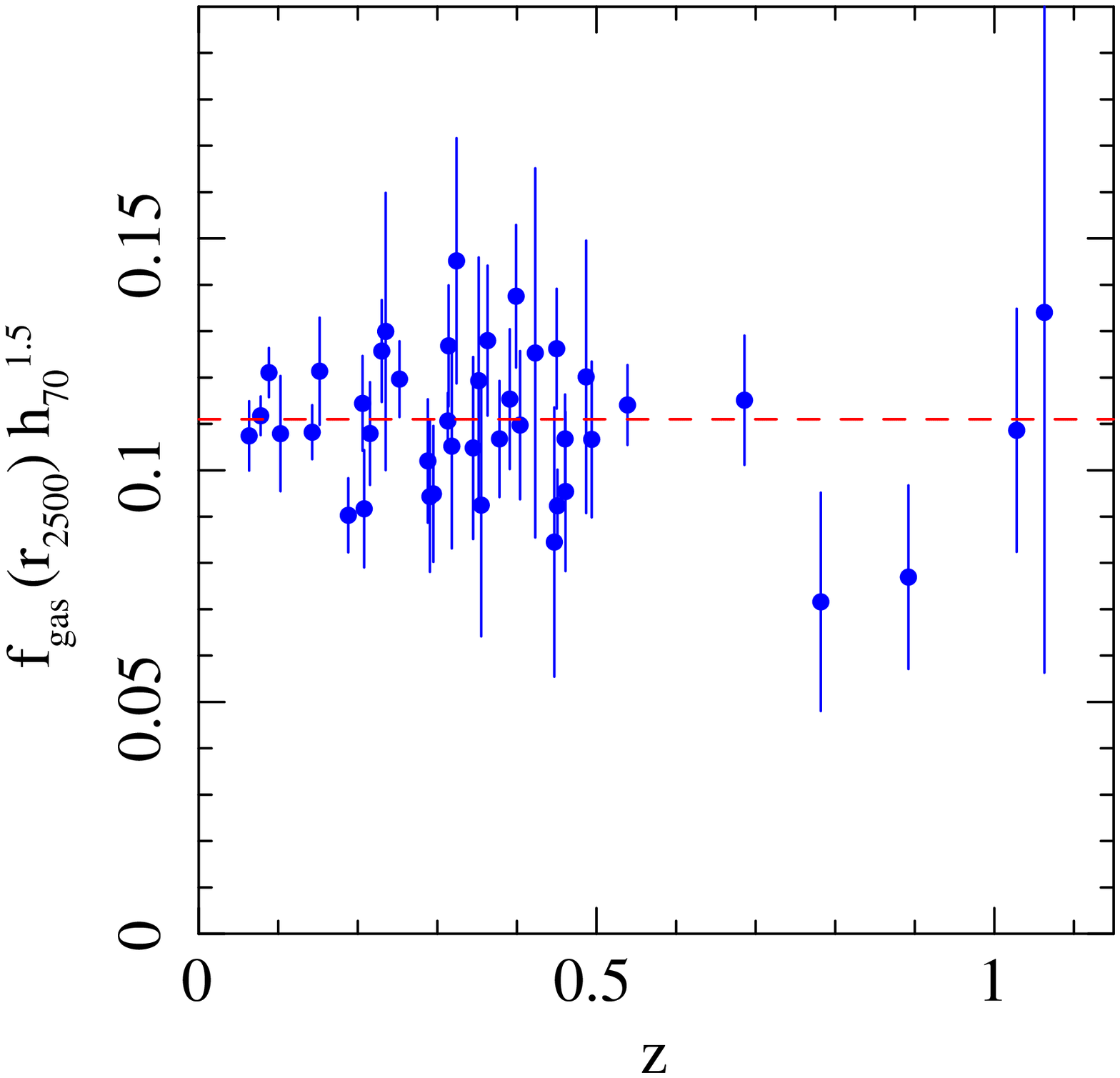}
    \hspace{6mm}
    \epsfxsize=6cm
    \epsfbox{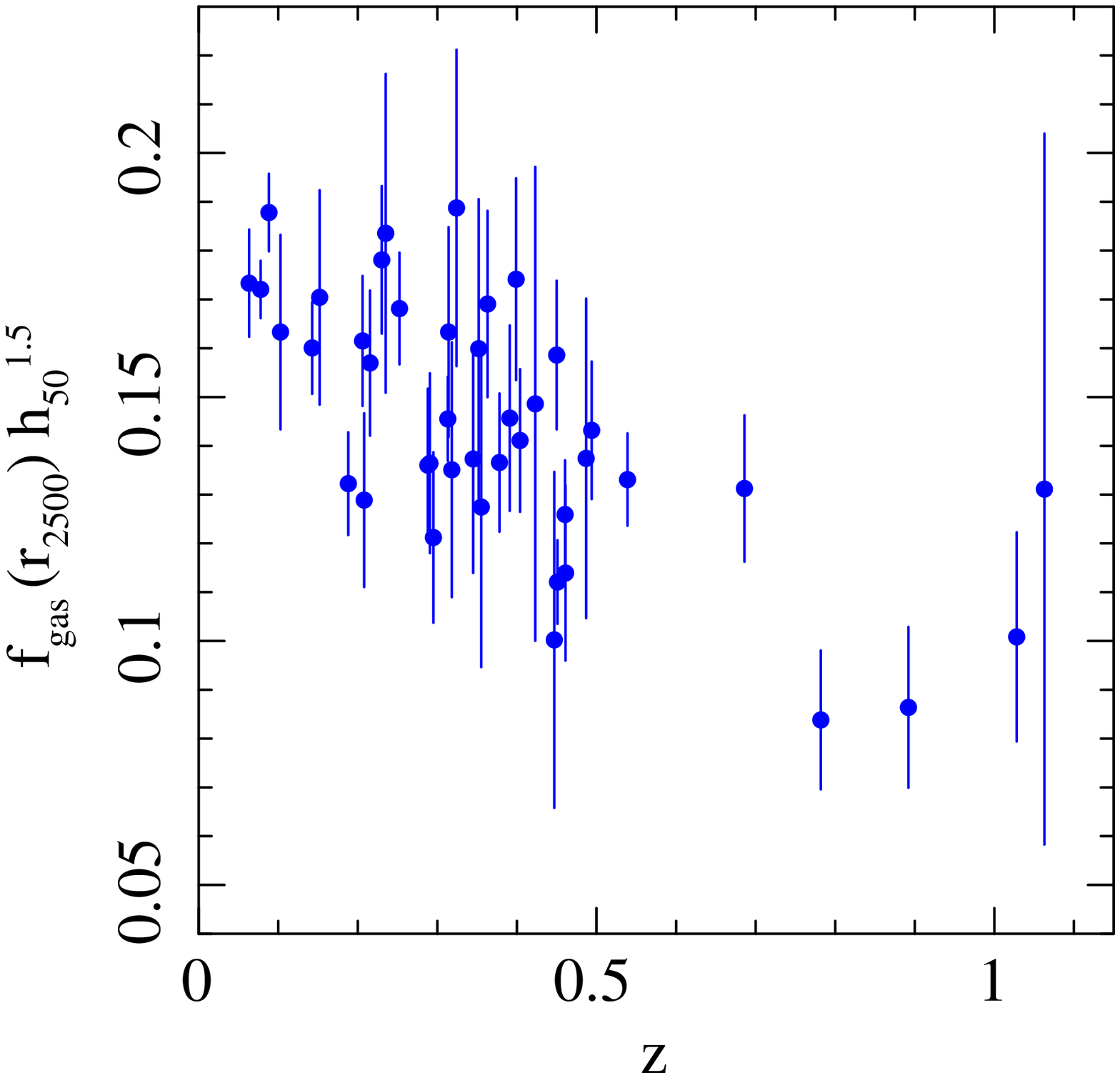}
  }
  \caption{Examples of cluster data used in recent cosmological work. {\it Top}: Measured mass functions of clusters at low and high redshifts are compared with predictions of a flat, \LCDM{} model and an open model without dark energy (from \citealt{Vikhlinin0812.2720}). {\it Bottom}: $\fgas(z)$ measurements for relaxed clusters are compared for a $\Omegam=0.3$, $\Omegal=0.7$, $h=0.7$ model ({\it left}, consistent with the expectation of no evolution) and a $\Omegam=1.0$, $\Omegal=0.0$, $h=0.5$ model ({\it right}; from \citealt{Allen0706.0033}). For purposes of illustration, cosmology-dependent derived quantities are shown (mass and \fgas{}); in practice, model predictions are compared with cosmology-independent measurements.}
  \label{fig:constraints:mfcn_fgas}
\end{figure}

The top panels of \figref{fig:constraints:mfcn_fgas} provide a simplified visualization of how constraints on dark energy arise from these data, comparing the observed mass function in two redshift ranges to model predictions for different cosmological models. In particular, an open universe with no dark energy clearly under-predicts the evolution of the mass function over the redshift range of the data.

The optically selected maxBCG sample \citep{Koester0701265} employed by \citet{Rozo0902.3702} probes a different part of the cluster population; it is restricted to lower redshifts than the X-ray samples described above ($0.1<z<0.3$), but extends to lower masses ($M_{500}>7\E{13}\Msun$). This lower effective mass limit, which changes less strongly with redshift compared to X-ray surveys, makes the maxBCG sample significantly larger than the others, with $>10^4$ clusters divided into 9 bins based on optical richness. Mean masses for 5 richness ranges were estimated through a weak gravitational lensing analysis of stacked clusters, providing information from which to constrain the richness--mass relation. The cosmological analysis accounts for the covariance between cluster counts in each richness bin and the mean lensing mass estimates.

The results obtained by these three groups on flat \LCDM{} and constant $w$ models are summarized in \tabref{tab:constraints:basic}. Note that, for the two works which fit $w$ models, the results on $\Omegam$ and $\sigma_8$ are dominated by the low-redshift data and so are not degraded noticeably by the introduction of $w$ as a free parameter; thus all three sets of constraints are directly comparable. The agreement between the different works, as well as others listed in \tabref{tab:constraints:basic}, is encouraging; in particular, the close agreement in the constraints on $\sigma_8$ reflects the relatively recent convergence in cluster mass estimates using different techniques, and our improved understanding of the relevant systematics (\secref{sec:obs_masses}; see also, e.g., \citealt{Henry0809.3832}). Importantly, the concordance \LCDM{} model provides an acceptable fit to the data in each case.

\begin{table}
  {
    \center
    \caption{Recent cosmological results from galaxy clusters$^{a,b}$}
    \vspace{1mm}
    \label{tab:constraints:basic}
   \begin{tabular}{ccccccc}
    \hline\hline \vspace{-2.5ex}\\
    Reference$^c$ & \multicolumn{1}{c}{Data} & $\sigma_8$ & \Omegam{} & $\mysub{\Omega}{DE}$ & $w$ & $h$ \\
    \hline \vspace{-2.25ex}\\
    \multicolumn{7}{c}{{\bf Local abundance and evolution}$^d$} \vspace{0.5ex} \\
    M10 & X-ray    & $0.82 \pm 0.05$        & $0.23 \pm 0.04$        & $1-\Omegam$            & $-1.01 \pm 0.20$        &                        \vspace{0.5ex} \\
    V09 & X-ray    & $0.81 \pm 0.04$        & $0.26 \pm 0.08$        & $1-\Omegam$            & $-1.14 \pm 0.21$        &                        \vspace{1.0ex} \\
    \multicolumn{7}{c}{\bf Local abundance only} \vspace{0.5ex} \\
    R10 & optical  & $0.80 \pm 0.07$        & $0.28 \pm 0.07$        & $1-\Omegam$            & $-1$                    &                        \vspace{0.5ex} \\
    H09 & X-ray    & $0.88 \pm 0.04$        & $0.3$                  & $1-\Omegam$            & $-1$                    &                        \vspace{1.0ex} \\
    \multicolumn{7}{c}{\bf Local abundance and clustering} \vspace{0.5ex} \\
    S03 & X-ray    & $0.71^{+0.13}_{-0.16}$ & $0.34^{+0.09}_{-0.08}$ & $1-\Omegam$            & $-1$                    &                        \vspace{1.0ex} \\
    \multicolumn{7}{c}{{\bf Gas-mass fraction}} \vspace{0.5ex} \\
    A08  & X-ray   &                        & $0.27 \pm 0.06$        & $0.86 \pm 0.19$        & $-1$                    &                        \vspace{0.5ex} \\
    A08  & X-ray   &                        & $0.28 \pm 0.06$        & $1-\Omegam$            & $-1.14^{+0.27}_{-0.35}$ &                        \vspace{0.5ex} \\
    E09 & X-ray    &                        & $0.32 \pm 0.05$        & $1-\Omegam$            & $-1.1^{+0.7}_{-0.6}$    &                        \vspace{0.5ex} \\
    L06 & X-ray+SZ &                        & $0.40^{+0.28}_{-0.20}$ & $1-\Omegam$            & $-1$                    &                        \vspace{1.0ex} \\
   
    \multicolumn{7}{c}{{\bf XSZ distances}} \vspace{0.5ex} \\
    B06 & X-ray+SZ &                        & $0.3$                  & $1-\Omegam$            & $-1$                    & $0.77^{+0.11}_{-0.09}$ \vspace{0.5ex} \\
    S04 & X-ray+SZ &                        & $0.3$                  & $1-\Omegam$            & $-1$                    & $0.69 \pm 0.08$        \vspace{0.5ex} \\
   \hline
    \end{tabular}
  }
  \vspace{1ex}\\
  $^a$ Entries $\mysub{\Omega}{DE}=1-\Omegam$ indicate the assumption of global spatial flatness (or use of \WMAP{} priors requiring flatness at the few percent level); other entries without error bars indicate parameters that were fixed in the corresponding analysis.

  $^b$ Error bars are marginalized (single-parameter) 68.3\% confidence intervals, and include each author's estimate of the systematic uncertainties (with the exception of S04).

  $^c$ A08~=~\citet{Allen0706.0033}; B06~=~\citet{Bonamente0512349}; E09~=~\citet{Ettori09}; H09~=~\citet{Henry0809.3832}; L06~=~\citet{LaRoque0604039}; M10~=~\citet{Mantz0909.3098}; R10~=~\citet{Rozo0902.3702}; S03~=~\citet{Schuecker0208251}; S04~=~\citet{Schmidt0405374}; V09~=~\citet{Vikhlinin0812.2720}.

  $^d$ Cluster surveys extending to redshifts $z\gtsim0.3$ are required to constrain $w$ from the evolution of the mass function. Note that the \Omegam{} and $\sigma_8$ constraints from these works are essentially unchanged, whether or not $w$ is allowed to differ from $-1$.

\end{table}

\begin{figure}
  \centerline{
    \epsfxsize=7cm
    \epsfbox{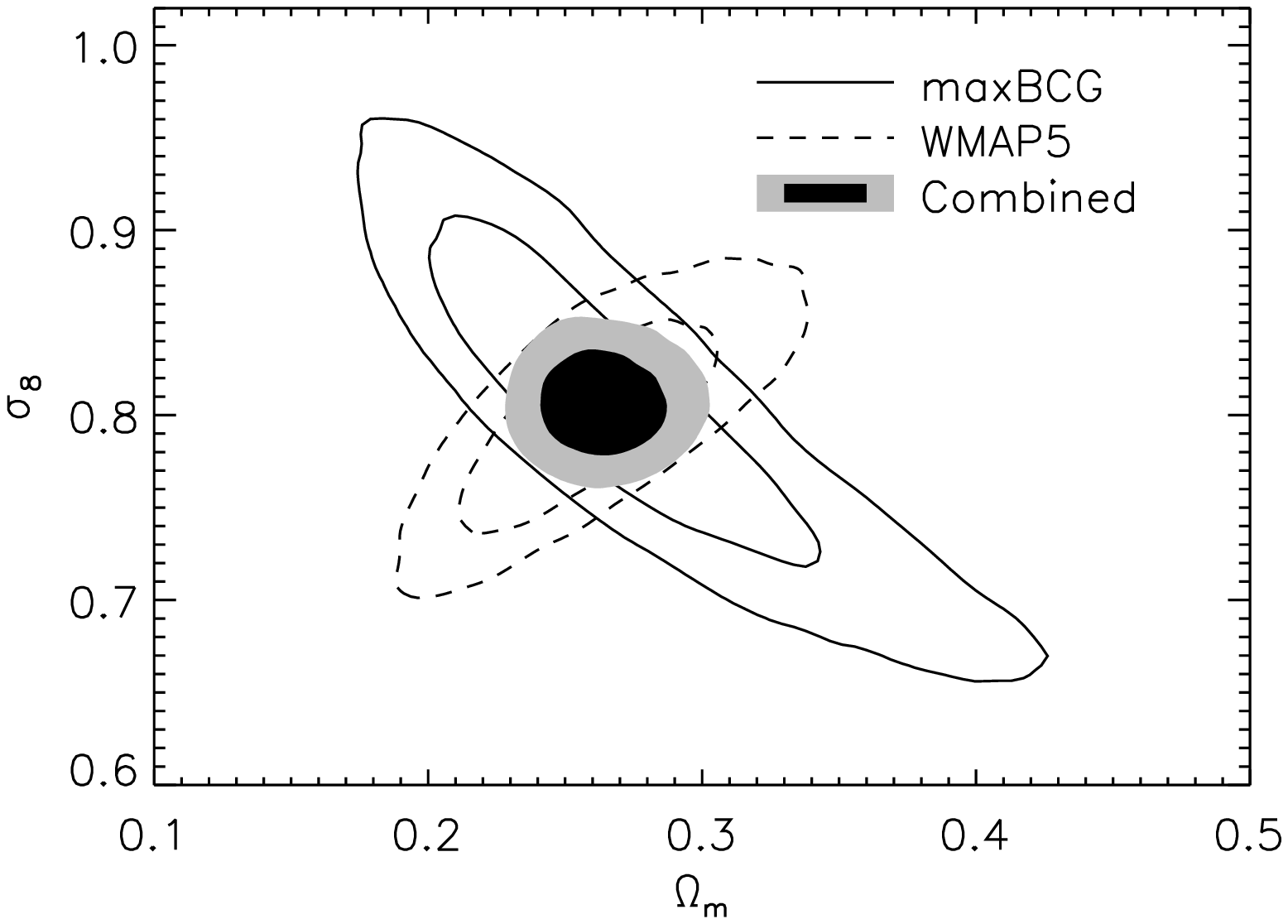}
    \hspace{7mm}
    \epsfxsize=5.5cm
    \epsfbox{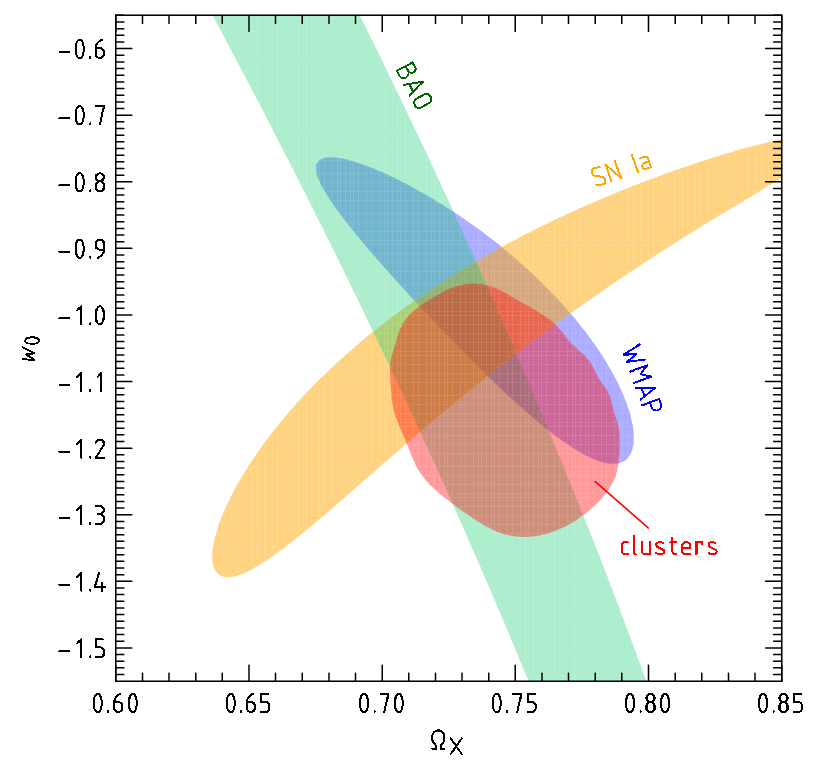}
  }
  \caption{{\it Left}: Joint 68.3\% and 95.4\% confidence regions for the mean matter density and perturbation amplitude from the abundance of clusters in the maxBCG sample ($z<0.3$) compared with those from \WMAP{} data \citep{Dunkley0803.0586} for spatially flat \LCDM{} models. The shaded region indicates the combination of the two data sets. From \citet{Rozo0902.3702}. {\it Right}: Constraints on the dark energy density and equation of state from the abundance and growth of clusters in the 400 Square Degree sample ($z<0.9$) compared with those from \WMAP{}, SNIa \citep{Davis0701510} and BAO \citep{Eisenstein0501171,Percival0608635} for spatially flat, constant $w$ models. {\it Note that, contrary to the convention followed in the other figures, the shaded regions in the right panel indicate only 39.3\% confidence.} The tight contraints from \WMAP{} compared with \figref{fig:constraints:Mantz_w} result from the fact that a simplified analysis was used, in particular neglecting the influence of dark energy on the Integrated Sachs-Wolfe effect (e.g. \citealt{Spergel0603449}). From \citet{Vikhlinin0812.2720}.}
  \label{fig:constraints:Rozo_Wm_s8}
\end{figure}

\figref{fig:constraints:Rozo_Wm_s8} (left panel) shows the joint constraints on $\Omegam$ and $\sigma_8$ obtained by \citet[][solid lines]{Rozo0902.3702}, which display the typical degeneracy between those parameters from cluster survey data. (The degeneracy can be broken, for example, by including cluster \fgas{} data; see \secref{sec:constraints:fgas}.) Also shown are results from 5 years of \WMAP{} data \citep[][dashed lines]{Dunkley0803.0586}, which are tight for the assumed flat \LCDM{} model. Nevertheless, it is evident that the combination of the two types of data (shaded regions) is significantly improved over either one individually.

\begin{figure}
  \centerline{
    \epsfxsize=6cm
    \epsfbox{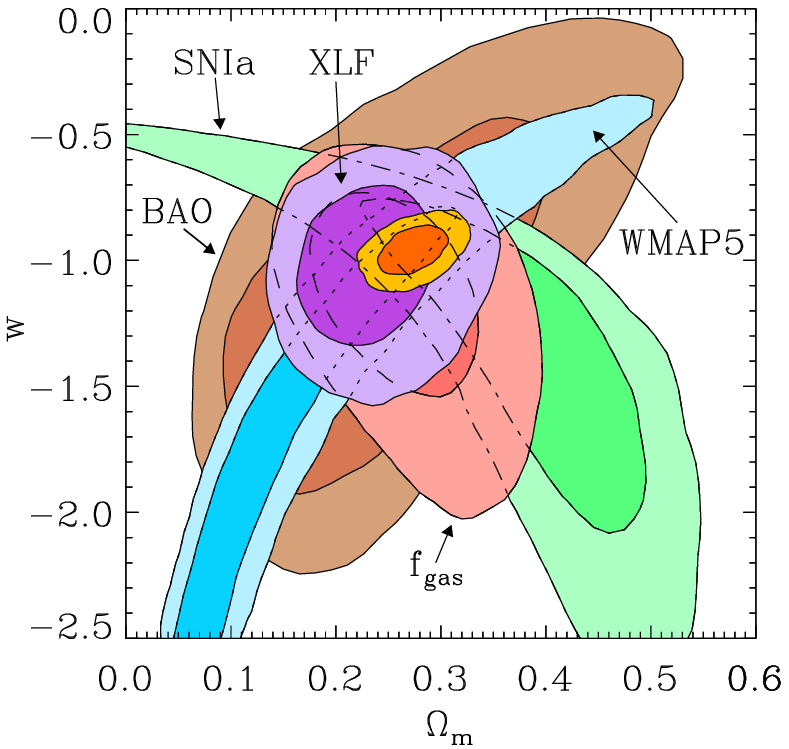}
    \hspace{7mm}
    \epsfxsize=6cm
    \epsfbox{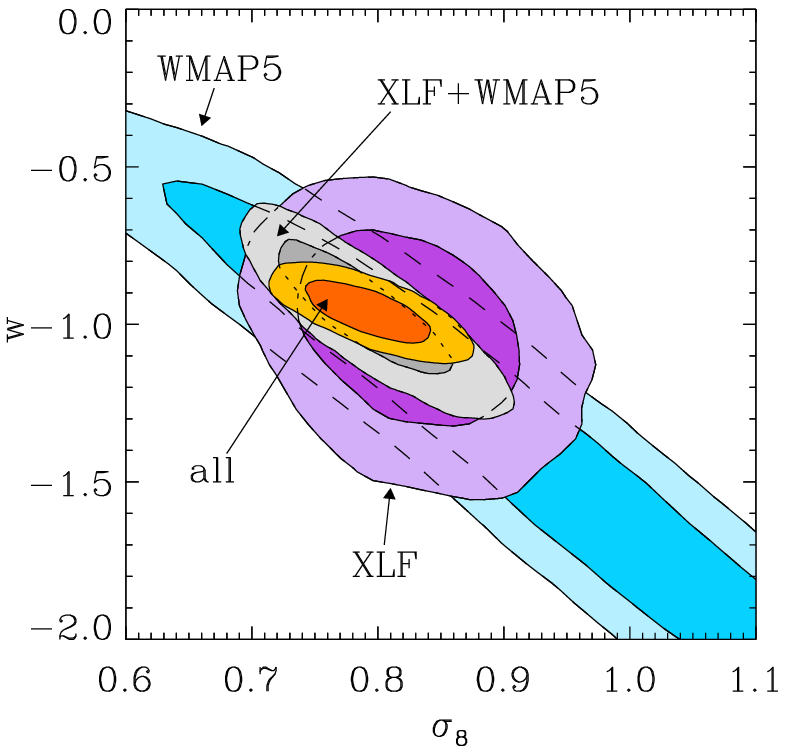}
  }
  \caption{Joint 68.3\% and 95.4\% confidence regions for the dark energy equation of state and mean matter density ({\it left}) or perturbation amplitude ({\it right}) from the abundance and growth of RASS clusters at $z<0.5$ (labeled XLF; \citealt{Mantz0909.3098}) and \fgas{} measurements at $z<1.1$ \citep{Allen0706.0033}, compared with those from \WMAP{} \citep{Dunkley0803.0586}, SNIa \citep{Kowalski0804.4142} and BAO \citep{Percival0907.1660} for spatially flat, constant $w$ models. Combined results from RASS clusters and \WMAP{} are shown in gray in the right panel; gold contours in both panels show the combination of all data sets. The BAO-only constraint differs from that in \figref{fig:constraints:Rozo_Wm_s8} due to the use of different priors. Adapted from \citet[][the BAO constraints in the left panel have been updated to reflect more recent data]{Mantz0909.3098}.}
  \label{fig:constraints:Mantz_w}
\end{figure}

The constraints on constant $w$ models from \citet{Vikhlinin0812.2720} and \citet{Mantz0909.3098} are respectively shown in \figrefs{fig:constraints:Rozo_Wm_s8} (right panel) and \ref{fig:constraints:Mantz_w}, along with results from various other cosmological data sets. The cluster results are in good agreement with one another, as well as with the other, independent data. The $\sim 20\%$ precision constraint on $w$ from cluster growth alone (including systematic uncertainties) is clearly competitive, and further constraining power may be available from theoretical advances (see \secref{sec:physics:lcdm}). The strong degeneracy in CMB constraints evident in \figref{fig:constraints:Mantz_w} also arises in many other models that are more complex than flat \LCDM{}; the ability of cluster data to break this degeneracy by providing precise and independent constraints on $\sigma_8$ (right panel of \figref{fig:constraints:Mantz_w}) makes the combination particularly powerful (e.g. \secref{sec:physics:nu}).

Although their respective surveys are not yet complete, preliminary results have been reported from SZ-detected clusters from the South Pole Telescope \citep[SPT;][]{Vanderlinde1003.0003} and Atacama Cosmology Telescope \citep[ACT;][]{Sehgal1010.1025}. As the number of cluster detections used in these works is small, respectively 21 and 9 at redshifts $0.16<z<1.2$, the data are not capable of producing interesting constraints on their own, but both groups demonstrate consistency with results from the \WMAP{} satellite \citep{Komatsu1001.4538}. Neither group has yet obtained simultaneous constraints on cosmology and the relevant SZ scaling relation, making their final error budgets strongly dependent on the priors chosen to constrain the scaling relation.

\subsubsection{SCALING RELATIONS} \label{sec:constraints:scaling}

To obtain scaling relations appropriate for constraining cosmology, the analysis of the two must be simultaneous and must account properly for the influence of the survey selection function (e.g. \citealt{Stanek0602324} and \citealt{Sahlen09}) and the cluster mass function, as in \citet[][see also \secrefs{sec:constraints:procedure} and \secref{sec:theory:selection}]{Mantz0909.3099}.
Some authors have included corrections for the expected Malmquist bias given a flux limit or selection function \citep{Pacaud0709.1950,Pratt0809.3784,Vikhlinin0805.2207}, but most analyses of scaling relations in the literature employs least-squares regression without detailed consideration of sample selection. Since various cluster samples, with different selection functions, have been used, it is not surprising that results on the slopes, scatters and evolution of the scaling relations have varied widely compared with the formal uncertainties. Another statistical issue is that the covariance that comes about when the scaling quantities are measured from the same observations, for example temperature and hydrostatic masses from X-ray data, is commonly ignored (see \secref{sec:modeling}).

The method used to estimate masses has also varied. Most authors have used the assumption of hydrostatic equilibrium applied to X-ray data, regardless of the dynamical state of the cluster, which must introduce spurious scatter due to departures from equilibrium and non-thermal support (e.g. \citealt{Nagai0609247}). More recently, mass proxies such as gas mass (e.g. \citealt{Mantz0909.3099}) and X-ray thermal energy ($\Yx$; e.g. \citealt{Maughan0703504,Pratt0809.3784,Vikhlinin0805.2207,Andersson1006.3068}), or gravitational lensing signal (e.g. \citealt{Hoekstra07,Johnston07,Rykoff0802.1069,Leauthaud10,Okabe10a}) have been employed.

These variations in mass estimation and analysis methods, in addition to changes to instrument calibration over the years, makes a comprehensive and fair census of scaling relation results problematic. Here we focus on the cosmological importance of scaling relation measurements, citing examples from recent work where the issues mentioned above are at least partially mitigated.

For X-ray and SZ observables, under the assumption of strict self-similarity (no additional heating or cooling), \citet{Kaiser1986MNRAS.222..323K} derived specific slopes and redshift dependences for the power-law form of \eqnref{eq:sbarmu}:
\begin{eqnarray}
  \label{eq:constraints:virial_scaling}
  \frac{\mysub{L}{bol}}{E(z)} & \propto & \left[ E(z)M \right]^{4/3}, \nonumber \\
  \mysub{kT}{mw} & \propto & \left[ E(z)M \right]^{2/3}, \nonumber \\
  E(z)\,Y & \propto & \left[ E(z)M \right]^{5/3},
\end{eqnarray}
where the factors of $E(z)=H(z)/H_0$ are appropriate for measurements made at a fixed critical-overdensity radius. The subscripts `bol' and `mw' reflect the fact that these predictions apply to the bolometric luminosity and mass-weighted temperature. 
Optical richness is more complex to predict, but empirical studies that map galaxies to sub-halos in simulations support a power law richness--mass relation for groups and clusters \citep{Conroy0805.3346}.

\begin{figure}
  \centerline{
    \epsfxsize=5.8cm
    \epsfbox{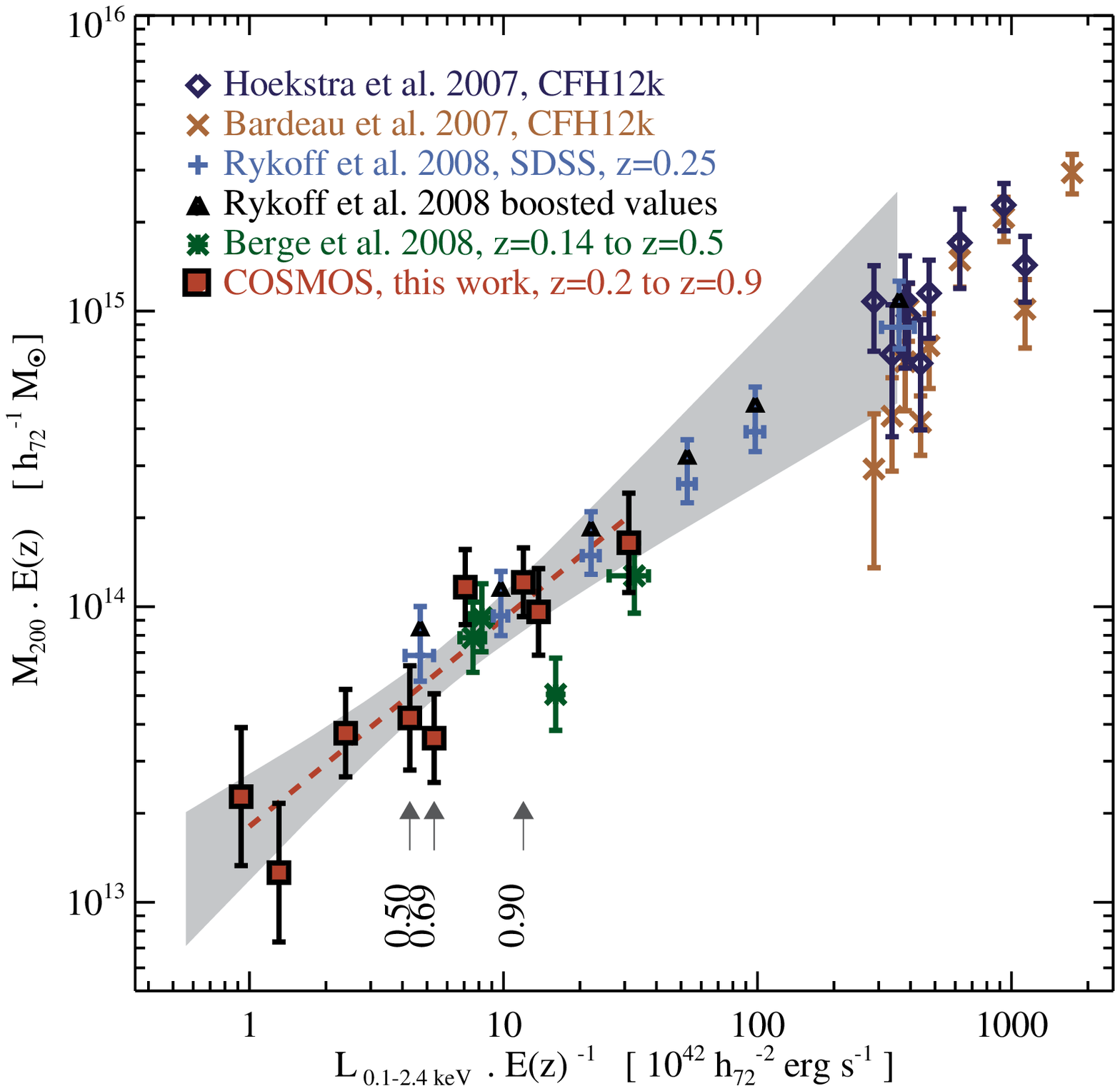}
    \hspace{7mm}
    \epsfxsize=5.8cm
   \epsfbox{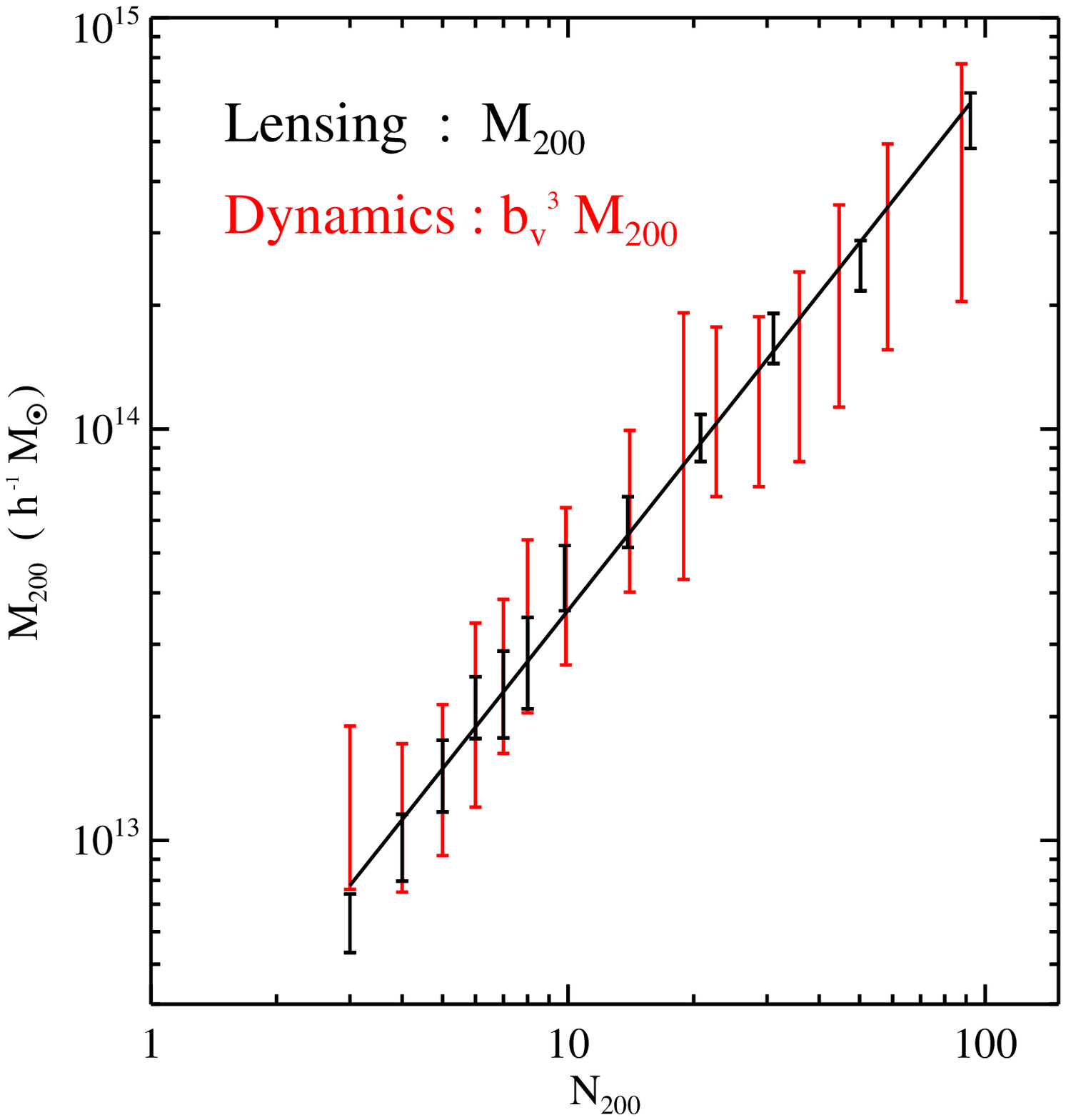}
 }
  \caption{Scaling relations using masses from stacked weak lensing observations. The vertical axes show mean mass derived from stacking clusters in each bin. {\it Left}: X-ray selected clusters from the COSMOS field binned in luminosity \citep[from][see references therein for other data sets plotted]{Leauthaud10}. The gray band corresponds to a 68\% confidence predictive region from a fit to the COSMOS data (brown squares). {\it Right}: Optically selected clusters from SDSS (a subset of the maxBCG catalog) binned in richness. Black points are masses from weak lensing, while red points show mass determinations from galaxy velocity dispersion measurements for the same clusters \citep{Becker0704.3614}. From \citet{Johnston07}.}
  \label{fig:constraints:scaling}
\end{figure}

\begin{figure}
 \centerline{
   \epsfxsize=5.9cm
  \epsfbox{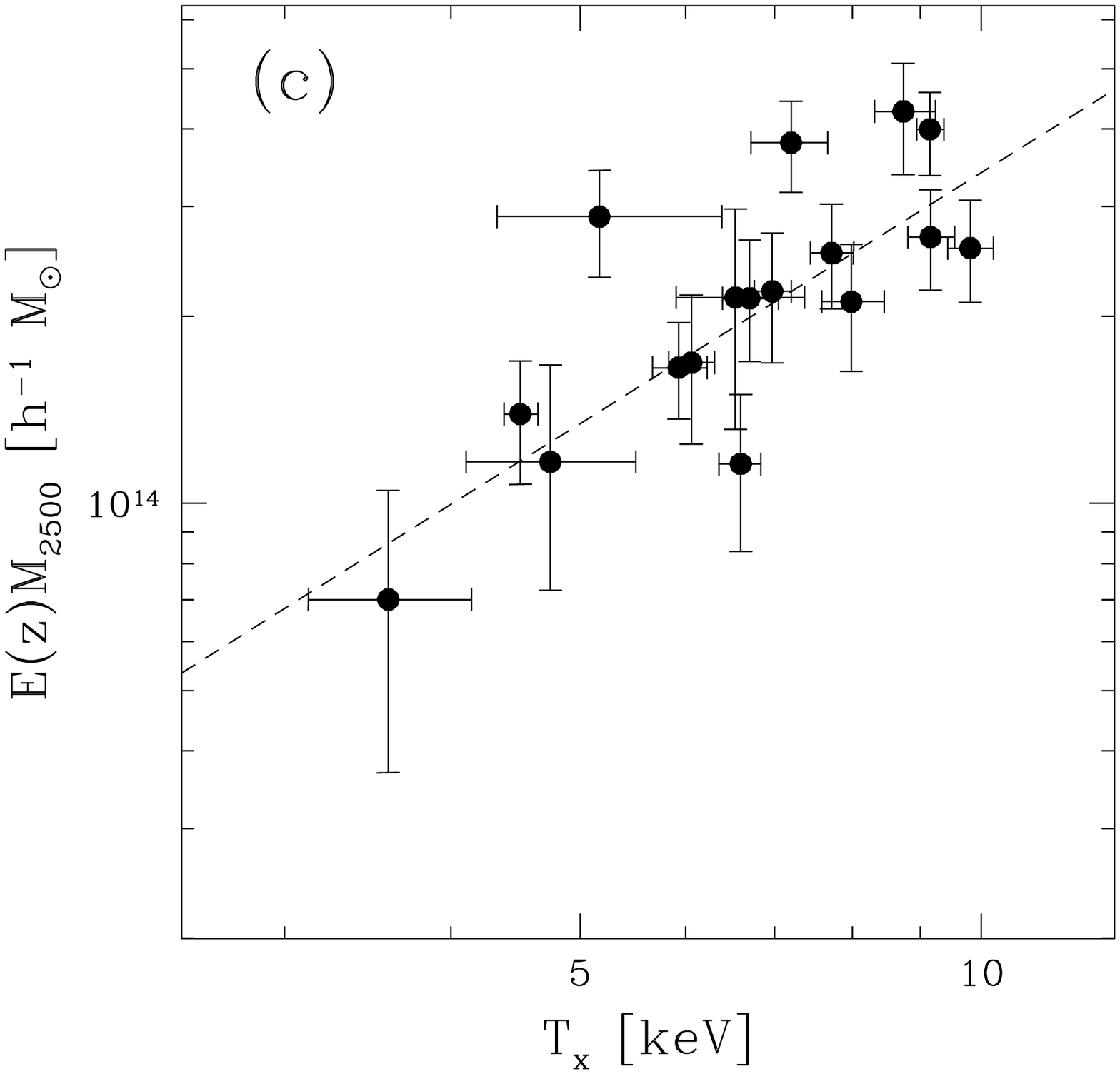}
  \hspace{6mm}
   \epsfxsize=8cm
   \epsfbox{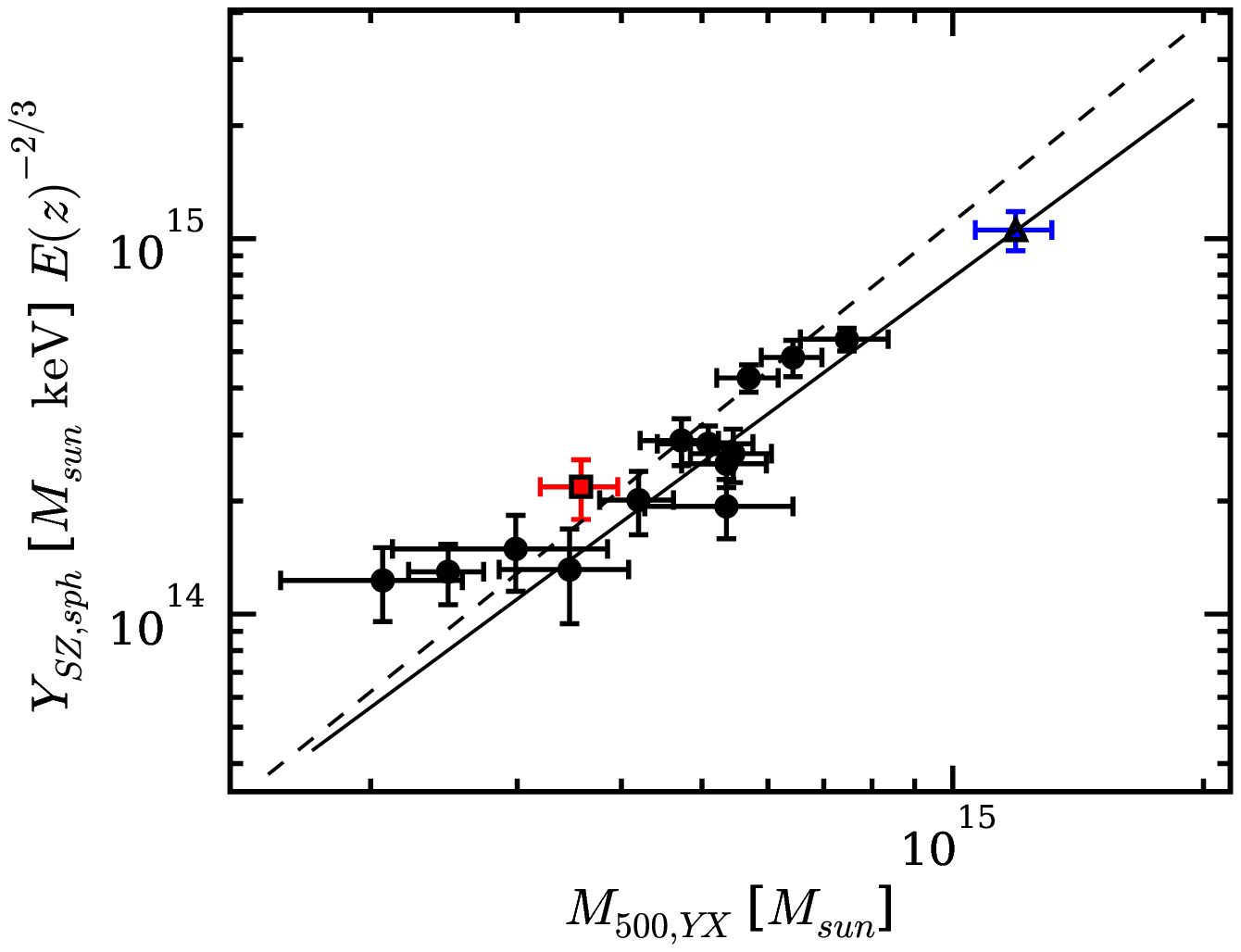}
 }
 \caption{Scaling relations using masses estimated for individual clusters. {\it Left}: weak lensing masses versus ICM temperature for an X-ray selected sample imaged by the Canada--France--Hawaii Telescope. From \citet{Hoekstra07}. {\it Right}: spherically-integrated SZ signal ($Y$) versus mass for SPT-detected clusters, where masses are estimated using X-ray thermal energy ($\Yx$) from \Chandra{} observations as a proxy. $Y$ is measured from the SPT survey data. The solid line is a fit to the data, while the dashed line shows a prediction based on the results of \citet{Arnaud10}. From \citet{Andersson1006.3068}.}
 \label{fig:constraints:other_scaling}
\end{figure}

\figrefs{fig:constraints:scaling} and \ref{fig:constraints:other_scaling} show a few examples of recent scaling relation measurements. \citet{Leauthaud10} present a $M$--$\Lx$ relation for X-ray selected clusters in the COSMOS field by measuring stacked weak lensing masses (left panel of \figref{fig:constraints:scaling}). Under the common assumption of symmetric scatter in the log, stacking on $\Lx$ allows the mean $M(\Lx)$ relation to be recovered, at the cost of losing information about the intrinsic scatter. As the authors note, transforming these results to an $\Lx(M)$ relation introduces a dependence on the mass function. Stacked lensing was also used to determine the mass--richness relation of optically selected clusters in SDSS by \citet{Johnston07}. Their results are shown in the right panel of \figref{fig:constraints:scaling}; red points in the figure show that compatible masses were derived from galaxy velocity dispersion measurements.

For more massive systems, lensing can provide mass estimates on a cluster-by-cluster basis. The left panel of \figref{fig:constraints:other_scaling} shows such a mass--temperature relation from \citet{Hoekstra07} for X-ray selected clusters. Since temperature only weakly influences X-ray detectability, selection bias should be relatively unimportant here, although intrinsic correlation with luminosity can still produces subtle biases relative to a mass-limited sample (\secref{sec:theory:selection}).

An SZ scaling relation from \citet{Andersson1006.3068}, using clusters detected by SPT, appears in the right panel of \figref{fig:constraints:other_scaling}; in this case, masses are estimated from measurements of the X-ray thermal energy, $\Yx$, and an assumed $\Yx$--$M$ relation. This case is instructive in that the plotted $Y$ values are derived from the survey data, with suggestions of Malmquist bias in the flattening observed at the lowest masses (compare with \figref{fig:constraints:selection}; this potential bias is accounted for in the fit in \figref{fig:constraints:other_scaling}).

\begin{figure}
 \centerline{
   \epsfxsize=7.5cm
   \epsfbox{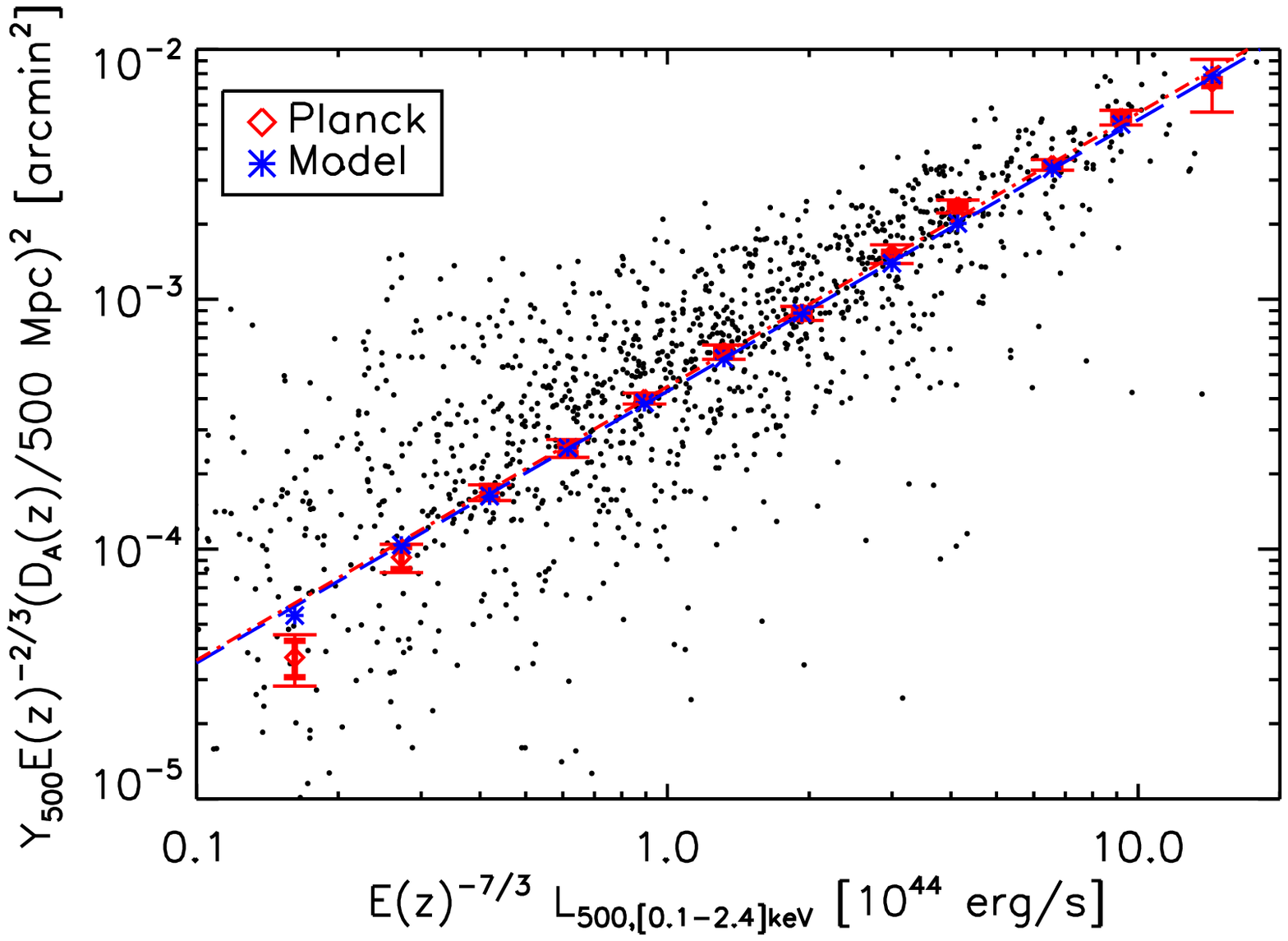}
   \hspace{6mm}
   \epsfxsize=7.5cm
   \epsfbox{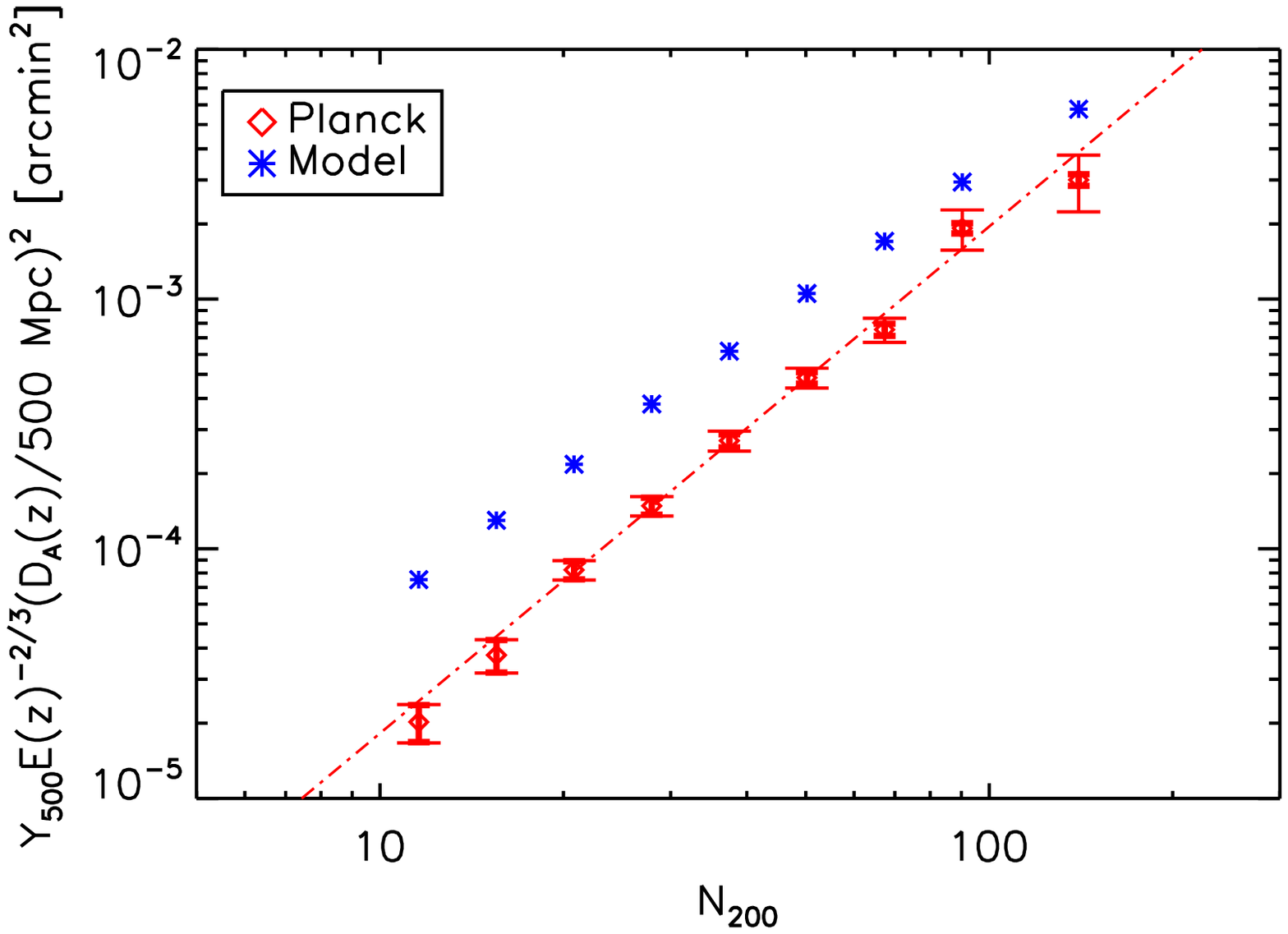}
 }
 \caption{Recent SZ scaling relations from \Planck{}. {\it Left}: scaled, intrinsic SZ signal vs. X-ray luminosity for a sample of X-ray selected clusters, with SZ extraction radii determined from $\Lx$ (see \citealt{Planckxraystack11} for details).  Red diamonds show the $\Lx$-stacked relation, while the blue asterisks show the predicted SZ signal based on X-ray scaling relations. For these clusters, the agreement of the model with measurements is good.   {\it Right}:  Scaled, stacked SZ signal vs. optical richness for optically selected maxBCG clusters (red diamonds) are compared to expectations based on the maxBCG mass--richness relation in combination with X-ray scaling relations (blue asterisks, see \citealt{Planckoptical11} for details).  The origin of the offset between the predictions and measurements is not yet understood.}
 \label{fig:constraints:planck_scaling}
\end{figure}

Recently, the \Planck{} Collaboration released an early SZ-selected
cluster catalog together with an analysis of SZ scalings for existing
X-ray and optical cluster samples \citep[][and references
therein]{Plancksurvey11}.  Measurements of the scaled SZ signal from \Planck{} as a function of
X-ray luminosity for $\sim1600$ X-ray selected galaxy clusters
are shown in the left panel of \figref{fig:constraints:planck_scaling}, along with the stacked, mean SZ signal
in luminosity bins and model expectations
derived from the observed X-ray properties
\citep{Planckxraystack11}. There is good agreement between the model expectations and
the \Planck{} observations, validating the simple model description of
the hot ICM over the mass and redshift range probed.

The right panel of \figref{fig:constraints:planck_scaling} shows the
result of a similar exercise, stacking the \Planck{} SZ signal as a
function of richness for the optically selected maxBCG catalog. Also
shown is the expected signal, based on the observed mean
mass--richness relation for the maxBCG sample \citep{Rozo0809.2794}
and a standard X-ray $Y$--$M$ scaling relation \citep{Planckoptical11}. 
 In this case, the observed \Planck{} SZ
signal lies well below model predictions.  For an X-ray detected
sub-sample of the maxBCG catalog, however, relatively good agreement
between the \Planck{} SZ measurements and model predictions is
observed \citep{Planckoptical11}.

An important difference between the $\Lx$ and richness comparisons is
that the centers and radii used to extract the clusters' SZ signals
from the (relatively low resolution) \Planck{} data were respectively
based on X-ray and optical estimates.  Besides optical center and
scale mis-estimates, other potential sources for the discrepancy in
\figref{fig:constraints:planck_scaling}b include lower than expected
purity in the optical sample (with lower mass halos being enhanced in
richness by projection); unmodeled biases in the scaling
relations used (e.g. the impact of non-thermal pressure support on the
X-ray relations); and the presence of an intrinsic anti-correlation
between the galactic and hot gas mass fractions in the clusters. These
and other possibilities are under investigation.

In general, most current scaling results are consistent with power-law relations over a wide range in mass (e.g. \citealt{Sun0805.2320,Rozo0902.3702}). However, the exponents often differ from the values in \eqnref{eq:constraints:virial_scaling}, due at least in part to the impact of astrophysical processes, mainly star/SMBH formation and associated feedback, discussed in \secref{sec:theory_astrophysics}. For example, the $\Lx$--$M$ slope is commonly found to exceed $4/3$, with the value depending on the mass range explored. There is less consensus on the $\Tx$--$M$ slope, with most estimates consistent with self-similarity (e.g. \citealt{Henry0809.3832}; $m_T=0.65\pm0.03$), but some finding a shallower slope (e.g. \citealt{Mantz0909.3099}; $m_T=0.48\pm0.04$); systematics related to correlated observables, mass estimation (see \secref{sec:modeling}), instrument calibration, selection, and differences in the mass range studied may play a role. There are fewer empirical estimates of the $Y$--$M$ or $\Yx$--$M$ slopes, but most are broadly compatible with the predicted value. We note that departure from self-similar scaling with mass does not necessarily imply departure from self-similar evolution with redshift; for example, the high-mass halos in the preheated simulations of \citet{Stanek0910.1599} have X-ray luminosities that evolve within $10\%$ of the self-similar expectation at $z \le 1$, despite having a slope with mass that is $50\%$ steeper than self-similar.

The marginal intrinsic scatter in each relation determines the degree of selection bias in surveys and each observable's usefulness as a mass proxy; current results are reviewed in \secref{sec:obs_masses_proxies}.

The evolution of the scaling relations is of great importance, since it is potentially degenerate with the cosmological signal of cluster growth. Because the impact of survey limits can vary strongly with redshift for X-ray surveys, it is particularly crucial to account for them in this context. Both \citet{Vikhlinin0805.2207} and \citet{Mantz0909.3099} investigate evolution in the $\Lx$--$M$ relation, finding only marginal $\sim 1\sigma$ departures from the self-similar redshift dependence in \eqnref{eq:constraints:virial_scaling}.
 \citet{Mantz0909.3099} additionally found no evidence for departures from self-similar evolution in the $\Tx$--$M$ or center-excised $\Lx$--$M$ relations (see \secref{sec:future:coreevol}). As current and upcoming surveys expand the reach of cluster samples to $z>1$, obtaining precise constraints on scaling relation evolution will be imperative (see also \secref{sec:modeling}).

The effects of survey bias on scaling relations can be {\it partially} mitigated by selecting clusters based on an observable other than the observable of immediate interest (e.g. \citealt{Rykoff0802.1069}). However, as \secrefs{sec:theory:mvscaling} and \ref{sec:theory:selection} make clear, the amount of residual bias depends on the intrinsic correlation between the two observables at fixed mass. To date, the only empirical estimates of such correlation are for mass and X-ray luminosity at fixed optical richness,
 $r(\Lx,M|\Ngal) \ge 0.85$ \citep{Rozo0809.2794}; and for soft-band X-ray luminosity and temperature at fixed mass, $r(\Lx,\Tx|M) = 0.09 \pm 0.19$ \citep{Mantz0909.3099}. Large, overlapping surveys, and/or surveys coupled with multi-wavelength follow-up campaigns, are required to better constrain the property covariance of clusters; ultimately, such constraints will provide a new level of robustness to cosmological work.

\subsubsection{MOST MASSIVE CLUSTER TESTS} \label{sec:current_most_massive}

In principle, the confirmed existence of even a single galaxy cluster of implausibly high mass would challenge the standard $\Lambda$CDM model with Gaussian initial conditions. Recently, the discovery of high redshift, massive systems such as XMMU\,J2235.3-2557 at $z\sim1.4$ \citep{Mullis0503004,Rosati0910.1716} and SPT-CL\,J0546-5345 at $z\sim1.1.$ \citep{Brodwin1006.5639} have led to reports of possible tension with Gaussian $\Lambda$CDM \citep{Jee09,Holz1004.5349,Hoyle1009.3884}.

Such a test is in some sense an attractively simple alternative to the more involved work discussed above. However, despite the fact that the test involves only one or a few very massive clusters detected in a survey rather than the complete sample, a robust assessment of the likelihood still requires a detailed understanding of the selection function and survey biases, as well as a full accounting for the effects of scatter in the mass--observable scaling relations (\secrefs{sec:theory:selection} and \ref{sec:constraints:procedure}). The accuracy and precision of the cluster mass measurements are also critical, due to the steepness of the high mass tail of the cluster mass function. Errors in mass measurements at the tens of per cent level, for example (as might be expected for weak lensing measurements of an individual cluster), can modify the likelihood of such a cluster being observed by up to an order of magnitude.

\citet{Mortonson1011.0004} estimate confidence limits for the exclusion of the Gaussian $\Lambda$CDM model based on the properties of the most massive galaxy cluster, or $N$ most massive galaxy clusters, detected in a given survey, employing constraints on the expansion history from current data. These authors conclude that none of the presently known high mass, high redshift clusters are in significant tension with the standard Gaussian $\Lambda$CDM paradigm.

\subsubsection{CLUSTERING} \label{sec:constraints:clustering}

The framework described in \secref{sec:constraints:procedure} for simultaneously constraining cosmology and scaling relations has yet to be applied to the spatial clustering of clusters. However, \citet{Schuecker0205342} have obtained cosmological constraints using a method based on the Karhunen-Lo\`eve eigenvectors of 428 clusters from the X-ray selected REFLEX sample above a luminosity of $5.1\E{42}h_{70}^{-2}\erg\second^{-1}$ (see \citealt{Vogeley9601185} for the theory underlying this method of estimating the power spectrum). The constraints are relatively weak compared with more recent results from other methods: $0.6<\sigma_8<2.6$ and $0.07<\Omegam<0.38$ at 95.4\% confidence, fixing $h=0.7$ and without including the systematic uncertainty due to cosmic variance. In \citet{Schuecker0208251}, the analysis was extended to include cluster abundance using an empirical X-ray luminosity--mass scaling relation from \citet{Reiprich0111285}, demonstrating that the combination of the two methods breaks the degeneracy between $\Omegam$ and $\sigma_8$, in particular reducing the uncertainty on $\sigma_8$ to the point where systematics related to mass estimation dominate (\tabref{tab:constraints:basic}).

In principle, the spatial distribution of clusters can be employed in a simpler way, as has been done with individual galaxies, by using the baryon acoustic oscillation signature in the power spectrum as a probe of cosmic distance. \citet{Estrada0801.3485} and \citet{Hutsi0910.0492} respectively analyzed the correlation function and power spectrum of clusters in the optically selected maxBCG catalog, finding a weak ($\sim 2\sigma$) detection of the BAO peak. More recently, \citet{Balaguera-Antolinez1012.1322} found no significant evidence for a BAO feature in the smaller, X-ray selected REFLEX~II catalog.

\subsection{Baryon Fractions} \label{sec:constraints:fgas}

Cluster gas mass fractions can be measured robustly using X-ray, or the combination of X-ray and SZ, data for dynamically relaxed clusters (\secref{sec:observations}). When measured from X-ray data, \fgas{} values within a given angular aperture depend on cosmology as $\fgas(z) \propto \dA(z)^{3/2}$, while the predicted \fgas{} for a given cosmology is given by \eqnref{eq:constraints:fgas_model}. (A more detailed expression for the comparison of measured and predicted \fgas{} values is given by \citealt{Allen0706.0033}, who include terms accounting for instrument calibration, non-thermal pressure, and the relationship between the characteristic radius of the model, e.g. $r_{2500}$, and the aperture of the measurement.) The apparent redshift dependence of \fgas{} measurements on the cosmological background is illustrated in the bottom panels of \figref{fig:constraints:mfcn_fgas}.

In principle, \fgas{} measurements can be made at radii corresponding to any overdensity. In practice, this overdensity should be low enough (i.e. the radius large enough) that non-gravitational feedback effects do not introduce prohibitive scatter; but not so small (i.e. the radius not so large) that the measurements become dominated by the systematic limitations of the instruments. A variety of simulations indicate that radii $\sim r_{2500}$ are sufficiently large to benefit from low \fgas{} scatter (e.g. \citealt{Borgani0906.4370}), while at radii $\gtsim r_{500}$ uncertainties in the X-ray background and the impact of gas clumping can become a concern (\citealt{Simionescu10}; see also \secref{sec:future_outskirts}).

\begin{figure}
  \centerline{
    \epsfxsize=5.5cm
    \epsfbox{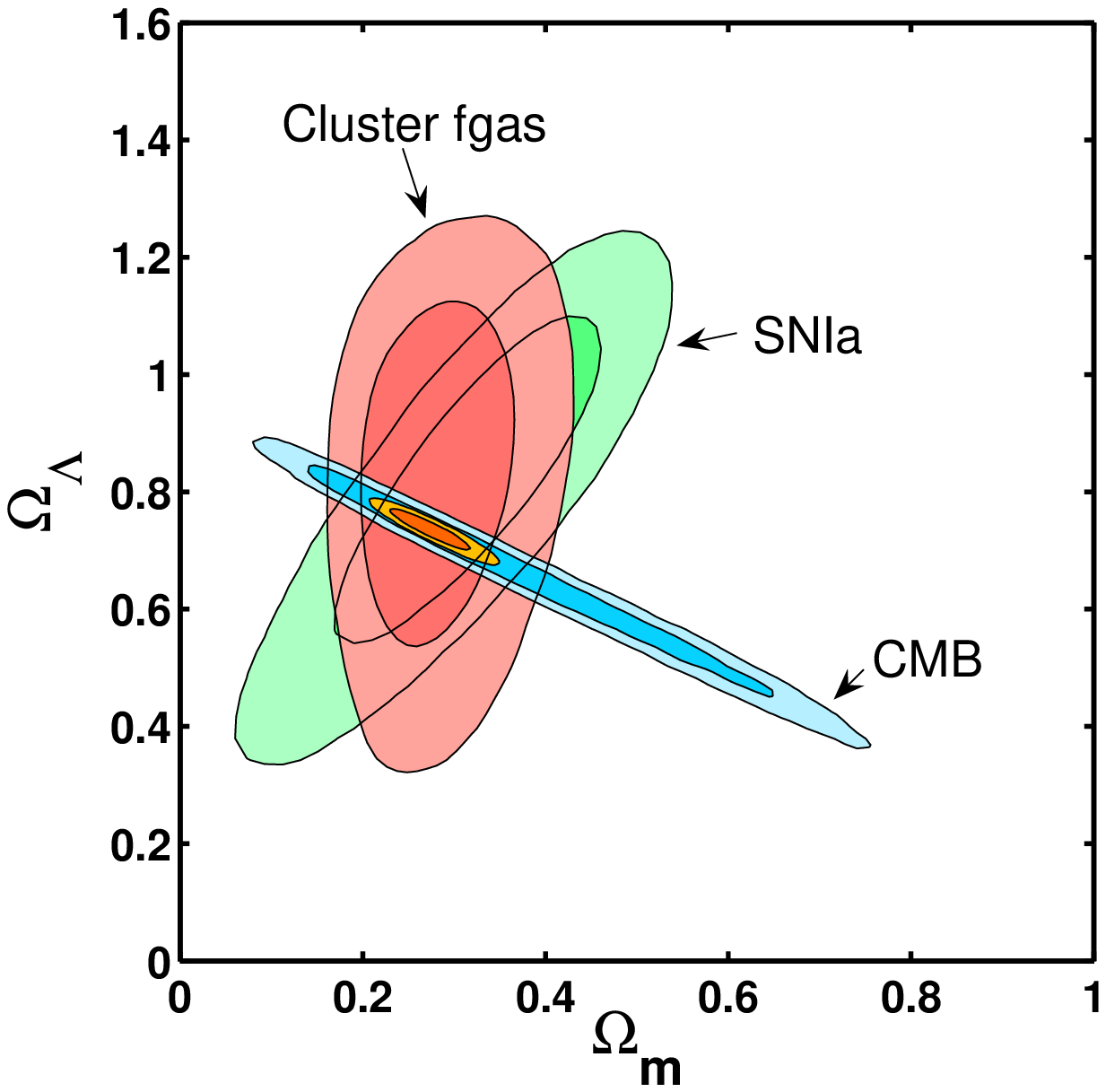}
    \hspace{7mm}
    \epsfxsize=5.5cm
    \epsfbox{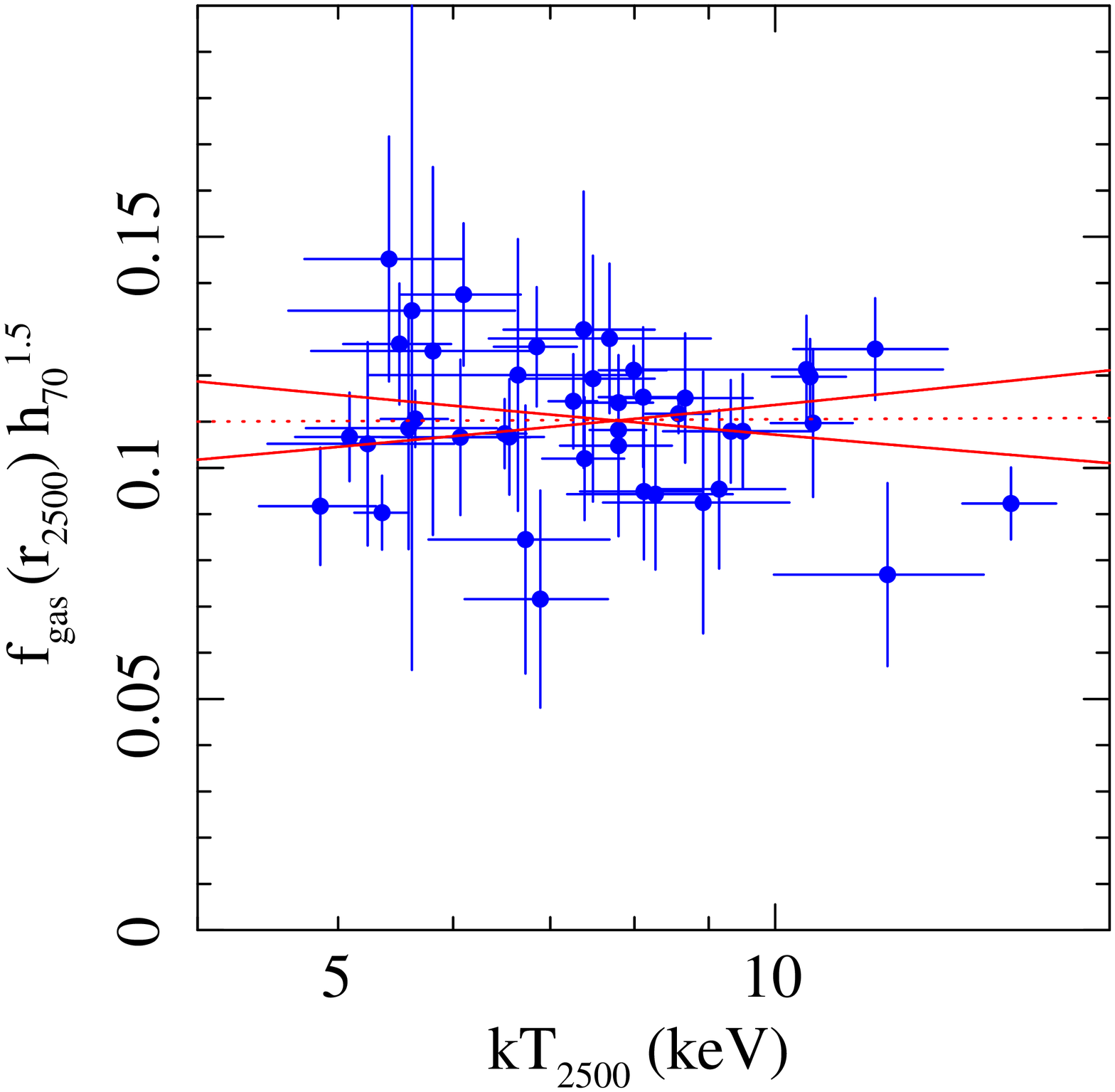}
  }
  \caption{{\it Left}: Joint 68.3\% and 95.4\% confidence regions on \LCDM{} models with curvature from cluster \fgas{} data at $z<1.1$, compared with those from CMB data \citep{Spergel0603449} and SNIa \citep{Davis0701510}. Inner, gold contours show results from the combination of these data. {\it Right}: \fgas{} values as a function of ICM temperature. The measurements are consistent with a constant value over the temperature range explored ($5\keV < k\Tx < 15\keV$). From \citet{Allen0706.0033}.}
  \label{fig:constraints:Allen_Wm_Wl}
\end{figure}

Using this method, \citet[][see also \citealt{Allen0405340}]{Allen0706.0033} obtained cosmological constraints using \fgas{} measurements at $r_{2500}$ from \Chandra{} observations of 42 hot ($k\Tx>5\keV$), relaxed clusters covering the redshift range $0.05<z<1.1$. Their analysis incorporated weak priors on $h$ and $\Omegab h^2$ from Hubble Key Project \citep{Freedman0012376} and big bang nucleosynthesis data, priors on the baryonic depletion and stellar mass fraction of clusters and their evolution (the term $\Upsilon(z)$ in \eqnref{eq:constraints:fgas_model}), and marginalized over systematic allowances accounting for instrument calibration and non-thermal pressure. From \eqnref{eq:constraints:fgas_model}, we see that the normalization of the $\fgas(z)$ curve, combined with the priors on $h$ and $\Omegab h^2$, provides a constraint on $\Omegam$; while the shape of $\fgas(z)$ allows dark energy parameters to be constrained via the apparent dependence of \fgas{} on distance. The results for spatially flat, constant $w$ models and non-flat \LCDM{} models are respectively shown in \figrefs{fig:constraints:Mantz_w} and \ref{fig:constraints:Allen_Wm_Wl}. For the non-flat models, the presence of dark energy is detected at high ($>99.99\%$) confidence, comparable to current SNIa results; the constraints, including systematic uncertainties, are $\Omegam=0.27 \pm 0.06$ and $\Omegal=0.86 \pm 0.19$, with the flat \LCDM{} model yielding an acceptable goodness of fit. For constant $w$ models, the \fgas{} data are again competitive with other cosmological results, obtaining $w=-1.14^{+0.27}_{-0.35}$. The systematic, cluster-to-cluster scatter in \fgas{} is small, $<7\%$, corresponding to only 5\% in distance; this high precision results from the restriction to hot, dynamically relaxed systems for which total masses can be accurately estimated (\secref{sec:obs_masses_xray}).

Interestingly, for clusters with $kT \gtsim 5\keV$, the measured \fgas{} values show no dependence on temperature (right panel of \figref{fig:constraints:Allen_Wm_Wl}), indicating that, for the most massive clusters, the self-similar expectation of constant \fgas{} with mass is realized. At lower temperature and mass (extending to the group scale), a trend of increasing \fgas{} with temperature and mass is observed (e.g. \citealt{Sun0805.2320}).

\citet{Ettori09} have also applied X-ray data to the \fgas{} test, using \Chandra{} measurements at $r_{500}$ for 52 clusters in the redshift range $0.3<z<1.3$, adopting similar priors on $h$ and $\Omegab h^2$ and marginalizing over a prior on $\Upsilon$ (assumed constant with redshift). However, their data set was not restricted to dynamically relaxed systems, resulting in significantly weaker constraints (\tabref{tab:constraints:basic}).

\citet{LaRoque0604039} employed \Chandra{} X-ray and OVRO/BIMA SZ observations of 38 clusters in the redshift range $0.14<z<0.89$ (with no restriction dynamically relaxed systems), finding \fgas{} values consistent with previous X-ray work. (Their analysis did not take advantage of the relative normalization of the X-ray and SZ signals to simultaneously provide a second distance constraint; see \secref{sec:constraints:xsz}.) They adopted the simpler approach of assuming constant \fgas{} and marginalizing over its value, incorporating a \WMAP{} prior on the total density, $\Omegam+\Omegal$. Although this explicitly ignores the information available from the normalization of $\fgas(z)$, their results clearly disfavor a dark matter dominated universe, preferring a low-density universe with dark energy (\tabref{tab:constraints:basic}).

\subsection{XSZ Distances} \label{sec:constraints:xsz}

The different dependence on distance of the gas density inferred from X-ray and SZ observations of clusters can be exploited in a conceptually similar way to \fgas{} data (\secref{sec:theory:xsz}). The most recent contribution is that of \citet{Bonamente0512349}, who measured distances to 38 clusters at redshifts $0.14<z<0.89$. This cosmological test is intrinsically less sensitive to distance than the \fgas{} test, with the signal proportional only to $\dA(z)^{1/2}$ (\eqnref{eq:dAXraySZ}), and currently can constrain only one free parameter. Assuming spatial flatness and fixing $\Omegam=0.3$, \citet{Bonamente0512349} obtained a constraint on the Hubble parameter, $h=0.77^{+0.11}_{-0.09}$, consistent with results from other data such as the Hubble Key Project \citep{Freedman0012376} or the combination of \fgas{} and CMB data \citep{Allen0706.0033}. We note that other works using the same method have typically found somewhat lower best fitting values ($h=0.6$--0.7; e.g. \citealt{Grainge2002MNRAS.333..318G,Schmidt0405374}), but these discrepancies are not significant given the systematic uncertainties.

\subsection{High-Multipole CMB Power Spectrum} \label{sec:constraints:tsz}

The CMB temperature power spectrum at multipoles $\ell \gtsim 1000$ encodes the thermal SZ signature of unresolved clusters at all masses and redshifts (\secref{sec:theory:tsz}). Although the primary CMB power decreases rapidly at these scales, extracting this cosmological information from the tSZ spectrum has proved challenging due to uncertainties in, e.g., the relevant observable--mass scaling relation at low masses and high redshifts; the population of infrared and radio point sources; the magnitude of the integrated kinetic SZ effect; and the form of the electron pressure profile at large cluster radii, where it is poorly constrained by current X-ray data (\citealt{Sehgal10}; \secref{sec:future_outskirts}). The \WMAP{}, SPT and ACT collaborations have all detected excess power at large multipoles, $\ell \gtsim 3000$. Subject to the systematic uncertainties mentioned, their results are broadly in agreement, and are consistent with estimates of $\sigma_8$ obtained from studies of resolved clusters and the primary CMB (\citealt{Dunkley1009.0866,Lueker0912.4317,Komatsu1001.4538}).

\subsection{Evolving Dark Energy Models} \label{sec:physics:lcdm}

\begin{figure}
  \centerline{
    \epsfxsize=6cm
    \epsfbox{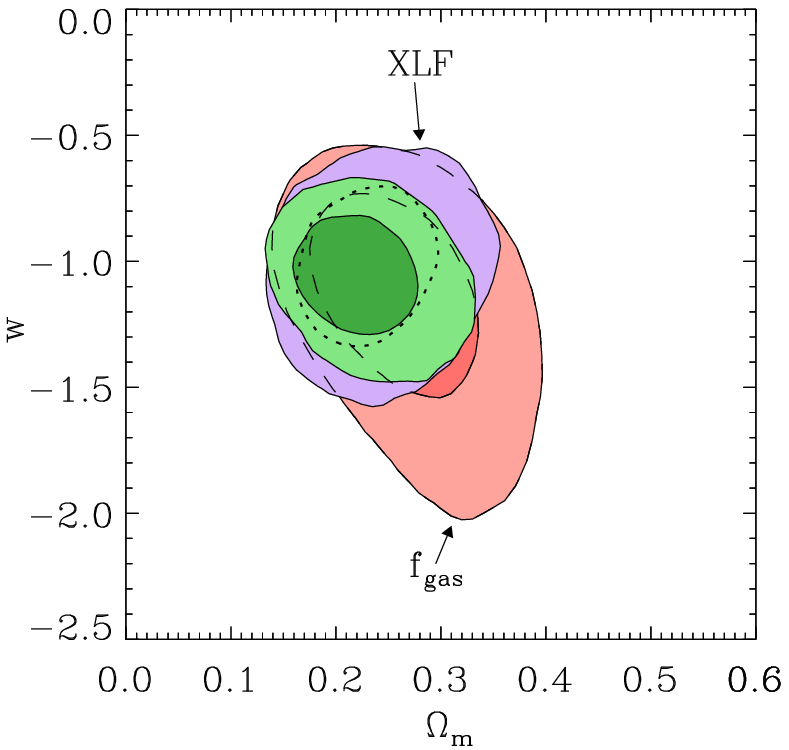}
    \hspace{7mm}
    \epsfxsize=6cm
    \epsfbox{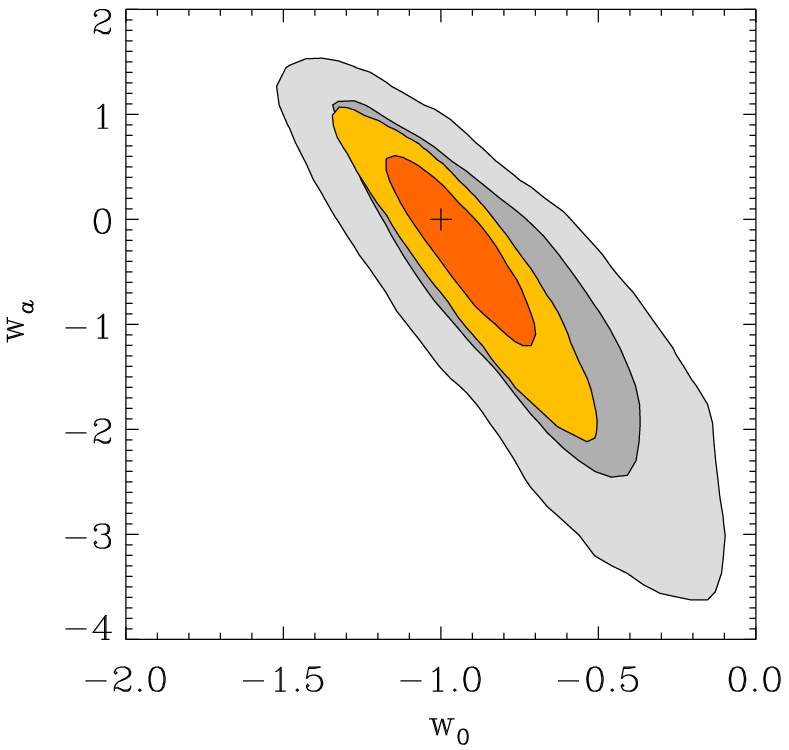}
  }	
  \caption{{\it Left}: Joint 68.3\% and 95.4\% confidence regions for constant $w$ models, using cluster growth (XLF; \citealt{Mantz0909.3098}) and \fgas{} \citep{Allen0706.0033} data and their combination (green contours). These cluster data provide a $15\%$ precision constraint on $w$ ($w=-1.06 \pm 0.15$) {\it without} incorporating CMB, SNIa or BAO data. {\it Right}: Constraints on parameters of the evolving $w$ dark energy model in \eqnref{eq:wwa} from the combination of cluster growth and \WMAP{} \citep{Dunkley0803.0586} data (gray), and with the addition of cluster \fgas{}, SNIa \citep{Kowalski0804.4142} and BAO \citep{Percival0608635} data (gold). The \LCDM{} model ($w(z)=-1$) corresponds to the black cross. From \citet{Mantz0909.3098}.}
  \label{fig:physics:wevol}
\end{figure}

As discussed above, current cluster growth and \fgas{} data can constrain spatially flat models with constant $w$, finding consistency with the cosmological constant model ($w=-1$). Constraints on constant $w$ models from the combination of these cluster data are shown in the left panel of \figref{fig:physics:wevol}. To go beyond this simple description of dark energy, it is necessary to include cosmological data from additional sources in the analysis.

\eqnref{eq:wwa} provides a simple and commonly adopted model of evolving dark energy, in which the equation of state takes the value $w_0$ at $z=0$ and approaches $w_0+w_a$ at high redshift. Constraints on this model were obtained from cluster growth data by \citet{Vikhlinin0812.2720} and \citet{Mantz0909.3098}, assuming spatial flatness and in combination with external CMB, SNIa and BAO (and \fgas{}, in the case of \citeauthor{Mantz0909.3098}) data. In both studies, the results are consistent with the \LCDM{} model ($w_0=-1$ and $w_a=0$; right panel of \figref{fig:physics:wevol}).

A slightly more general model due to \citet{Rapetti0409574},
\begin{equation}
  \label{eq:physics:lcdm_wetmodel}
  w(z) = \frac{\wet z + w_0 \zt}{z + \zt},
\end{equation}
also makes a smooth transition from one value at the present day ($w_0$) to another at early times (\wet{}), but has the advantage that the transition redshift, \zt{}, can be marginalized over. (When $\zt=1$, \eqnref{eq:physics:lcdm_wetmodel} reduces to \eqnref{eq:wwa} with $w_a=\wet-w_0$.) \citet{Mantz0909.3098} used the data sets above to obtain $w_0=-0.88 \pm 0.21$ and $\wet=-1.05^{+0.20}_{-0.36}$, again consistent with \LCDM{}.

Absent a concrete physical model for dark energy, phenomenological models of evolving dark energy can take any form; for example, one possibility that has not yet been investigated using cluster data expands $w(z)$ in principal components (e.g. \citealt{Mortonson0912.3816}). Although these ad-hoc descriptions of dark energy are perfectly straightforward to apply to measurements of the expansion history, their applicability to an analysis based on the cluster mass function or power spectrum is less clear. Generically, descriptions of dark energy as a fluid with $w \neq -1$ should include the effects of spatial variations in dark energy density (e.g. \citealt{Hu0410680}). However, the results from cluster growth so far have made use of mass functions from \LCDM{} simulations, accounting for the value of $w$ only in the expansion history (\citealt{Vikhlinin0812.2720}) or, at most, including the effect of density variations on the linear matter power spectrum (\citealt{Mantz0709.4294,Mantz0909.3098}). In principle, the cluster mass function must also be adjusted to account for the behavior of fluid dark energy on smaller scales and in higher density environments. Encouragingly, preliminary work in this area suggests that the influence of dark energy on the mass function might be readily measurable, resulting in additional constraining power from clusters 
(e.g. \citealt{Creminelli0911.2701}).
Such improvements, however, might come at the cost of requiring additional sophistication in the theoretical description of dark energy (e.g. the dark energy sound speed and viscosity; \citealt{Mota0708.0830}).

\section{OTHER CONTRIBUTIONS TO FUNDAMENTAL PHYSICS} \label{sec:physics}

\subsection{Dark Matter}
\label{sec:physics_dm}

The $\Lambda$CDM paradigm, while providing an excellent model for the
large scale structure of the Universe, incorporates little information
on the physical nature of dark matter. It assumes only that dark
matter is non-baryonic; that it interacts weakly with baryonic matter
and itself; that it emits and absorbs no detectable electromagnetic
radiation; and that the dark matter particles move at sub-relativistic
speeds.

\begin{figure}
  \epsfxsize=12cm
  \centerline{\epsfbox{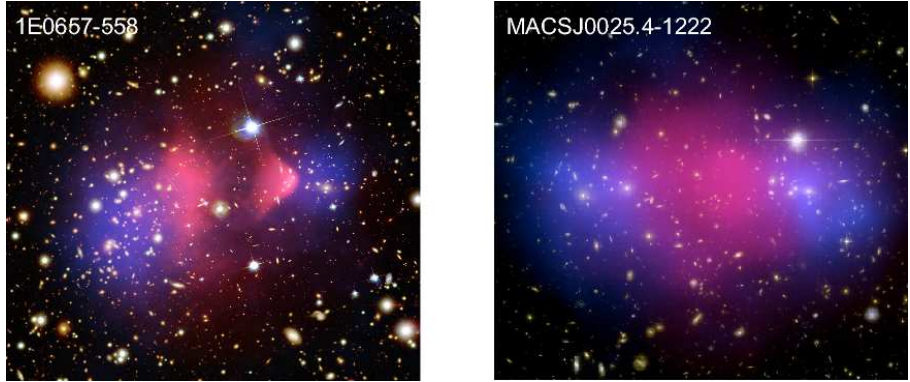}}
  \caption{Hubble Space Telescope optical images of the massive,
merging clusters 1E0657-558 ($z=0.30$) and MACSJ0025.4-1222
($z=0.54$), with the X-ray emission measured with \Chandra{} overlaid in
pink and total mass reconstructions from gravitational lensing data in
blue. The separations of the X-ray and lensing peaks, and the
coincidence of the lensing and optical centroids, imply that the dark
matter has a small self-interaction cross-section. Figure credits:
{\it Left}: X-ray: NASA/CXC/CfA/M.Markevitch et al.; Optical: NASA/STScI;
Magellan/U.Arizona/D.Clowe et al.; Lensing Map: NASA/STScI; ESO WFI;
Magellan/U.Arizona/D.Clowe et al.; {\it Right}: X-ray:
NASA/CXC/Stanford/S.Allen; Optical/Lensing: NASA/STScI/UC Santa
Barbara/M.Brada{\v c}.}
  \label{fig:bullet_macs0025}
\end{figure}

As clusters merge under the pull of gravity, their dark matter halos
and X-ray emitting gas can become separated temporarily, as the gas
experiences ram pressure and is slowed. The observed offsets between
the dark matter and X-ray peaks in the Bullet Cluster (1E\,0657-558;
\citealt{Bradac0608408,Clowe0608407}) and MACS\,J0025.4-1222
(\citealt{Bradac0806.2320}; \figref{fig:bullet_macs0025}), 
both massive merging systems with relatively
simple geometries, require conservatively that the scattering depth
for the merging dark matter cannot be greater than one. Using
gravitational lensing data to estimate the dark matter column
densities through these clusters, \citet{Markevitch04} and
\citet{Bradac0806.2320} use the observed dark matter and X-ray peak
separations to derive limits on the velocity-independent dark matter
self-interaction cross-section per unit mass of
$\sigma/m<5$\,cm$^{2}$\,g$^{-1}$ and $\sigma/m<4$\,cm$^{2}$\,g$^{-1}$ for
1E\,0657-558 and MACS\,J0025.4-1222, respectively. \citet{Randall08}
additionally use the non-detection of an offset between the lensing
peaks and the galaxy centroids for the Bullet Cluster to refine this
constraint to $\sigma/m<1.5$\,cm$^{2}$\,g$^{-1}$. Upcoming surveys
(Section~\ref{sec:new_surveys}), should provide hundreds of similar
examples, removing the current systematic limitations set by small
number statistics.  In combination with multiwavelength follow-up
observations and improved numerical simulations
\citep[e.g.][]{Forero-Romero10}, this should allow the properties of
dark matter in merging clusters to be studied in a robust, statistical
manner.

One of the most remarkable predictions of the CDM model is that the
density profiles of relaxed dark matter halos, on all resolvable mass
scales, can be approximated by a simple, universal profile 
(\eqnref{eq:NFW}) with an inner, density slope $\rho_{\rm DM} \proptosim
r^{-1}$.  In contrast, for dark matter models with significant
self-interaction cross sections, halos are expected to exhibit
flattened, quasi-isothermal cores \citep{Spergel00,Yoshida00}. In the
absence of significant rotational support, these cores are also 
expected to be approximately spherical.

Using Chandra X-ray data for a sample of massive, dynamically
relaxed galaxy clusters, \citet{Schmidt07} measure a mean central
density slope of $-0.88\pm0.29$ (95 per cent confidence limits), in
good agreement with $\Lambda$CDM. Detailed strong-plus-weak lensing
analyses for a subset of these systems yields consistent
constraints \citep{Bradac0711.4850,Newman09,Newman11}. On smaller scales,
using stellar velocity dispersion profiles for the dominant cluster
galaxies, \citet{Sand08} (see also \citealt{Newman11}) 
infer a possible flattening of the central
dark matter halos in Abell 383 and MS\,2137.3-2353. However, on these
small scales, the impact of baryonic physics becomes important.

\citet{Arabadjis02} use the lack of a dark matter core in X-ray and
lensing data for the relaxed cluster MS\,1358+6245, to place a limit on
the velocity independent dark matter particle-scattering cross section
$\sigma/m<0.1$\,cm$^2$\,g$^{-1}$.  \citet{Miralda-Escude02} use constraints
on the ellipticity in the central regions of MS\,2137.3-2353 to infer
$\sigma/m< 0.02$\,cm$^2$\,g$^{-1}$.  To improve these constraints, combined
multiwavelength observations for large samples of relaxed clusters,
coupled with improved simulations modeling fully the interactions
between dark matter and baryons, are required.

Certain dark matter candidates, including sterile neutrinos, possess a
two-body radiative decay channel that produces a photon with
energy $E_\gamma = M_{\rm DM}/2$, where $M_{\rm DM}$ is the dark
matter particle mass \citep[e.g.][]{Feng10}. Galaxy clusters have been
the targets of searches for emission lines associated with such
decays. The soft X-ray (keV) regime is particularly interesting,
marking the lower limit of masses consistent with constraints from
large scale structure formation. To date, all searches for
monochromatic X-ray emission lines associated with non-baryonic matter
in clusters (as well as other dark matter rich objects) have proved
negative \citep[e.g.][]{Boyarsky06, Riemer-Sorensen07, Boyarsky08}.
Gamma-ray observations of clusters with the Fermi Gamma-ray
Space Telescope have been used to place interesting upper 
limits on dark matter annihilation, and the lifetimes of particles 
for a range of masses and decay final states \citep{Ackermann10, Dugger10}.

\subsection{Gravity}  \label{sec:physics_gr}

Dark energy, though a key component of the standard cosmological
model, provides by no means the only possible explanation for cosmic
acceleration. Various non-standard gravity models can also produce
acceleration on cosmological scales (for reviews, see
\citealt{Copeland0603057, Frieman0803.0982}). These include frameworks
that consistently parametrize departures from General Relativity (GR;
\citealt{Hu0708.1190,Amin0708.1793,Daniel1002.1962}); full,
alternative theories such as the Dvali-Gabadadze-Porrati (DGP)
braneworld gravity \citep{Dvali0005016}; $f(R)$ modifications of the
Einstein-Hilbert action \citep{Carroll0306438}; and modifications of
gravity based on the mechanism of ghost condensation
\citep{Arkani-Hamed0312099}. A critical requirement for any modified
gravity model is that it should mimic GR in the relatively small
scale, high density regime where GR has been tested precisely.
Thus, in addition to investigating whether dark energy is well
described by a cosmological constant, we are simultaneously interested
in asking whether GR provides the correct description of gravity, and
indeed whether dark energy is needed at all.

To discriminate among these possibilities, and between particular dark
energy and modified gravity models, it is important to combine
expansion history data with measurements of the growth and
scale-dependence of cosmic structure. 
Galaxy clusters provide some of our strongest constraints 
on structure growth. To utilize these constraints robustly, however,
accurate predictions for the halo mass function are required. 
Recently, a few mass
functions for specific modified gravity models have been constructed
and calibrated using N-body simulations. These include the
self-accelerated branch \citep{ Chan0906.4548,Schmidt0905.0858} and
normal branch \citep{Schmidt0910.0235} of DGP gravity, and an $f(R)$
model \citep{Schmidt0812.0545}. Constraints on the latter model using
the observed local cluster abundance and other data are presented by
\citet{Schmidt0908.2457}.

\begin{figure}
  \centerline{
    \epsfxsize=6cm
    \epsfbox{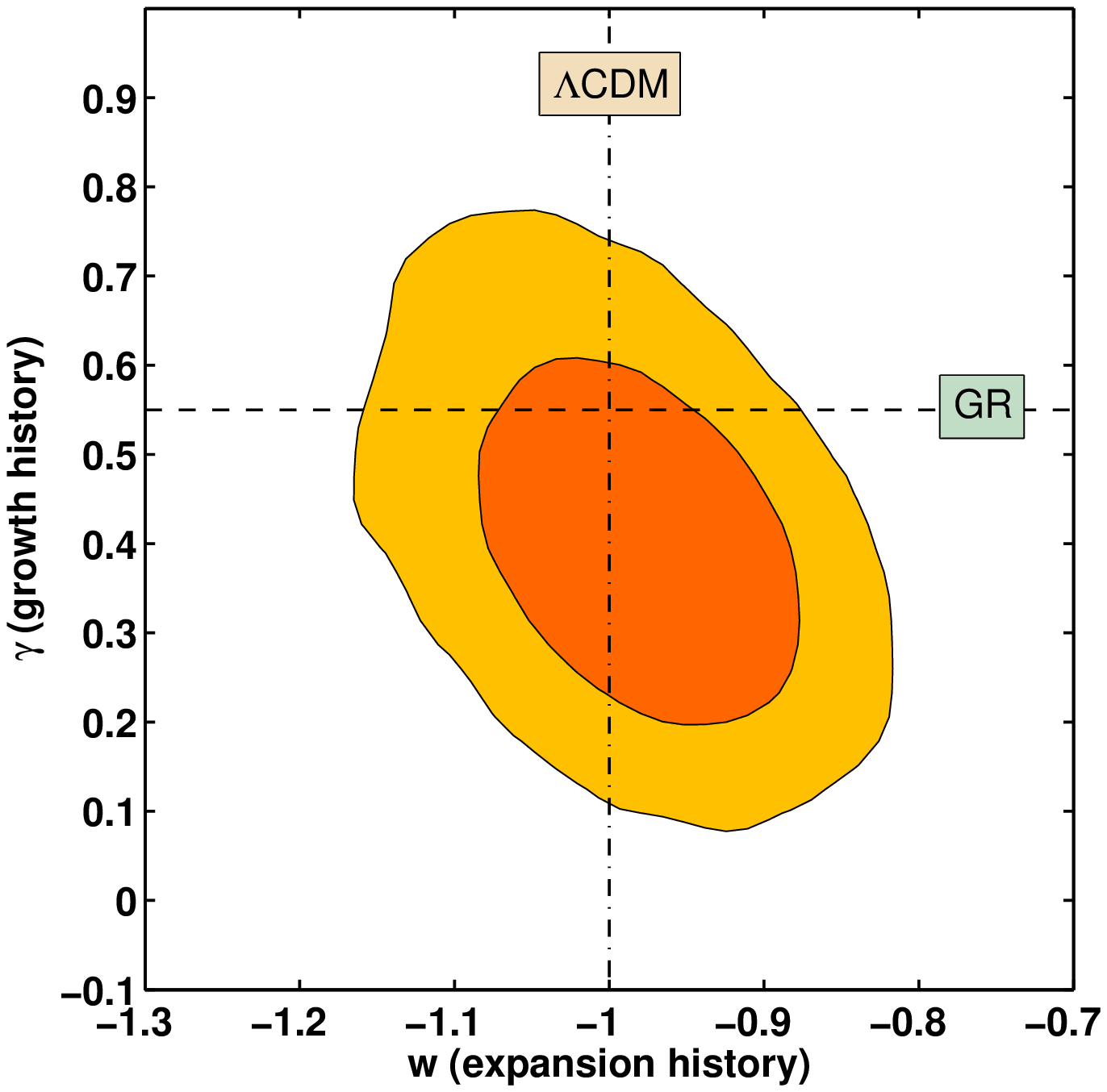}
    \hspace{7mm}
    \epsfxsize=6cm
    \epsfbox{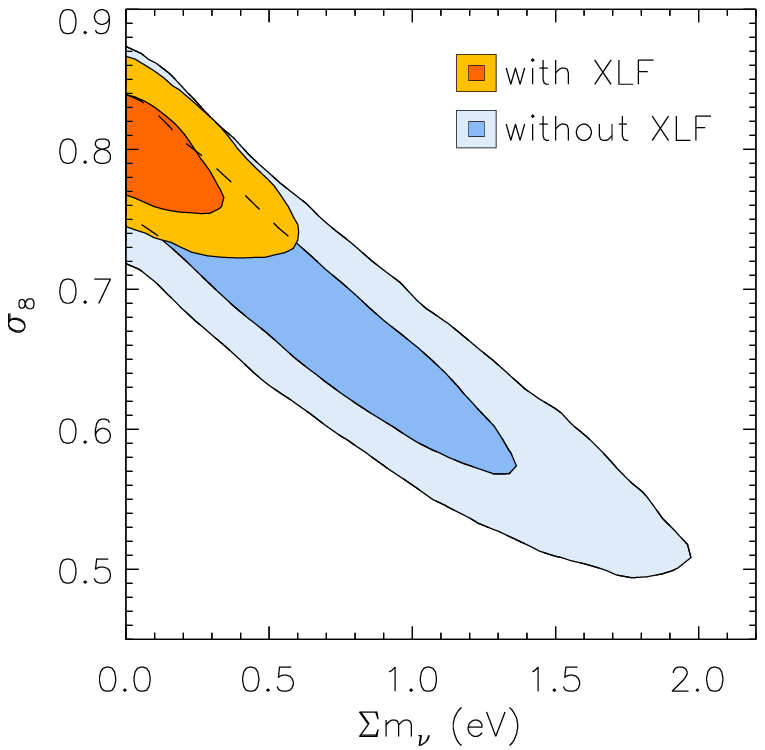}
  }
  \caption{{\it Left}: Joint 68.3\% and 95.4\% confidence regions for departures from a General Relativistic growth history, parameterized by $\gamma$, and a \LCDM{} expansion history, parameterized by $w$. The analysis uses a combination of cluster growth (XLF; \citealt{Mantz0909.3098}), \fgas{} \citep{Allen0706.0033}, \WMAP{} \citep{Dunkley0803.0586}, SNIa \citep{Kowalski0804.4142} and BAO \citep{Percival0608635} data. From \citet{Rapetti0911.1787}.
  {\it Right}: Constraints on neutrino mass and the amplitude of density perturbations for \LCDM{} models, including global curvature and marginalized over the amplitude and spectral index of primordial tensor perturbations. Gold contours correspond to the same combination of data as in the left panel; blue contours show the strong degeneracy between neutrino mass and $\sigma_8$ that exists when cluster growth data are not included in the analysis. From \citet{Mantz0911.1788}.
  }
  \label{fig:physics:gamma_s8}
\end{figure}

An alternative to evaluating specific gravity theories is to adopt a
convenient, parameterized description for the growth of
structure. This can then be used to constrain departures from the
predictions of \LCDM{}+GR \citep{Nesseris0710.1092}.  At late-times,
the linear growth rate can be simply parametrized as
(e.g. \citealt{Linder0507263})
\begin{equation} \label{eq:physics:gr_gamma}
  \frac{d\ln\delta}{d\ln a} = \Omegam(a)^\gamma,
\end{equation}
where $\delta$ is the density contrast and $\gamma$ the growth
index. Conveniently, GR predicts a nearly constant and
scale-independent value of $\gamma \approx 0.55$ for models consistent
with current expansion data. As in the case of $w$ for dark energy
models, constraining $\gamma$ constitutes a phenomenological approach
to studying gravity.  \citet{Rapetti0812.2259,Rapetti0911.1787} report
constraints on departures from GR on cosmic scales using 
this parameterization with cluster data.
Their results are simultaneously consistent with GR ($\gamma \sim 0.55$) and
\LCDM{} ($w=-1$) at the 68 per cent confidence level (left panel of \figref{fig:physics:gamma_s8}).

\subsection{Neutrinos} \label{sec:physics:nu}

The mass of neutrinos directly influences the growth of cosmic structure, since any particle with non-zero mass at some point cools from a relativistic state, in which it effectively suppresses structure formation, to a non-relativistic state, in which it actively participates in the growth of structure (details are reviewed in \citealt{Lesgourgues0603494}). In the standard scenerio where the neutrino species have approximately degenerate mass, the species-summed mass, \Mnu{}, is sufficient to describe their cosmological effects.

Although current data lack the precision to directly detect the effect of neutrino mass on the time-dependent growth of clusters, cluster data do play a key role in cosmological constraints on neutrinos when combined with CMB observations. On its own, the CMB can place only a relatively weak upper bound on the mass, $\Mnu<1.3\eV$ at 95\% confidence for a \LCDM{} model (e.g. \citealt{Dunkley0803.0586}; works discussed here used 5 years of \WMAP{} data, but their conclusions are not significantly changed by the 7-year update).
 Incorporating cosmic distance measurements improves this to $\Mnu<0.61\eV$, with the results displaying a strong degeneracy between \Mnu{} and the value of $\sigma_8$ predicted from the amplitude of the primordial power spectrum, due to the integrated effect of neutrinos on the growth of structure. Cluster data  at  low redshift provide a direct measurement of $\sigma_8$, breaking this degeneracy (right panel of \figref{fig:physics:gamma_s8}), improving the upper limit by a further factor of two, to $\Mnu<0.33\eV$ (\citealt{Vikhlinin0812.2720,Mantz0911.1788,Reid0910.0008}). The degeneracy-breaking power of cluster observations also significantly improves the robustness of neutrino mass limits to the assumed cosmological model (e.g. marginalizing over global curvature; \citealt{Mantz0911.1788,Reid0910.0008}). In combination with \Planck{} data, expected near-term improvements in cluster mass measurements from high-quality lensing data could reduce the limit on $\Mnu$ to the point of distinguishing between the normal and inverted neutrino mass hierarchies.

\section{OPPORTUNITIES} \label{sec:opportunities}

\subsection{Cluster Surveys on the Near and Mid-Term Horizons}
\label{sec:new_surveys}

In the near future (over the next 2--3 years), the completion of the
SPT, ACT and \Planck{} SZ catalogs will extend our detailed,
statistical knowledge of galaxy clusters out to $z>1$. Together, these
projects expect to find $\sim 1000$ new clusters, mostly at
intermediate-to-high redshifts
\citep{Marriage10,Vanderlinde1003.0003}. Used in combination with
existing low-redshift X-ray and optical catalogs, they should provide
significant improvements in our knowledge of cluster growth, and
corresponding improvements in the constraints on dark energy 
and gravity models. As discussed in Section~\ref{sec:obs_surveys_sz},
challenges for these surveys will be defining and calibrating the
optimal survey mass proxies, and understanding the impact of
contamination from associated infrared sources and AGN. Looking
further ahead, the development of experiments with improved spatial
resolution, expanded frequency coverage and improved sensitivity, such
as the Cerro Chajnantor Atacama Telescope (CCAT), will be of
advantage.

At optical and near-infrared wavelengths, a suite of powerful, new
ground-based surveys are about to come on-line. These include the
Panoramic Survey Telescope and Rapid Response System (Pan-STARRS); the
Dark Energy Survey (DES); the KIlo-Degree Survey (KIDS) and
complementary VISTA Kilo-degree INfrared Galaxy survey (VIKING); the
Subaru Hyper Suprime-Cam survey (HSC); and, later, the Large Synoptic
Survey Telescope (LSST). These experiments offer significant potential
for finding clusters, and will provide critical photometric redshift
and lensing data.  A primary challenge in constructing optical cluster
catalogs will be the definition of robust mass proxies with minimal,
well-understood scatter across the mass and redshift ranges of
interest.  Planned, space-based survey missions such as the Wide Field
Infrared Survey Telescope (WFIRST) and Euclid also offer outstanding
potential for cluster cosmology, complementing the ground-based
surveys in providing lensing masses and photometric redshifts, and
extending cluster search volumes out to higher redshifts.

In the near term, X-ray cluster samples constructed from serendipitous
detections in \Chandra{} and \XMM{} observations offer the potential
for important gains \citep[e.g.][]{Fassbender08,Sahlen09}. The main
advances at X-ray wavelengths, however, will be provided by the
eROSITA telescope on the Spektrum-Roentgen-Gamma Mission.  Scheduled
for launch in 2012, eROSITA will perform a four year all-sky survey
that should detect an estimated 50,000--100,000 clusters with
excellent purity and completeness.  As discussed in
Section~\ref{sec:obs_surveys_xray}, the unambiguous detection of
clusters at X-ray wavelengths requires angular resolution to
distinguish point sources from extended cluster emission. At modest
redshifts and high fluxes, this should be straightforward for eROSITA;
at high redshifts ($z>1$) and faint fluxes, however, separating
cool-core clusters from AGN will be challenging and will require
follow-up observations.  Here, we note more than a dozen known
examples of powerful AGN surrounded by X-ray bright clusters at
intermediate-to-high redshifts (e.g. \citealt{Belsole07,
Siemiginowska10}). Looking further ahead, the development of improved
X-ray mirrors with high spatial resolution across a wide field of
view, such as those proposed for the Wide Field X-ray Telescope, would
be a major advance, rendering trivial the removal of contaminating AGN
emission and allowing surveys to take full advantage of the
center-excised X-ray luminosity as a low-scatter mass proxy
(\secref{sec:future:coreevol}).

For all of these surveys, accurate calibration will be important; this
is a particular challenge for space-based missions. In comparison to
other cosmological probes, the demands on photo-$z$ calibration will
be relatively modest. This is due both to the pronounced 4000 \AA\
break in early-type galaxy spectra and the ability to combine
measurements for many galaxies per cluster. Fisher matrix studies by
\citet{Cunha1003.2416} suggest targets of $<0.003$ for bias error and
$<~0.03$ for error in the scatter in surveys with $\sim \! 10^5$
clusters.  For surveys with fewer counts, shot noise dominates, and
photo-$z$ errors become less important.

Extensive programs of follow-up observations using high resolution,
high throughput telescopes will also be essential.  X-ray
observatories like \Chandra{}, \XMM{}, \Suzaku{} and {\it ASTRO-H} are
likely to remain cornerstones of this work, providing excellent,
low-scatter mass proxy measurements for individual clusters. We note
that such information need only be gathered for a fraction of the
clusters in a survey to gain a significant boost in constraining power
with respect to self-calibration alone
\citep[e.g.][]{Wu10}. Statistical calibration of the mean masses of
clusters in flux and redshift intervals from weak lensing measurements
will also be critical: in order for the intrinsic power of large
surveys not to be impacted severely, calibration of the mean mass at
the few per cent level is required \citep{Wu10}. To achieve this
accuracy, detailed redshift information for the lensed, background
sources will be needed; indeed, the use of full redshift probability
density functions rather than simple color cuts may be
required. Detailed simulations will also be needed to probe the
systematic limitations of these measurements, and to advise on the
best analysis approaches (Section~\ref{sec:future_sims}).

\subsection{Footprints of Inflation: Primordial Non-Gaussianites}
\label{sec:physics_inflation}

Inflation predicts a near scale invariant power spectrum and nearly
Gaussian distribution for the primordial curvature inhomogeneities
that seed LSS. For slow-roll, single-field inflation, departures from
Gaussianity are currently unobservable (by at least four orders of
magnitude; \citealt{Acquaviva0209156,Maldacena0210603}). However,
other multi-field and single-field inflation models predict observable
non-Gaussianity (NG). Examples include certain brane models such as
the Dirac-Born-Infeld (DBI) inflation; single-field models with
non-trivial kinetic terms; ghost inflation; models in which density
perturbations are generated by another field such as the curvaton; and
models with varying inflation decay rate. (See \citealt{Chen1002.1416}
and \citealt{Komatsu1003.6097} for recent reviews.) Any detection of
NG would provide critical information about the physical processes
taking place during inflation. In particular, a convincing detection
of NG of the `local' type (referring to particular configurations in
Fourier space) would rule out not only slow-roll but {\it all} classes
of single field inflation models \citep{Creminelli0407059}.

Current measurements of CMB anisotropies and LSS are consistent with
Gaussianity \citep{Slosar0805.3580,Komatsu1001.4538}. In principle,
measurements of the clustering and abundance of galaxy clusters can
also be used to place powerful, complementary constraints.  Galaxy
clusters trace the rare, high-mass tail of density perturbations in
the Universe (\secref{sec:theory}) and are uniquely sensitive to
NG. N-body simulations have been used to study cluster formation under
non-Gaussian initial conditions \citep[e.g.][]{Decjaques0811.2748,
 Grossi0902.2013, Pillepich0811.4176}. Analytical mass functions have
also been calculated using the Press-Schechter formalism
\citep{LoVerde0711.4126, D'Amico1005.1203} or excursion set theory
\citep{Maggiore0910.5125}.

To date, CMB and LSS studies have only been used to place
scale-independent constraints on NG.  However, certain models that
produce relatively large NG (see above) also present strong
scale-dependence of the signal, due to, e.g., a changing (effective or
otherwise) sound speed \citep{LoVerde0711.4126}. The addition of
measurements at the smaller, galaxy cluster scale should provide
important constraints on such models \citep{Riotto1009.3020,
Shandera1010.3722}. Several recent works \citep{Oguri0905.0920,
 Cunha1003.2416, Sartoris1003.0841} have investigated quantitatively
the potential of future optical and/or X-ray cluster surveys for
constraining NG.

\subsection{The Thermodynamics of Cluster Outskirts} \label{sec:future_outskirts}

To date, robust thermodynamic measurements have, in general, only been
possible for the inner parts of clusters ($r \ltsim r_{500}$), where
the X-ray emission is brightest and the SZ signal strongest; a large
fraction of their volumes remain practically unexplored. Precise,
accurate measurements of the density, temperature, pressure and
entropy out to the virial radii of clusters provide important insight
into the physics of clusters and (ongoing) large-scale structure
formation, and can improve the precision and robustness of
cosmological constraints.

Recently, the Japanese-US \Suzaku{} satellite has opened a new window
onto the outskirts of clusters. Due to its lower instrumental
background than flagship X-ray observatories like \Chandra{} and
\XMM{}, which orbit beyond the Earth's protective magnetic fields,
\Suzaku{} can study the faint, outer regions of clusters more
reliably.  The primary challenge with \Suzaku{} is the relatively
large point spread function of its mirrors, which limits detailed,
spatially-resolved studies to systems at modest redshift ($z \ltsim
0.1$).

Over the past two years, a series of ground breaking measurements of
the outskirts of clusters with \Suzaku{} have been reported
\citep{George09,Reiprich09,Bautz09,Hoshino10}. These have confirmed
the presence of smoothly decreasing density and temperature profiles
out to large radii, as was qualitatively expected from theoretical
models and earlier data
\citep[e.g.][]{Markevitch9711289,Frenk99}. Interestingly, these
observations have also suggested a possible flattening of the outer
entropy profiles.

\begin{figure}
  \epsfxsize=7cm
  \centerline{\epsfbox{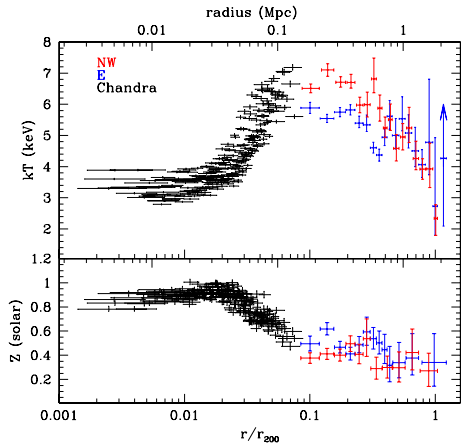}}
  \caption{The observed, projected temperature ($kT$) and metallicity
($Z$) profiles in the Perseus Cluster, the nearest, massive galaxy
cluster and brightest, extended extragalactic X-ray source. \Suzaku{}
results for the northwestern (NW) arm of the cluster are shown in red;
and for the eastern (E) arm (approximately aligned with the major
axis) in blue \citep{Simionescu10}.  \Chandra{} measurements of the
inner regions \citep{Sanders07} are shown in black.}
  \label{fig:suzaku_kt}
\end{figure}

Of particular interest are the advances made with the \Suzaku{} Key
Project study of the Perseus Cluster ($z=0.018$), the nearest,
massive galaxy cluster and brightest, extended extragalactic X-ray
source. The temperature and metallicity profiles for the northwestern
(NW) and eastern (E) arms of Perseus measured with \Suzaku{} are shown
in \figref{fig:suzaku_kt} \citep{Simionescu10}.  Also plotted, for
comparison, are the results from earlier, deep \Chandra{} observations
of the cluster core \citep{Sanders07}. The \Suzaku{} and \Chandra{}
data show excellent agreement where they meet, and measure the
thermodynamic structure of the ICM over three decades in radius, out
to $r_{200}$.

\begin{figure}
 \centerline{
    \epsfxsize=6.5cm
    \epsfbox{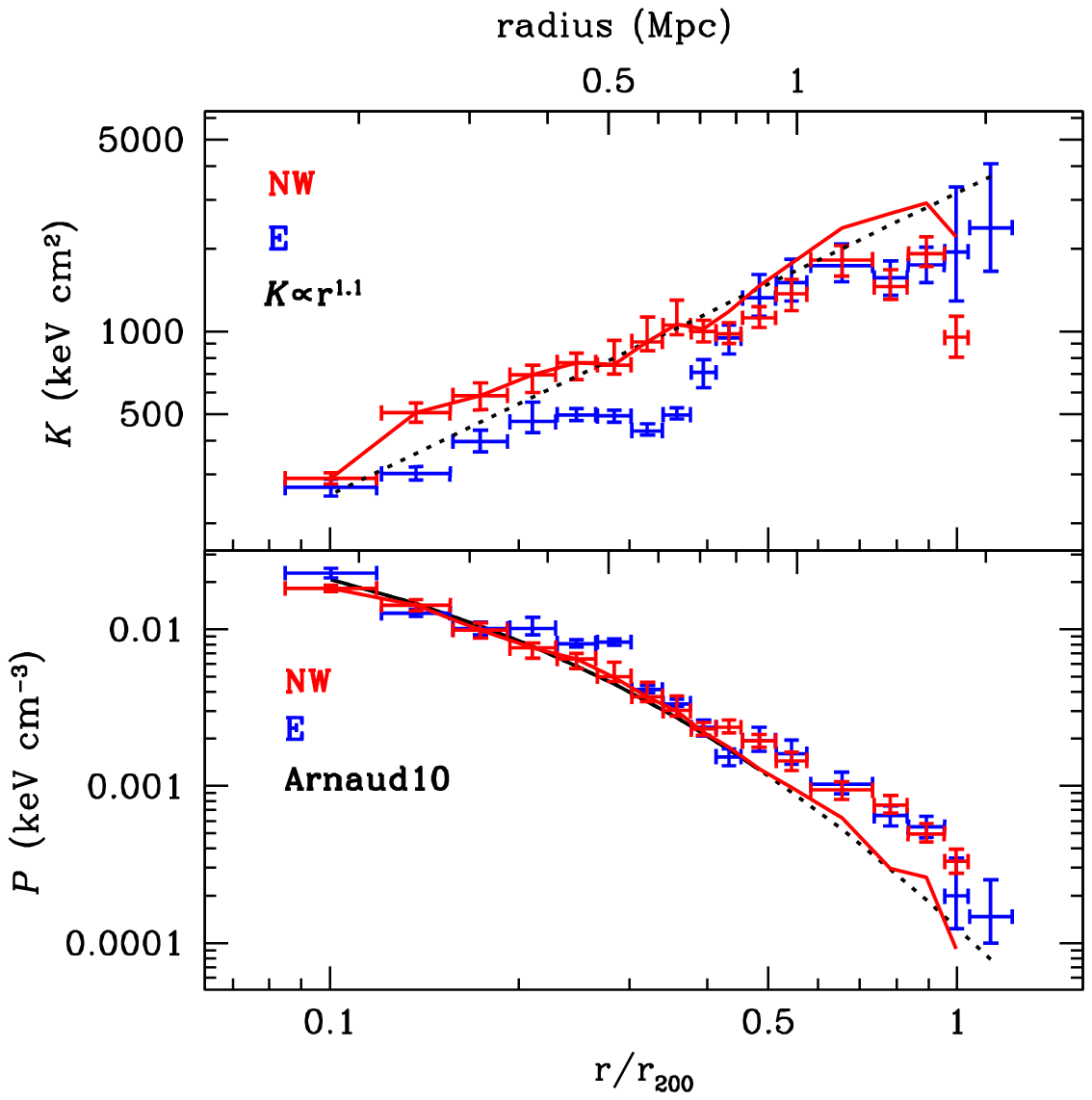}
    \epsfxsize=6.5cm
    \epsfbox{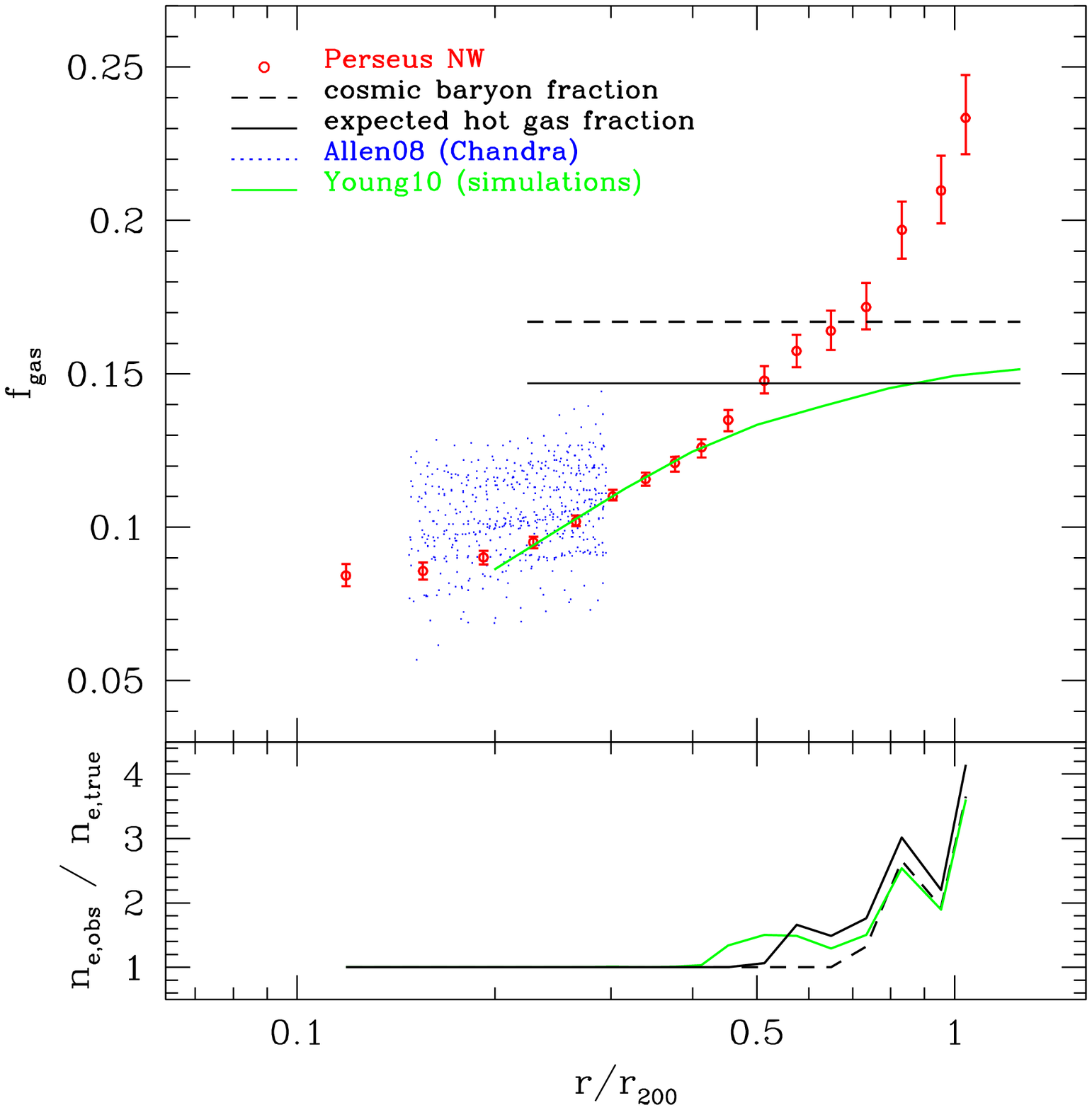}
  }
  \caption{{\it Left}: Deprojected entropy ($K$) and pressure ($P$)
profiles out to $r_{200}$ for the NW (red) and E (blue) arms of
Perseus. In the upper panel, the black dotted line shows the expected,
power-law entropy profile determined from simulations of
non-radiative, hierarchical cluster formation: $K \propto r^{1.1}$
\citep[e.g.][]{Voit05}. In the lower panel, the black solid line shows
a parameterized pressure model, motivated by simulations \citep{Nagai0703661}, 
that was fitted to the inner regions 
($r\ltsim0.5r_{200}$) of clusters studied 
with \XMM{} \citep{Arnaud10}. The dotted curve is the extrapolation of
this model to larger radii.  The solid, red curves show the entropy
and pressure profiles measured for the NW arm by \Suzaku{} {\it after} corrections for gas 
clumping, which agree with the model predictions. {\it Right}: The
integrated gas mass fraction profile for the NW arm of the Perseus
Cluster. The dashed black line denotes the mean cosmic baryon fraction
measured by WMAP7 \citep{Komatsu1001.4538}. Accounting for 12\% of the
baryons being in stars \citep{Lin04, Gonzalez07, Giodini09} gives the
expected fraction of baryons in the hot gas phase, marked with the
solid black line. Previous results at $r \leq r_{2500}$ from \Chandra{}
\citep{Allen0706.0033} are shown in blue. Predictions from numerical
simulations \citep{Young1007.0887} incorporating a simplified AGN feedback
model are shown in green. The bottom panel shows the overestimation of
the electron density as a function of radius due to gas clumping. See
\citet{Simionescu10} for details.}
  \label{fig:suzaku_thermo}
\end{figure}

Models of large scale structure formation show that gas is shock
heated as it falls into clusters. Entropy is an important tracer of
this virialization process. Numerical simulations predict that in the
absence of non-gravitational processes such as radiative cooling and
feedback, the entropy, $K$, should follow a power-law with radius, $K
\proptosim r^{\beta}$, with $\beta \sim 1.1$--$1.2$
\citep[e.g.][]{Voit05}. With the exception of a cold front seen along
the eastern arm at $r \sim 0.3r_{200}$, the entropy profile in Perseus
roughly follows the expected trend out to $r\sim 0.6 r_{200}$
(\figref{fig:suzaku_thermo}). Beyond this radius, however, both arms
flatten away from the predicted power-law shape. The pressure profile
shows good agreement between the NW and E arms.  Within $r\ltsim
0.5r_{200}$, the \Suzaku{} pressure results show good agreement with
the predictions from numerical simulations \citep{Nagai0703661}, and
previous measurements with \XMM{} \citep{Arnaud10}. At larger radii,
however, the observed pressure profile is shallower than a simple
extrapolation of the \XMM{} results.

Accurate estimates of the gas masses and total masses out to large
radii are of particular importance for cosmological studies.  The
\Suzaku{} observations of the Perseus Cluster provide the first such
measurements for a massive cluster.  The best-fit NFW mass model
determined from a hydrostatic analysis of the (relatively relaxed) NW
arm has parameters in good agreement with the predictions from
cosmological simulations ($c=5.0\pm0.5$,
$M_{200}=6.7\pm0.5\times10^{14}\Msun$) and provides a good description
of the data \citep{Simionescu10}.  Of particular interest is the
cumulative \fgas{} profile, shown in the right panel of
\figref{fig:suzaku_thermo}. Within $r_{2500}$ ($r\ltsim0.3r_{200}$),
the observed \fgas{} profile is consistent with previous \Chandra{}
and SZ measurements for other massive, relaxed clusters
\citep{Allen0405340,Allen0706.0033,LaRoque0604039}.  From
$0.2$--$0.45r_{200}$ (i.e. excluding the central cooling core) the
\fgas{} profile is also consistent with the predictions from recent
numerical simulations, incorporating a simple model of AGN feedback
\citep{Young1007.0887}.  At $r\sim0.6r_{200}$, the enclosed \fgas{}
value approximately matches the mean cosmic baryon fraction, as
measured from the CMB \citep{Komatsu1001.4538}.  Beyond $r \sim
2/3r_{200}$, however, where the entropy also flattens away from the
expected power-law behavior, the  \fgas{} apparently {\it
exceeds} the mean cosmic value. The most plausible explanation for the
apparent excess of baryons at large radii is gas clumping: in X-rays,
the directly measurable quantity from the intensity of the emission is
the average of the square of the electron density, rather than the
average electron density itself. If the density is not uniform,
i.e. the gas is clumpy, then the average electron density estimated
from the X-ray intensity will overestimate the truth; this will lead
to an overestimate of the gas mass and gas mass fraction, and will
flatten the apparent entropy and pressure profiles.

The amount of gas clumping in Perseus, as inferred from comparison of
the observed and expected \fgas{} profiles, is shown in the lower
panel of the right of \figref{fig:suzaku_thermo}. Using this clumping
profile to correct the pressure and entropy measurements gives the
solid red curves in the left panel. The clumping-corrected entropy
profile shows good agreement with the expected power-law form out to
$r_{200}$. Likewise, the clumping-corrected pressure profile matches
the form predicted by simulations \citep{Nagai0703661}.

Importantly, the \Suzaku{} results provide no evidence for the
puzzling deficit of baryons at $r \sim r_{500}$ inferred from some
previous studies of massive clusters using lower quality X-ray data
(at large radii) and stronger modeling priors
\citep[e.g.][]{Vikhlinin06,Gonzalez07,McCarthy07,Ettori09}.  The
Perseus data suggest that beyond the innermost core but within $r 
\ltsim 0.5r_{200}$, X-ray measurements for massive, relaxed clusters
can be used simply and robustly for cosmological work. At larger
radii, the effects of gas clumping become increasingly important and
must be accounted for.

In principle, the combination of X-ray and SZ observations can also be
used measure gas clumping, offering an important cross-check of the
Suzaku results. The origin of these density fluctuations is also an
important, open question. Although numerical simulations predict that
the gas in the outskirts of clusters should be clumpy
\citep[e.g.][]{Mathiesen99}, the degree of inhomogeneity
predicted depends in detail on a range of uncertain physical
processes, including cooling, conduction, viscosity, and the impact of 
magnetic fields.

Over the next few years, \Suzaku{} and SZ studies of other bright,
nearby clusters, complemented later by higher spectral resolution
X-ray data from {\it ASTRO-H}, can be expected to stimulate
significant progress.  In particular, \fgas{} measurements out to
large radii for a statistical sample of nearby clusters, and
measurements of the dispersion in this and other properties from
region-to-region and system-to-system, should provide a robust
low-redshift anchor for cosmological work and powerful constraints on
astrophysical models.

\subsection{Evolution of Cluster Cores} \label{sec:future:coreevol}

Cluster cores -- typically describing the central region of
$r<50$--100\kpc{} radius -- often host a variety of astrophysical
processes including efficient radiative cooling, AGN outbursts, modest
star formation, and sloshing or other bulk motions (see, e.g.,
\citealt{McNamara0709.2152} for a review).  In the cores of even the
most dynamically relaxed clusters, small-scale departures from
hydrostatic equilibrium are often apparent
(\citealt{Fabian03,Markevitch07, Allen0706.0033}).

The astrophysics of cluster cores is especially important for
cosmological studies at X-ray wavelengths. In particular, clusters for
which radiative cooling in the core is efficient form an easily
distinguishable sub-population with significantly enhanced central gas
density and luminosity, also commonly accompanied by a drop in ICM
temperature \citep{Fabian1994MNRAS.267..779F,Peterson0512549}. The
sharp, central density peaks of these `cool core' clusters enhance
their detectability at X-ray wavelengths. However, without spatial
resolution $\ltsim10$~arcsec or additional follow-up data, cool-core
clusters at high redshift lying close to the survey flux limit cannot
be distinguished easily from non-cluster X-ray point sources. (This
selection bias has affected some studies in the literature.) The
prevalence and evolution of cool cores in the cluster population thus
plays a significant role in determining the shape and evolution of the
scatter in X-ray luminosity at fixed mass.

\begin{figure}
  \centerline{
    \epsfxsize=6cm
    \epsfbox{MantzII_Lce.eps}
    \hspace{7mm}
    \epsfxsize=6.7cm
    \epsfbox{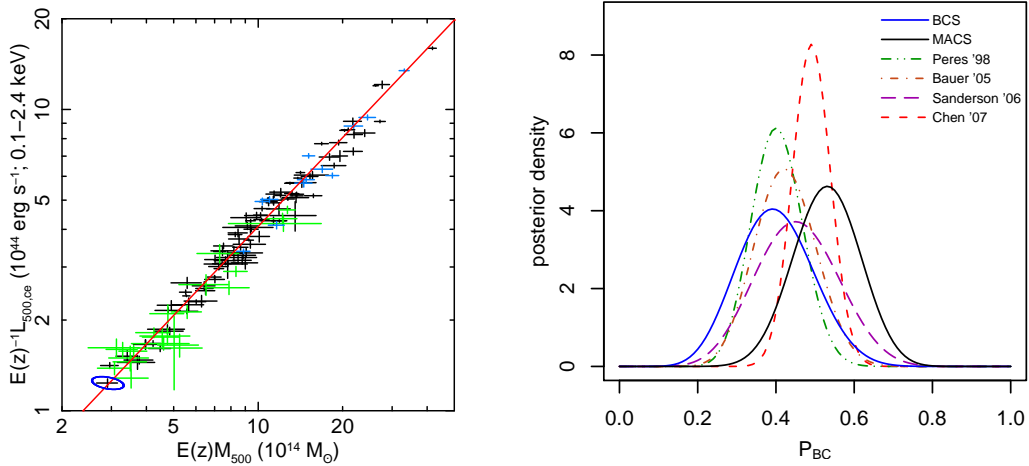}
  }
  \caption{{\it Left}: Center-excised luminosity--mass relation, in
which emission within $0.15r_{500}$ of cluster centers is excluded
from the luminosity measurements. Error bars in the plot show
statistical uncertainties only. The intrinsic scatter in this relation
is $<10\%$, significantly reduced from $\sim40\%$ for the full
luminosity--mass relation. Gray points are at redshifts $z<0.5$, blue points are at $0.5<z< 0.7$, and green points are at $0.35<z<0.9$. The blue ellipse shows the typical
correlation of measurement errors, which is accounted for in the
fit. The negative sense of the measurement correlation results from
details of the center excision: larger mass implies larger $r_{500}$
and thus a larger excised region. Since more flux is excised than is
gained at large radius when $r_{500}$ increases, there is a net
reduction of luminosity with larger mass. From
\citet{Mantz0909.3099}. {\it Right}: Posterior probability
distributions for the fraction of cool-core clusters based on the
number identified in various X-ray selected samples at $z<0.5$. Due to
Malmquist bias, these must overestimate the fraction in the full
cluster population. From \citet[][see also references
therein]{Mantz2009PhDT........18M}.}
  \label{fig:corescatter}
\end{figure}

Confirmation of this fact can be found in the dramatic reduction in
scatter obtained when emission from a central region of radius
$0.15r_{500}$ is excised from luminosity measurements in forming the
$\Lx$--$M$ scaling relation, from $\sim40\%$ to $<10\%$
(\figref{fig:corescatter}; see also
\citealt{Markevitch9802059,Allen9802218,Zhang0702739,Maughan0703504,Mantz0909.3099}).
(While the excision radius of $0.15r_{500}$ has become somewhat conventional, a similar reduction in scatter is evident
when excising a fixed metric radius of similar size, e.g. 150\kpc.)
\citet{Mantz2009PhDT........18M}
found that roughly half of the intrinsic scatter can be attributed to
radii $<0.05r_{500}$, the typical scale of cool, dense cluster cores,
with most of the remainder due to variations in the gas density
profile at $0.05<r/r_{500}<0.15$. The fraction of the total flux
coming from $r<0.05r_{500}$ can be as large as 50\%, and correlates
strongly with other observable signatures of cool cores such as
central cooling time and cuspiness of the surface brightness profile
(\citealt{Mantz2009PhDT........18M}; see also
\citealt{Andersson0902.0003}). Interestingly, the fractions of
cool-core systems identified using this criterion are comparable in
X-ray flux-selected samples, at least within $z<0.5$ ($\sim 40\%$; see
\figref{fig:corescatter}; \citealt{Peres9805122,Bauer0503232,Sanderson0608423,Chen0702482,Mantz2009PhDT........18M}),
although these fractions are biased
relative to the full population due to selection effects.
Cool cores are also common in X-ray selected samples at high redshift
(\citealt{Santos0802.1445}; $0.7<z<1.4$).

Although the number of massive clusters found in RASS is too small to
detect evolution or departures from log-normality in the
luminosity--mass scatter, future, deeper X-ray surveys such as eROSITA
will provide the larger samples necessary for these
investigations. Conversely, a better understanding of cluster cores
will be required to fully exploit these surveys for cosmology. Such
studies will need to employ the full statistical apparatus described
in \secref{sec:constraints:procedure} to obtain unbiased results.

Due to the weaker dependence of the SZ signal on the central gas
density (Section~\ref{sec:obs_observations}), SZ surveys are expected
to be less sensitive to the presence of cool cores than
X-ray surveys. This is consistent with the relatively low
fraction of cool core systems observed among the newly discovered
clusters in the \Planck{} Early Release cluster catalog
\citep{Plancksurvey11,PlanckXMMfollowup11}. We note, however, that current SZ surveys are not necessarily immune to biases
associated with cool cores; depending on the selection
techniques employed, infrared emission associated with star formation in the
central regions of cool core clusters may actually diminish the measured
SZ signal.

\subsection{Improved Simulations}
\label{sec:future_sims}

Unleashing the full statistical power of upcoming surveys will require
careful control of theoretical uncertainties.  For cluster abundance
and evolution tests, uncertainties in scaling relation parameters are
currently more important than uncertainties in the halo mass and bias
functions \citep{Cunha0908.0526}.  With aggressive mass calibration
efforts to reduce the former, the mass function and bias errors will
need to be limited to the percent level in order to avoid significant
degradation in cosmological parameter constraints \citep{Wu0910.3668}.
Simulation campaigns will be needed to address this challenge.

Over the next decade, increased computing power will enable models
with new capabilities (e.g. multi-scale simulation in place of
sub-grid models) and will vastly expand the size of current simulation
suites.  By densely sampling a large control space of cosmological and
astrophysical parameters, simulation ensembles can support survey
analysis via functional interpolation, a method termed emulation
\citep{Habib0702348}.  An initial application of this technique uses
38 $10^9$-particle N-body simulations to predict the non-linear matter
power spectrum to $1\%$ accuracy (for $k \ltsim 1\ h/\mpc$) in a
5-dimensional space of cosmological parameters
\citep{Lawrence0912.4490}.

Precise N-body calibrations must be treated with some caution, as the
baryons representing $17\%$ of the total mass undergo different
small-scale dynamics than CDM.  Back-reaction effects of cooling and
star formation could be important \citep{Rudd0703741, Stanek0809.2805}
and should be systematically investigated.  As discussed in
\secrefs{sec:physics:lcdm} and \ref{sec:physics}, studies of LSS in
non-standard cosmologies, including those for which dark energy may
cluster or in which weak-field gravity is non-Newtonian, should also
be pursued.

Full solution of the galaxy formation problem from first principles
remains challenging. Direct simulation methods will continue to
improve, though not likely to the point where multi-fluid simulations
will offer definitive predictions for next-generation survey analysis.
Semi-analytic methods are increasingly informative; for example, the
frantic early merging that forms the central galaxies in clusters is
becoming understood \citep{DeLucia0509725, Weinmann0912.2741}.  But
many elements required to predict the spectrophotometric properties of
galaxies remain poorly understood, including stellar population
synthesis and dust evolution \citep{Conroy0904.0002}.  Empirically
tuned statistical approaches, essentially assigning galaxy properties
to halos or sub-halos in a manner informed by observed clustering as a
function of luminosity and/or color, will continue to provide a
valuable complement to physical methods.

Upcoming surveys will require sophisticated, automated reduction and
analysis pipelines for their large data streams, along with quality
assurance to validate accuracy.  Simulations of sky expectations,
often referred to as `mock' or `synthetic' surveys, will be required
to provide realistic testbeds in which the answers are known.
Mock surveys provide key quality assurance support by: (i)
incorporating line-of-sight projection effects on measured properties;
(ii) including distortions and noise from telescope/camera optics and
other sources; and (iii) validating image processing and data
management pipelines.
Since the effort involved in producing such data is non-trivial, mechanisms to publish and enable their broader use within the community need to be pursued.

\section{MODELING CONSIDERATIONS}
\label{sec:modeling}

With such significant improvements in data quality expected and such
profound questions to be addressed, modeling considerations will
become increasingly important. The key issues will be modeling and
mitigating all important sources of bias and systematic error in the
analyses, and using the information efficiently.

For tests based on the mass function and clustering of clusters and
their associated mass--observable scaling relations, the impact of
survey biases must be accounted for
(Section~\ref{sec:theory:selection}).  Typically this requires
simultaneous modeling of the cluster population and scaling relations
in a single likelihood function. Such an approach also facilitates
understanding the covariance between model parameters, and provides a
structure within which to examine the impact of residual systematic
uncertainties which often correlate with model parameters. To a large
degree, the statistical frameworks required for such analyses have
been developed (\secref{sec:constraints:procedure}) and their
application to existing data are discussed in this review.

The application of judicious, blind (see below) cuts can improve the
balance of statistical versus systematic uncertainties.  For example,
with the \fgas{} test (Section~\ref{sec:constraints:fgas}),
cosmological constraints are best derived from the most massive,
dynamically relaxed clusters, which can be identified easily from
short, snapshot \Chandra{} or \XMM{} X-ray observations.  These
clusters are also ideal targets for the XSZ test
(Section~\ref{sec:constraints:xsz}).  For the mass function and
clustering tests, one must determine the optimal mass/flux/redshift
range over which to compare the data and models, given the
characteristics of the survey and follow-up data and one's
understanding of mass--observable scaling relations.

Strong priors should be used with caution and their implications
understood.  In cluster cosmology, common assumptions include
log-normal scatter in the mass--observable scaling relations, and
negligible evolution of this scatter with redshift. When measuring
masses from X-ray or optical dynamical data, the use of parameterized
temperature, velocity and gas density models is also common. Where such
strong assumptions are not necessary, they should be
avoided.  Where priors do
become necessary, for example in parameterizing the impacts of known
astrophysical effects, the validity of these assumptions should be
checked empirically, where possible.

The covariance between different measured quantities is an important
consideration that is often wrongly neglected. The simplest example of
this is simply the statistical correlation of quantities measured from
the same observation (e.g. due to Poisson noise). A more subtle case
follows from the definition of cluster radius for scaling relations in
terms of the mean density enclosed: because the measured mass and
radius covary, other quantities measured within that radius
necessarily covary with mass, even if measured from independent
observations.
 Fortunately, within the framework of Bayesian analysis, it is
straightforward to account for such measurement correlations. For
example, given a set of scaling relation parameters to be tested, the
data likelihood can be integrated over all possible true values of,
e.g., the mass and temperature, with one of the terms in the integrand
being the probability of the true mass and temperature values
producing the observed values (the sampling distribution; see
\secref{sec:constraints:procedure}). The form of this probability
density can be as general as is required -- a multidimensional
gaussian or log-gaussian in the simplest case -- and in particular may
have non-zero correlation. \citet[][see also
\citealt{Gelman2004BayesianDataAnalysis}]{Kelly0705.2774} discusses
this general approach in the astrophysical context, and provides
useful tools for Bayesian linear regression.

In other situations, unnecessary covariance can be introduced by the model applied to the data. An example is the historically common practice of fitting parametric temperature and gas density profiles to X-ray data, and inferring overdensity radii and hydrostatic masses from these model profiles. Here, the procedure introduces an artificial covariance between mass and temperature (i.e. a prior on the scaling relation) which could be avoided simply by modeling the temperature and gravitating mass profiles independently.

Where binning will result in a loss of relevant information, it should
be avoided. For example, the binning of X-ray surface brightness data
to determine an integrated X-ray luminosity results in a significant
loss of information; the center-excised luminosity provides a tighter
mass proxy (Section~\ref{sec:future:coreevol}).

Hypothesis testing will remain a critical element of future analyses.
Wherever a model is fitted to data, one should check that it provides
an adequate description (i.e. that the goodness of fit is acceptable);
if it doesn't, then the model can be ruled out. This is particularly
important in the context of cosmological surveys and scaling relations
that are subject to selection bias, since no straightforward, visual
check of the goodness of fit is typically possible. Where the simplest
models fail to describe the data, one must evaluate carefully the
additional degrees of freedom needed to alleviate the tension, in
particular considering both astrophysical and cosmological
possibilities; simulations will play a critical role in motivating and validating alternatives to the simplest scaling models.
  A related situation arises in the combination of
constraints from independent experiments: before combining, one should
ensure that the model in question provides an adequate description of
the data sets individually, and that the parameter values are mutually
consistent, i.e. that their multi-dimensional confidence contours
overlap. Here it can be helpful to include the impact of all known
systematics in the contours; where contours do not overlap, the
model is incomplete and/or unidentified systematic errors are
present. Historically, the combination of mutually {\it inconsistent}
data sets has sometimes led to unphysically tight formal constraints.

Training sets can be useful to tune analyses. However, the biases
introduced by this training must be understood and accounted for. 
The inclusion of sufficient redundancy into experiments
(i.e. having more than one independent way to make a measurement) can
also enhance significantly the robustness of conclusions. In
cosmological studies, for example, we require more than a single way
to measure both the growth and expansion histories, e.g. clusters and
lensing, and SNIa and BAO, respectively.

A final modeling consideration is the impact of experimenter's bias:
the subjective, subconscious bias towards a result expected by an
experimenter. This is important where we are seeking to measure a
quantity for which prior expectations exist, e.g. $w$ or the growth
index, $\gamma$.  Evidence for experimenter's bias can be found
readily in the literature, e.g. in historical measurements of the
speed of light or the neutron lifetime (see \citealt{klein05} for a
review). The best approach to overcoming experimenter's bias is blind
analysis.

A range of blind analysis techniques are commonly employed in the
particle and nuclear physics communities. These are used to mask the
ability of scientists to determine quantities of interest until all
methods are finalized, the systematic errors that can be identified
have been, and the final measurement is ready to be made.  A technique
with good applicability to cosmological studies is {\it hiding the
answer}, wherein a fixed (unknown to the experimenter) offset is added
to the parameter(s) of interest, and only revealed once the final
measurements are made. Before un-blinding, it is advisable to think
through how to proceed afterward, and what additional checks to
employ.  Double-blind analyses, where two independent teams repeat the
same process (often used in biomedical research) target
additional sources of error. Such techniques can be used powerfully,
where resources allow.

\section{CONCLUSIONS}  \label{sec:conclusions}

Our article has
summarized the status of cluster cosmology and the methods used to
extract cosmological information from galaxy cluster observations. These
methods have been applied successfully to samples of tens to thousands
of massive clusters. Much of this work has been pioneered at X-ray and
optical wavelengths, but SZ surveys and gravitational lensing studies
are also now poised to play central roles.

The availability of powerful, new cluster surveys will help to address
profound mysteries such as the origin of cosmic acceleration,
inflation and the nature of gravity.  This work will require a
multiwavelength approach, combining the strengths of the available
techniques for finding clusters, calibrating their masses and
obtaining low-scatter mass proxies.  The analysis of these data will
require dedicated efforts by large teams of researchers. The demands
for follow-up observations will be high, and will require the support
of time allocation committees.

Building on the progress made, we can expect clusters to remain at the
forefront of cosmological work through the next decade.  By
combining cluster measurements with other, powerful probes such as
SNIa, CMB, BAO and cosmic shear, we can be optimistic of having the
precision, complementarity and redundancy required to allow robust
conclusions to be drawn.

\section*{Acknowledgments}

We are grateful to our colleagues for their many valuable insights. 
In particular, we thank 
Roger Blandford,
Hans B\"ohringer,
Marusa Brada{\v c},
Joanne Cohn,
Carlos Cunha, 
Megan Donahue,
Harald Ebeling, 
Andy Fabian,
Pat Henry,
Dragan Huterer, 
Andrey Kravtsov,
Dan Marrone,
Tim McKay,
Glenn Morris, 
Daisuke Nagai,
Aaron Roodman,
Eduardo Rozo,
Robert Schmidt, 
Tim Schrabback, 
Neelima Sehgal,
Aurora Simionescu, 
Alexey Vikhlinin,
Mark Voit,
Anja von der Linden, 
Risa Wechsler
and Norbert Werner. 
We particularly thank David Rapetti for his detailed input to
the material on gravity tests and non-gaussianities.  SWA was
supported in part by the U.S. Department of Energy under contract
number DE-AC02-76SF00515.  AEE acknowledges support from NASA
Astrophysics Theory Grant NNX10AF61G.  ABM was supported by an
appointment to the NASA Postdoctoral Program at the Goddard Space
Flight Center, administered by Oak Ridge Associated Universities
through a contract with NASA.  This research was supported in part by
the National Science Foundation under Grant No. NSF PHY05-51164.

\def \aap {A\&A} 
\def \aapr {A\&AR} 
\def \statisci {Statis. Sci.} 
\def \physrep {Phys. Rep.} 
\def \pre {Phys.\ Rev.\ E} 
\def \sjos {Scand. J. Statis.} 
\def \jrssb {J. Roy. Statist. Soc. B} 
\def \pan {Phys. Atom. Nucl.} 
\def \epja {Eur. Phys. J. A} 
\def \epjc {Eur. Phys. J. C} 
\def \jcap {J. Cosmology Astropart. Phys.} 
\def \ijmpd {Int.\ J.\ Mod.\ Phys.\ D} 
\def \nar {New Astron. Rev.}

\def \araa {ARA\&A}
\def \aj {AJ}
\def \aar {A\&AR}
\def \apj {ApJ}
\def \apjl {ApJL}
\def \apjs {ApJS}
\def \asl {Adv. Sci. Lett.} 
\def \mnras {MNRAS}
\def \nat {Nat}
\def \pasj {PASJ}
\def \pasp {PASP}
\def \science {Science}

\def \gca {Geochim.\ Cosmochim.\ Acta}
\def \npa {Nucl.\ Phys.\ A}
\def \plb {Phys.\ Lett.\ B}
\def \prc {Phys.\ Rev.\ C}
\def \prd {Phys.\ Rev.\ D}
\def \prl {Phys.\ Rev.\ Lett.}
\setlength{\bibsep}{0ex}

\end{document}